\pdfoutput=1
\documentclass[11pt,a4paper]{article}
\usepackage{jheppub}
\usepackage{graphicx}
\usepackage{subfigure}
\usepackage{lineno}
\usepackage{array}
\usepackage{amsmath}

\usepackage{preprintcover}  
\PreprintCoverPaperTitle{Measurement of the electroweak production of dijets in association with a $\boldsymbol{Z}$-boson and distributions sensitive to vector boson fusion in proton-proton collisions at $\boldsymbol{\sqrt{s} = 8}$~TeV using the ATLAS detector}  
\PreprintIdNumber{CERN-PH-EP-2013-227}  
\PreprintCoverAbstract{Measurements of fiducial cross sections for the electroweak production of two jets in association with a \Z-boson are presented. The measurements are performed using $20.3\,{\rm fb}^{-1}$ of proton-proton collision data collected at a centre-of-mass energy of $\sqrt{s}=8$~TeV by the ATLAS experiment at the Large Hadron Collider. The electroweak component is extracted by a fit to the
dijet invariant mass distribution in a fiducial region chosen to
enhance 
the electroweak contribution 
over the dominant background in which the jets are produced via the strong interaction. The electroweak cross sections measured in two fiducial regions are in good agreement with the Standard Model expectations and the background-only hypothesis is rejected with significance above the 5$\sigma$ level.
The electroweak process includes the vector boson fusion production of
a \Z-boson and the data are used to place limits on anomalous triple gauge boson couplings.
In addition, measurements of cross sections and differential distributions for inclusive
\Z-boson-plus-dijet production are performed in five fiducial regions,  each with different sensitivity to
the electroweak contribution. 
The results are corrected for detector effects and compared to predictions from the \sherpa\ and \powheg\ event generators.}  
\PreprintJournalName{JHEP}  

\def\W{\ensuremath{W}}
\def\Z{\ensuremath{Z}}
\def\Zg{\ensuremath{Z/\gamma^\ast}}
\def\Zdijet{\ensuremath{Zjj}}

\def\ifb{\ensuremath{\rm fb^{-1}}}
\def\ee{\ensuremath{e^+e^-}}
\def\mm{\ensuremath{\mu^+\mu^-}}
\def\Zee{\ensuremath{\Z\to\ee}}
\def\Zmm{\ensuremath{\Z\to\mm}}

\def\diww{\ensuremath{WW}}
\def\ttbar{\ensuremath{t\bar{t}}}
\def\tW{\ensuremath{t\W}}
\def\Wjets{\ensuremath{W$+$\mathrm{jets}}}

\def\lowpt{{\it baseline}}
\def\highpt{{\it high-\pt}}
\def\highmass{{\it high-mass}}
\def\search{{\it search}}
\def\control{{\it control}}

\def\mll{\ensuremath{m_{\ell\ell}}}
\def\mll{\ensuremath{m_{\ell\ell}}}
\def\dr{\ensuremath{\Delta R}}
\def\drjl{\ensuremath{\dr_{j,\ell}}}
\def\pt{\ensuremath{p_{\rm T}}}

\def\mjj{\ensuremath{m_{jj}}}
\def\deltay{\ensuremath{|\Delta{y}|}}
\def\ptsc{\ensuremath{\pt^{\rm balance}}}

\def\ptsceff{\ptsc\ {\it cut efficiency}}
 
\def\dphijj{\ensuremath{\Delta\phi(j,j)}} 
\def\gapfrac{{\it jet veto efficiency}}

\def\ptscfrac{\ptsc\ {\it cut efficiency}}
\def\avgnjet{\ensuremath{\langle N_{\rm jet}^{\rm gap}{\rangle}}}
\def\njet{\ensuremath{N_{\rm jet}^{\rm gap}}}
\def\dsdmjj{\ensuremath{\frac{1}{\sigma} \cdot \frac{{\rm d}\sigma}{{\rm d}\mjj}}} 
\def\dsddeltay{\ensuremath{\frac{1}{\sigma} \cdot \frac{{\rm d}\sigma}{{\rm d}\deltay}}} 
\def\dsddphijj{\ensuremath{\frac{1}{\sigma} \cdot \frac{{\rm d}\sigma}{{\rm d}|\dphijj|}}} 
 
\def\dsdptsc{\ensuremath{\frac{1}{\sigma} \cdot
    \frac{{\rm d}\sigma}{{\rm d}\ptsc}}} 
\def\dnjet{\ensuremath{\frac{1}{\sigma} \cdot \frac{{\rm d}\sigma}{{\rm d}\njet}}}

\def\antikt{anti-\ensuremath{k_{t}}}
\def\ctten{{\tt CT10}}
\def\nnpdf{{\tt NNPDF2.3}}
\def\mstw{{\tt MSTW2008nlo}}
\def\sherpa{{\tt Sherpa}} 
\def\alpgen{{\tt ALPGEN}} 
\def\herwig{{\tt HERWIG}} 
\def\jimmy{{\tt JIMMY}} 
 
\def\powheg{{\tt Powheg}}
\def\powhegbox{{\tt Powheg Box}}  
\def\pythia{{\tt PYTHIA 8}} 
\def\pythiasix{{\tt PYTHIA 6}}
\def\newk{\ensuremath{N_{\rm EW}}}
\def\nbck{\ensuremath{N_{\rm bkg}}}

\def\sewk{\ensuremath{\sigma_{\rm EW}}}
\def\sigmafid{\ensuremath{\sigma_{\rm fid}}}
\def\nobs{\ensuremath{N_{\rm obs}}}

\def\lint{\ensuremath{\int L \, {\rm d} t}}
\def\c{\ensuremath{\cal{C}}} 

\def\resultCaption{The data are shown as filled (black) circles. The vertical error bars show the size of the total uncertainty on the measurement, with tick marks used to reflect the size of the statistical uncertainty only. Particle-level predictions from \sherpa\ and \powheg\ are shown for combined strong and electroweak \Zdijet\ production (labelled as QCD$+$EW) by hatched bands, denoting the model uncertainty, around the central prediction, which is shown as a solid line. The predictions from \sherpa\ and \powheg\ for strong \Zdijet\ production (labelled QCD) are shown as dashed lines.}

\def\resultCaptionTwo{The data and theoretical predictions are presented in the same way as in figure~\ref{fig:unfolding-Mjj-Final}.}

\title{Measurement of the electroweak production of dijets in association with a $\boldsymbol{Z}$-boson and distributions sensitive to vector boson fusion in proton-proton collisions at $\boldsymbol{\sqrt{s} = 8}$~TeV using the ATLAS detector}
\author{ATLAS Collaboration}
\abstract{
Measurements of fiducial cross sections for the electroweak production of two jets in association with a \Z-boson are presented. The measurements are performed using $20.3\,{\rm fb}^{-1}$ of proton-proton collision data collected at a centre-of-mass energy of $\sqrt{s}=8$~TeV by the ATLAS experiment at the Large Hadron Collider. The electroweak component is extracted by a fit to the
dijet invariant mass distribution in a fiducial region chosen to
enhance 
the electroweak contribution 
over the dominant background in which the jets are produced via the strong interaction. The electroweak cross sections measured in two fiducial regions are in good agreement with the Standard Model expectations and the background-only hypothesis is rejected with significance above the 5$\sigma$ level.
The electroweak process includes the vector boson fusion production of
a \Z-boson and the data are used to place limits on anomalous triple gauge boson couplings.
In addition, measurements of cross sections and differential distributions for inclusive
\Z-boson-plus-dijet production are performed in five fiducial regions,  each with different sensitivity to
the electroweak contribution. 
The results are corrected for detector effects and compared to predictions from the \sherpa\ and \powheg\ event generators.
}

\begin{document}
\maketitle

\section{Introduction}
\label{sec:intro}

\begin{figure}
\centering
\subfigure[vector boson fusion]{\label{fig:vbfz}
\includegraphics[width=0.31\linewidth,height=0.25\linewidth]{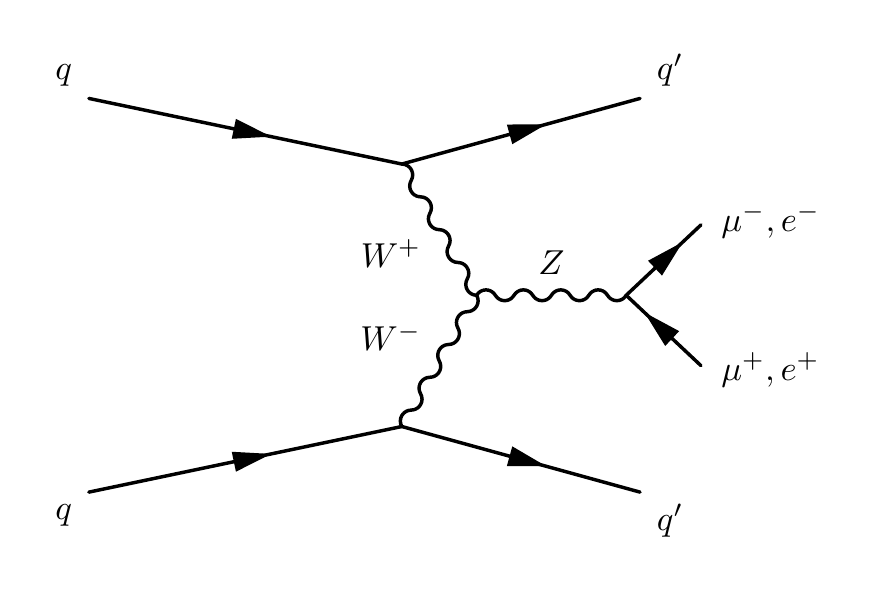}}
\subfigure[\Z-boson bremsstrahlung]{\label{fig:ewkz2}
\includegraphics[width=0.31\linewidth,height=0.25\linewidth]{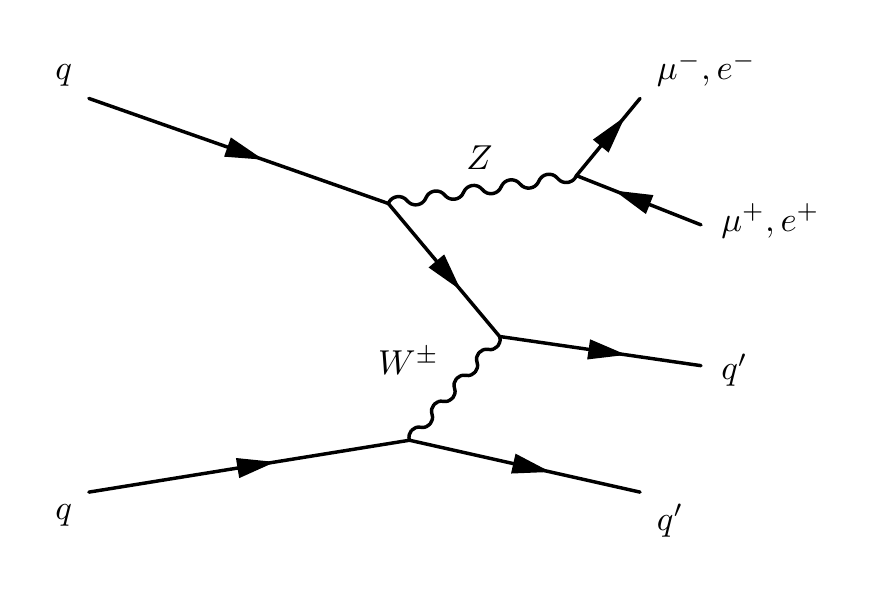}}
\subfigure[non-resonant $\ell^+ \ell^- jj$]{\label{fig:offShellZ}
\includegraphics[width=0.31\linewidth,height=0.25\linewidth]{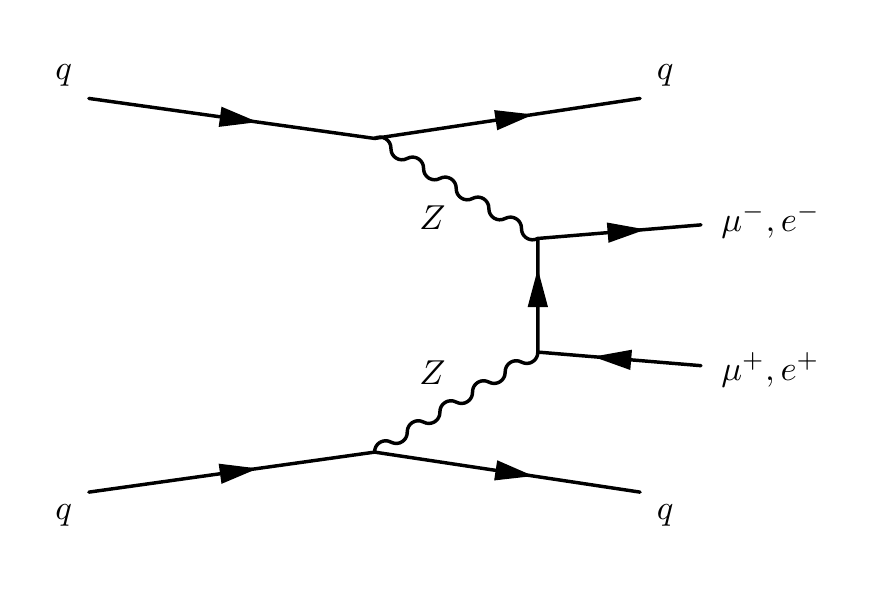}}

\caption{Representative leading-order Feynman diagrams for electroweak \Zdijet\ production at the LHC: (a) vector boson fusion (b) \Z-boson bremsstrahlung and (c) non-resonant $\ell^+ \ell^- jj$ production.}
\label{fig:ewkz}
\end{figure}

The dominant production mechanism for a leptonically decaying \Zg-boson\footnote{The contribution from $\gamma^*$ production in association with two jets is substantially reduced in this analysis by an invariant mass cut on the \Zg\ decay products.}  in association with two jets (\Zdijet) at the Large Hadron Collider (LHC) is  
via the Drell--Yan process, 
with the additional jets arising as a result of the strong interaction.
Production of \Zdijet\ events via the $t$-channel exchange of an electroweak gauge boson is a purely electroweak process and is therefore much rarer. 
Electroweak \Zdijet\ production in the leptonic decay channel is defined to include all contributions to $\ell^+ \ell^- jj$ production for which there is a $t$-channel exchange of an electroweak gauge boson~\cite{Oleari:2003tc,Arnold:2011wj}. These contributions include \Z-boson production via vector boson fusion (VBF), \Z-boson bremsstrahlung and non-resonant production, as shown in  
 figure~\ref{fig:ewkz}. The VBF process is of particular interest because of the similarity to the VBF production of a Higgs boson and the sensitivity to anomalous $WWZ$ triple gauge couplings.\footnote{The VBF process cannot be isolated due to a large destructive interference with the electroweak \Z-boson bremsstrahlung process. 
The contribution to the electroweak cross section from non-resonant $\ell^+ \ell^- jj$ production is less than 1\% after applying the selection criteria used in this analysis.} 

This paper presents two measurements of \Zdijet\ production using 20.3~\ifb\ of 
proton-proton collision data collected by the ATLAS experiment \cite{cite:atlas} at a centre-of-mass energy of $\sqrt{s}=8$~TeV:
\begin{enumerate}
  \item Measurements of fiducial cross sections and differential distributions of inclusive \Zdijet\ production. These measurements are performed in five fiducial regions with different sensitivity to the electroweak component. Inclusive \Zdijet\ production is dominated by the strong production process, an example of which is shown in figure \ref{fig:feyn}(a). The data therefore provide important constraints on the theoretical modelling of QCD-initiated processes that produce VBF-like topologies.\footnote{Inclusive \Zdijet\ production contains a small (percent-level) contribution from diboson events  (figure~\ref{fig:feyn}(b)).} 
  \item  Observation of electroweak \Zdijet\ production and
  measurements of the cross section in two fiducial regions. 
  Limits are also placed on anomalous $WWZ$ couplings.
\end{enumerate}
These measurements are performed using a combination of the \Zee\ and \Zmm\ decay channels. 

Using electroweak \Zdijet\ production as a probe of colour-singlet exchange and as a validation of the vector boson fusion process has been discussed extensively in the literature~\cite{Oleari:2003tc,Rainwater:1996ud,Khoze:2002fa}. 
A previous measurement by the CMS Collaboration showed evidence for electroweak \Zdijet\ production using proton-proton collisions at  $\sqrt{s}=7$~TeV \cite{Chatrchyan:2013jya}.  
However, due to large experimental and theoretical uncertainties associated with the modelling of strong \Zdijet\ production, the background-only hypothesis could be rejected only at the 2.6$\sigma$ level. The measurement presented in this paper constrains the modelling of strong \Zdijet\ production using a data-driven technique. This allows the background-only hypothesis to be rejected at greater than 5$\sigma$ significance and leads to a more precise cross section measurement for electroweak \Zdijet\ production.

\begin{figure}
  \centering
    \label{fig:qcdz}
         \subfigure[strong \Zdijet\ production]{
      \includegraphics[width=0.31\linewidth,height=0.25\linewidth]{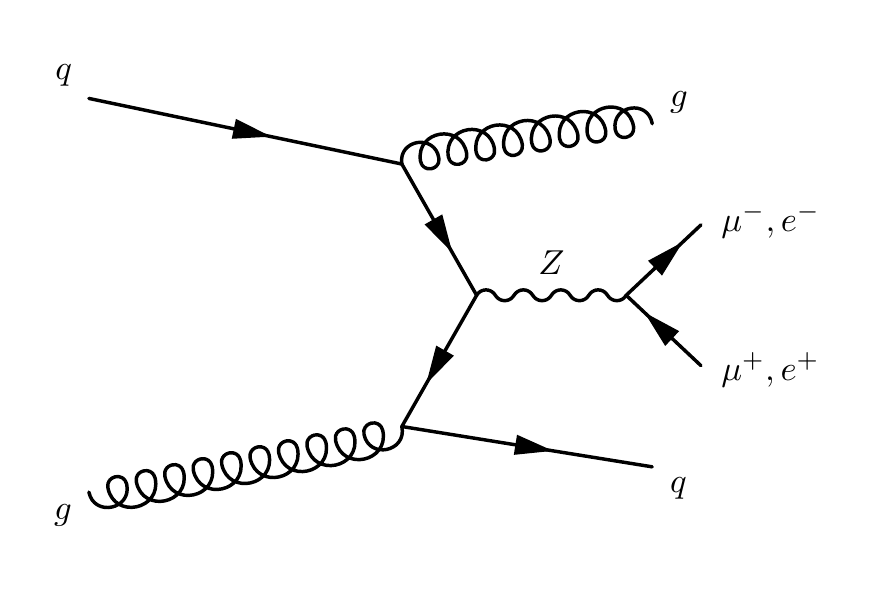}}\qquad\quad
          \subfigure[diboson-initiated]{
      \includegraphics[width=0.31\linewidth,height=0.25\linewidth]{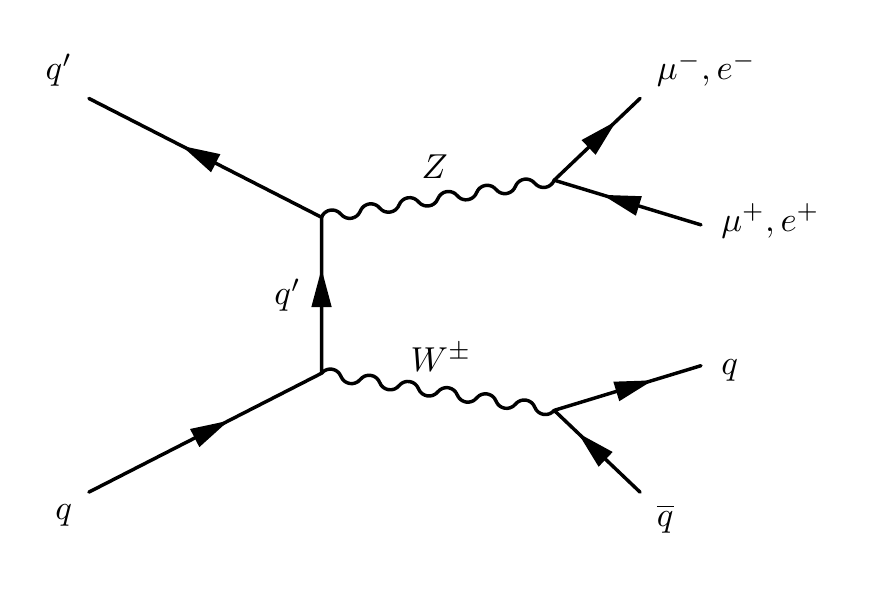}}
    \caption{ Examples of leading-order Feynman diagrams for (a) strong \Zdijet\ production and (b) diboson-initiated \Zdijet\ production.}
    \label{fig:feyn}
\end{figure}

\section{The ATLAS detector}
\label{sec:atlas}
The ATLAS detector is described in detail in ref.~\cite{cite:atlas}. Tracks and interaction vertices are reconstructed with the 
inner detector tracking system, which consists of a silicon pixel detector, a 
silicon microstrip detector and a transition radiation tracker, all 
immersed in a 2~T axial magnetic field, providing charged-particle
tracking in the pseudorapidity range $|\eta| < 2.5.$\footnote{ATLAS uses a right-handed coordinate system with its origin at the nominal interaction point (IP) in the centre of the detector and the $z$-axis along the beam pipe. The $x$-axis points from the IP to the centre of the LHC ring, and the $y$-axis points upward. Cylindrical coordinates $(r,\phi)$ are used in the transverse plane, $\phi$ being the azimuthal angle around the beam pipe. The pseudorapidity is defined in terms of the polar angle $\theta$ as $\eta=-\ln\tan(\theta/2)$. The rapidity is defined as $y=0.5\ln\left(\left( E + p_{\rm z} \right)/ \left( E - p_{\rm z} \right)\right)$, where $E$ and $p_{\rm z}$ refer to energy and longitudinal momentum, respectively.}
The ATLAS calorimeter system  provides fine-grained measurements of shower energy depositions over a wide range of $\eta$. 
An electromagnetic liquid-argon sampling
calorimeter covers the region  \mbox{$|\eta|<3.2$}. 
It is divided into a barrel part ($|\eta|
< 1.475$) and an endcap part ($1.375 < |\eta| < 3.2$). 
The hadronic barrel calorimeter \mbox{($|\eta|<1.7$)}  consists of
steel absorbers and active scintillator tiles. 
The hadronic endcap calorimeter \mbox{($1.5<|\eta|<3.2$)} and forward
 electromagnetic and hadronic calorimeters \mbox{($3.1<|\eta|<4.9$)}
use liquid argon as the active medium. 
The muon spectrometer comprises 
separate trigger and high-precision tracking chambers measuring the deflection of muons
in a magnetic field generated by superconducting air-core toroids. The precision chamber
system covers the region $|\eta| < 2.7$ with three layers of monitored drift tube chambers, complemented
by cathode strip chambers in the forward region, where the background is highest. The
muon trigger system covers the range $|\eta| < 2.4$ with resistive plate chambers in the barrel,
and thin gap chambers in the endcap regions. A three-level trigger system is used to select
interesting events \cite{Aad:2012xs}. The Level-1 trigger reduces the event rate to less than 75~kHz using hardware-based trigger algorithms acting on a subset of detector information. Two software-based trigger levels further reduce the event rate to about
400~Hz using the complete detector information.

\section{Event reconstruction and selection}
\label{sec:selection}

The measurement is performed using proton-proton collision data recorded at $\sqrt{s}=8$~TeV. The data were collected between April and December 2012 and correspond to an integrated luminosity of 20.3~\ifb. 
Events containing a \Z-candidate in the $\mu^+ \mu^-$ decay channel were retained for further analysis using a single-muon trigger, with muon transverse momentum, \pt, greater than 24~GeV or 36~GeV (isolation criteria are applied at the lower threshold). 
Events containing a \Z-candidate in the $e^+ e^-$ decay channel were retained  using a dielectron trigger
with both electrons having $\pt > 12$~GeV. 

In both decay channels, events are required to have a reconstructed collision vertex, defined by at least three associated inner detector tracks with {$\pt > 400$~MeV}. The primary vertex for each event is then defined as the collision vertex with the highest sum of squared transverse momenta of associated inner detector tracks. Finally, the event is required to be in a data-taking period in which the detector was fully operational.

Muon candidates are identified as tracks in the inner detector matched
and combined with track segments in the muon
spectrometer~\cite{Aad:2010yt}. They are required to have $\pt > 25$~GeV and 
 $|\eta|< 2.4$.  
In order to suppress backgrounds, track quality
requirements are imposed for muon identification, and impact parameter requirements ensure that the muon candidates originate
from the primary vertex. The muon candidates are also required to be isolated: the scalar sum of the
$\pt$ of tracks with $\dr < 0.2$ around the muon track is required to
be less than 10\%\ of the \pt\ of the muon. The radius parameter is defined as $(\dr)^2=(\Delta\eta)^2+(\Delta\phi)^2$. 

Electron candidates are reconstructed from clusters of energy in the
electromagnetic calorimeter matched to inner detector tracks. They
are required to have $\pt > 25$~GeV and $|\eta| < 2.47$, but excluding the transition 
regions between the barrel and endcap electromagnetic calorimeters,
$1.37 < |\eta| < 1.52$. 
The electron candidates must satisfy a set of `medium' selection 
criteria~\cite{Aad:2011mk} that have been reoptimised for the higher rate of proton-proton collisions per beam crossing (pileup) observed in the 2012 data.   
Impact parameter requirements ensure the electron candidates originate from the primary vertex.

Jets are reconstructed with the \antikt\ jet algorithm~\cite{Cacciari:2008gp} with a jet-radius
parameter of 0.4.
The input objects to the jet algorithm are three-dimensional topological clusters of
energy in the calorimeter~\cite{cite:clustering}.
The resultant jet energies are initially corrected to account for soft energy arising from
pileup \cite{jet-pile}. The energy and direction of each jet is then corrected
for calorimeter non-compensation, detector material and the transition between calorimeter regions, 
using a combination of MC-derived calibration
constants and in situ data-driven calibration constants \cite{Aad:2011he,jes2011}.
Jets are required to have $\pt>25$~GeV and $|y| < 4.4$, where $y$ is the rapidity. 
Additional data quality requirements are imposed to minimise the effect of noisy calorimeter cells.
To suppress
jets from overlapping proton-proton collisions, the jet
vertex fraction (JVF) is used to identify jets
from the primary interaction. Tracks are associated with jets using ghost-association \cite{Cacciari:2008gn}, where tracks are assigned negligible momentum and clustered to the jet using the \antikt\ algorithm. The JVF is subsequently defined as
the scalar summed transverse momentum of associated
tracks from the primary vertex divided by the summed
transverse momentum of associated tracks from all vertices. Each jet with $\pt<50$~GeV and $|\eta|<2.4$ is required to have $\rm{JVF}>0.5$. Finally, jets are required to be well separated from any of the selected leptons (jets within a cone of radius $\dr < 0.3$ in $\eta$--$\phi$ space around any lepton are removed from the analysis).

\section{Theoretical predictions}
\label{sec:theory}

Theoretical predictions for strong and electroweak \Zdijet\ production are obtained using the \powhegbox\ \cite{Nason:2004rx,Frixione:2007vw,Alioli:2010xd} and \sherpa\ v1.4.3 \cite{Gleisberg:2008ta} event generators. The small contribution from diboson events is estimated using \sherpa.

\sherpa\  is a matrix-element plus parton-shower generator that provides $Z+n$-parton predictions ($n=0,1,2...$)  at leading-order (LO) accuracy in perturbative QCD. The CKKW method is used to combine the various final-state topologies and match to the parton shower \cite{Catani:2001cc}. 
Electroweak \Zdijet\ production is accurate at LO for two and three partons in the final state. Strong \Zdijet\ production is accurate at LO for two, three and four partons in the final state, and the \Z-boson plus zero and one parton configurations are also produced (at LO accuracy) to allow contributions from double parton scattering to be included. Diboson-initiated \Zdijet\ production ($ZV$) is generated with up to three partons in addition to the partonically decaying boson. For all production channels, parton-shower, hadronisation and multiple parton interaction (MPI) algorithms create the fully hadronic final state. 
The \sherpa\ predictions are produced using the \ctten~\cite{Lai:2010vv}  parton distribution functions (PDFs) and the default generator tune for underlying event activity.

The \powhegbox\  provides \Zdijet\ predictions at next-to-leading-order (NLO) accuracy in perturbative QCD for both electroweak and strong production \cite{Jager:2012xk,Schissler:2013nga,Re:2012zi,Campbell:2013vha}. The fully hadronic final state is produced by interfacing the \powhegbox\ to \pythiasix~\cite{Sjostrand:2006za}, which provides parton showering, hadronisation and MPI. These particle level-predictions are referred to as \powheg\ in the remainder of this paper. The \powheg\ predictions are produced using the \ctten\ PDFs and the Perugia 2011 tune~\cite{Skands:2010ak} for underlying event activity. The strong \Zdijet\ sample was generated with the MiNLO  feature~\cite{Hamilton:2012np}, which also produces \Z\ plus zero and one jet events at LO accuracy and allows contributions to \Zdijet\ production from double parton scattering to be evaluated. 

Theoretical uncertainties are estimated for the strong and electroweak \Zdijet\ predictions from \sherpa\ and \powheg. 
Scale uncertainties on all theoretical predictions are estimated by varying the renormalisation and factorisation scales (separately) by a factor of 0.5 and 2.0. Additional modelling uncertainties in the \sherpa\ prediction arise from the choice of CKKW matching scale, the choice of parton-shower scheme, and the MPI-modelling.\footnote{The uncertainty in the CKKW matching is determined by increasing the matching scale by a factor of two. Uncertainties associated with the parton shower are estimated by changing the recoil strategy for dipoles with initial-state emitter and final-state spectator, from the default \cite{Hoeche:2009xc} to that proposed in ref. \cite{Schumann:2007mg}. The uncertainty due to a potential mismodelling of the underlying event is estimated by increasing the MPI activity uniformly by 10\% \cite{Hoeche:2012fm}, or changing the shape of the MPI spectrum such that more jets from double parton scattering are produced. The parameter variations for the latter are:  {\tt SIGMA\_ND\_FACTOR=0.14} and {\tt SCALE\_MIN=4.0}} Similar modelling uncertainties in the \powheg\ prediction are estimated using the suite of Perugia 2011 tunes \cite{Skands:2010ak}, with the largest effects coming from those tunes with increased/decreased parton-shower activity or increased MPI activity.

The use of independent strong and electroweak \Zdijet\ samples relies on the fact that interference between the two processes is colour and kinematically suppressed, and therefore negligible. Interference between the strong and electroweak processes has been proven to be negligible for the production of the Higgs boson in association with two jets~($Hjj$)~\cite{Andersen:2006ag,Ciccolini:2007ec,Andersen:2007mp,Bredenstein:2008tm}. Although no such studies have been performed for the electroweak production of a \Zdijet\ system, the interference effects arise from the same sources as $Hjj$ production and should therefore be small. The assumption of negligible interference is checked for this measurement using a combined strong/electroweak \sherpa\ sample that is accurate to leading order for \Zdijet\ production. This combined sample includes electroweak and strong \Zdijet\ matrix elements at the amplitude level and thereby calculates the interference between them. The interference contribution is established by subtracting the strong-only and electroweak-only \Zdijet\ components. The impact of interference on inclusive \Zdijet\ cross sections and distributions is found to be negligible. The impact of interference on the extraction of the electroweak \Zdijet\ component is at the few-percent level and is discussed in more detail in section~\ref{sec:fits}.

\section{Monte Carlo simulation}
\label{sec:mc}

Event generator samples are passed through \texttt{GEANT4} \cite{geant-1,geant-2} for a full simulation~\cite{Aad:2010ah} of the ATLAS detector and 
reconstructed with the same analysis chain as used for the data.  
Pileup is simulated by overlaying inelastic proton-proton interactions produced with \pythia~\cite{Sjostrand:2007gs}, tune
A2~\cite{cite:atwo} with the MSTW2008LO PDF set~\cite{Martin:2009iq}.

Strong and electroweak \Zdijet\ simulated events are produced using the \sherpa\ samples discussed in section~\ref{sec:theory}. The samples are normalised to reproduce the NLO calculations for \Zdijet\ production obtained from \powheg; the NLO K-factors are 1.23 and 1.02 for the strong and electroweak samples, respectively. The contribution from $ZV$ events is also produced using \sherpa. To cross-check aspects of the theoretical modelling of strong \Zdijet\ production at the detector level, a small simulated sample of \Zdijet\ events is produced using \alpgen~\cite{Mangano:2002ea}. \alpgen\ is a leading-order matrix-element generator that produces \Z-boson events with up to five additional partons in the final state and is interfaced to \herwig~\cite{Corcella:2000bw,Corcella:2002jc} and \jimmy~\cite{Butterworth:1996zw} to add the parton shower, hadronisation and MPI (AUET2 tune \cite{ue-tune}).

Background events stemming from \ttbar\ and single-top production are produced using \texttt{MC}@\texttt{NLO} v4.03~\cite{Frixione:2002ik} interfaced to \herwig\ and \jimmy\ (AUET2 tune). The generator modelling of \ttbar\ events is cross-checked with a simulated sample produced using the \powhegbox\ interfaced to \pythiasix\ (Perugia 2011 tune). The \ttbar\ samples are normalised to a next-to-next-to-leading-order (NNLO) calculation in QCD including resummation of next-to-next-to-leading-logarithmic (NNLL) soft gluon terms \cite{Czakon:2011xx}.  The backgrounds arising from $WW$ and \Wjets\  events are produced using \sherpa.

\section{Fiducial cross-section measurements of inclusive \Zdijet\  production}
\label{sec:total}

The cross section for inclusive \Zdijet\ production, $\sigmafid$, is  defined by
\begin{equation} 
   \sigmafid = \frac{\nobs - N_{\rm bkg}}{\lint \, \, {\cdot} \, \, \c}\label{eqn:xsecDefn}
\end{equation}
where $\nobs$ is the number of events observed in the data passing the
reconstruction-level selection criteria, $N_{\rm bkg}$ is the expected
number of background events, 
$\lint$ is the integrated luminosity and $\c$ is a correction factor
accounting for differences in event yields at reconstruction and particle level due to detector inefficiencies and resolutions.

The particle-level prediction is constructed using final-state particles with mean lifetime ($c\tau$) longer than 10~mm. Leptons are defined as objects constructed from the four-momentum combination of an electron (or muon) and all nearby photons in a cone of radius $\dr = 0.1$ in $\eta$--$\phi$ space centred on the lepton (so-called `dressed leptons'). Leptons are required to have $\pt > 25$ GeV and  $|\eta| < 2.47$. Jets are reconstructed using the \antikt\ algorithm with a jet-radius parameter of 0.4. Jets are required to have $\pt>25$~GeV, $|y| < 4.4$ and $\drjl \geq 0.3$, where \drjl\ is the distance in $\eta$--$\phi$ space between the jet and the selected leptons.

The  cross section for inclusive \Zdijet\ production is measured in five fiducial regions, each with different sensitivity to the electroweak component of \Zdijet\  production. A summary of the selection criteria for each fiducial region is given in table \ref{tab:criteria}. The \search\ region is chosen to optimise the expected significance when extracting the electroweak \Zdijet\ component, and is defined as:
\begin{itemize}
\item A $Z$-boson candidate, defined as exactly two oppositely charged, same-flavour leptons with a dilepton invariant mass of $81\leq \mll < 101$~GeV. 
\item The transverse momentum of the dilepton pair must satisfy $p_{\rm T}^{\ell\ell}  > 20$~GeV.
\item At least two jets that satisfy $p_{\rm T}^{j_1} > 55$~GeV, $p_{\rm T}^{j_2}  > 45$~GeV, where ${j_1}$ and ${j_2}$ label the highest and second highest transverse momentum jets in the event. 
\item The invariant mass of the two leading jets is required to satisfy $\mjj > 250$~GeV. 
\item No additional jets with $\pt > 25$~GeV in the rapidity interval between the two leading jets.
\item  The normalised transverse-momentum balance between the two leptons and the two highest transverse momentum jets, $\ptsc$, is required to be less than 0.15. The \ptsc\ is defined as
\begin{equation}\label{eq:ptsc}
\ptsc = \frac{ \left|\vec{p}^{\, \ell_1}_{\rm T} + \vec{p}^{\, \ell_2}_{\rm T} + \vec{p}^{\, j_1}_{\rm T} + \vec{p}^{\, j_2}_{\rm T}\right|}{\left|\vec{p}^{\, \ell_1}_{\rm T}\right| + \left|\vec{p}^{\, \ell_2}_{\rm T}\right| + \left|\vec{p}^{\, j_1}_{\rm T}\right| + \left|\vec{p}^{\, j_2}_{\rm T}\right|}, 
\end{equation}
where $\vec{p}^{\, i}_{\rm T}$ is the transverse momentum vector of object $i$, and $\ell_1$ and $\ell_2$ label the two leptons that define the \Z-boson candidate.
\end{itemize}
The tight cut on the dilepton invariant mass is chosen to suppress backgrounds from events that do not contain a \Z-boson. The high-\pt\ requirement on the two leading jets and the veto on additional jet activity preferentially suppress strong \Zdijet\ production with respect to electroweak \Zdijet\ production. The dijet invariant mass criterion removes a large fraction of diboson events. The \ptsc\ and $p_{\rm T}^{\ell\ell}$ requirements reduce the impact of those events containing jets that originate from pileup interactions or multiple parton interactions. Events with poorly measured jets are also removed by the \ptsc\ requirement.

\begin{table}[t]
\caption{Summary of the selection criteria that define the fiducial regions. `Interval jets' refer to the selection criteria applied to the jets that lie in the rapidity interval bounded by the dijet system. }
 \label{tab:criteria}
\newlength{\critlen}\settowidth{\critlen}{$\pt^{\rm balance,3}<0.15$}
\resizebox{0.95\textwidth}{!}{\renewcommand{\arraystretch}{1.65}
\begin{tabular}{@{} c *{5}{|>{\centering\arraybackslash\rule{0mm}{0mm}}m{\critlen}} }
\hline\hline
 Object & \lowpt & \highmass & \search & \control & \highpt \\
\hline
 Leptons       & \multicolumn{5}{c}{$|\eta^\ell|<2.47$, $\pt^\ell>25$~GeV} \\
\hline
 ~Dilepton pair~ & \multicolumn{5}{c}{$81\leq m_{\ell\ell}\leq 101$~GeV} \\
\cline{2-6}
               & \multicolumn{2}{c}{---} & \multicolumn{2}{|c|}{$\pt^{\ell\ell}>20$~GeV} & --- \\
\hline
 Jets          & \multicolumn{5}{c}{$|y^j|<4.4$, $\Delta R_{j,\ell}\geq 0.3$} \\
\cline{2-6}
               & \multicolumn{4}{c|}{$\pt^{j_1}>55$~GeV} & $\pt^{j_1}>85$~GeV \\
\cline{2-6}
               & \multicolumn{4}{c|}{$\pt^{j_2}>45$~GeV} & $\pt^{j_2}>75$~GeV \\
\hline
 Dijet system  & --- & $m_{jj}>1$~TeV & \multicolumn{2}{c|}{$m_{jj}>250$~GeV} & --- \\
\hline
 Interval jets & \multicolumn{2}{c|}{---} & $N_{\rm jet}^{\rm gap}=0$ & $N_{\rm jet}^{\rm gap}\geq 1$ & --- \\
\hline
 \Zdijet\ system & \multicolumn{2}{c|}{---} & $\pt^{\rm balance}<0.15$ & $\pt^{\rm balance,3}<0.15$ & --- \\
\hline\hline
    \end{tabular}}
\end{table}

The \control\ region criteria are chosen in order to suppress the electroweak \Zdijet\ contribution, allowing the theoretical modelling of strong \Zdijet\ production to be evaluated. The selection criteria are similar to the \search\ region, with two modifications: (i) at least one additional jet with $\pt > 25$~GeV must be present in the rapidity interval between the two leading jets. (ii) the transverse-momentum balancing variable is redefined to use the two leptons, the two highest transverse momentum jets, and the highest transverse momentum jet in the rapidity interval bounded by the two leading jets. This variable, $\pt^{\rm balance,3}$, is defined in an analogous way to the \ptsc\ variable in eq.~(\ref{eq:ptsc}), but incorporating the additional jet in the numerator and denominator.

The remaining three fiducial regions are chosen with fewer selection criteria, in order to study inclusive \Zdijet\ production in simpler topologies. The \lowpt\ region is defined as containing a \Z-boson candidate plus at least two jets with $p_{\rm T}^{j_1} > 55$~GeV and $p_{\rm T}^{j_2}  > 45$~GeV. This is the most inclusive fiducial region examined and contains the events in all other fiducial regions. The \highmass\ region is chosen as the subset of these events that have $\mjj > 1$~TeV. The \highpt\ region is defined as containing a \Z-boson candidate plus at least two jets with $p_{\rm T}^{j_1} > 85$~GeV and $p_{\rm T}^{j_2}  > 75$~GeV. The \highmass\ and \highpt\ regions are useful to probe the impact of the electroweak \Zdijet\ process, which produces a harder jet transverse momentum and harder dijet invariant mass than the strong \Zdijet\ process.

The simulation-based correction factor (\c) used to correct the measurement to the particle level is estimated using the \sherpa\ \Zdijet\ samples. The correction factor is found to lie between 0.80 and 0.92 in the muon channel, and between 0.64 and 0.71 in the electron channel, depending on the fiducial region. The difference between the channels arises primarily from the different efficiency in reconstructing and identifying electrons and muons in the detector.

\subsection{Backgrounds}
\label{sec:backgrounds}

The contributions from the \ttbar, \diww,  \tW\ and \Wjets\ background processes are obtained by applying the analysis chain to the dedicated simulated  samples introduced in section~\ref{sec:mc}. The multijet background contributes if two jets are misidentified as leptons or contain leptons from $b$- or $c$-hadron decays. A multijet sample is obtained from the data by reversing some of the electron identification criteria for the analysis in the electron channel, or reversing the muon isolation criteria for the analysis in the muon channel. The normalisation of the multijet sample in each fiducial region is then obtained by a two-component template fit to the dilepton invariant mass distributions, using the multijet template and a template formed from all other processes.

Table~\ref{tab:mc-predictions} shows the composition by
percentage of the predicted signal and background processes in each of
the five fiducial regions. The event sample is dominated by processes producing a \Z-boson in the final state. The dominant  background to inclusive \Zdijet\ production is from \ttbar\ production.

\begin{table}
\centering
\caption{Process composition (\%) for each fiducial region  for the combined muon and electron channels. The  strong \Zdijet, electroweak \Zdijet, diboson, \ttbar, \Wjets\ and \tW\ rates are estimated by running the analysis chain over MC samples fully simulated in the ATLAS detector. The multijet background is estimated using a data-driven technique.}
  \label{tab:mc-predictions}
  \begin{tabular}{@{} l r@{.}l r@{.}l r@{.}l r@{.}l c r@{}l}
  \hline\hline
  ~ & \multicolumn{11}{c}{Composition (\%)}\rule{0mm}{4mm} \\
  ~Process & \multicolumn{2}{c}{\lowpt} & \multicolumn{2}{c}{\highpt} & \multicolumn{2}{c}{\search} & \multicolumn{2}{c}{\control} & \multicolumn{3}{c}{\highmass}\rule{0mm}{4mm} \\
  \hline
  ~Strong \Zdijet\     &   95 & 8 &   94 & 0 &   94 & 7 &   96 & 0 & &       85 & \rule{0mm}{4mm} \\
  ~Electroweak \Zdijet &    1 & 1 &    2 & 1 &    4 & 0 &    1 & 4 & &       12 & \rule{0mm}{4mm} \\
  ~$WZ$ and $ZZ$       &    1 & 0 &    1 & 3 &    0 & 7 &    1 & 4 & &        1 & \rule{0mm}{4mm} \\
  \hline
  ~\ttbar              &    1 & 8 &    2 & 2 &    0 & 6 &    1 & 0 & &        2 & \rule{0mm}{4mm} \\
  ~Single top          &    0 & 1 &    0 & 1 & $<0$ & 1 & $<0$ & 1 & & $<0.1$ & \rule{0mm}{4mm} \\
  ~Multijet            &    0 & 1 &    0 & 2 & $<0$ & 1 &    0 & 2 & & $<0.1$ & \rule{0mm}{4mm} \\
  ~\diww, $\Wjets$            & $<0$ & 1 & $<0$ & 1 & $<0$ & 1 & $<1$ & 1 & & $<0.1$ & \rule{0mm}{4mm} \\
 \hline\hline
 \end{tabular}
\end{table}

\subsection{Systematic uncertainties}
\label{sec:systematics}

The systematic uncertainties on the lepton reconstruction, identification, isolation and trigger efficiencies, as well as the lepton momentum scale and resolution, are defined in refs.~\cite{Aad:2011mk,cite:muon-eff}. The total impact of the lepton-based systematic uncertainties on the cross-section measurement in each fiducial region is typically 3\% in the electron channel and 2\% in the muon channel. The uncertainty on the integrated luminosity is estimated to be 2.8\%, using the methodology detailed in ref.~\cite{Aad:2013ucp} for beam-separation scans performed in November 2012.

The jet energy scale (JES) and jet energy resolution (JER) uncertainties account for differences between the calorimeter response in simulation and data \cite{Aad:2011he,Aad:2012ag,jes2011}. The JES uncertainty for 2012 data includes components for the soft-energy pileup corrections, the MC-based/data-driven calibration constants, the calibration of forward jets, and the unknown jet flavour.\footnote{The jet flavour uncertainty refers to the different calorimeter response for quark-initiated and gluon-initiated jets.} The uncertainty due to JES is the dominant systematic uncertainty, ranging from 7.5\% in the \search\ region to 19\%  in the \highmass\ region. The uncertainty due to JER is much smaller, ranging from 0.1\% in the \highpt\ region to 5\% in the \highmass\ region.

The JVF cut removes a fraction of the jets associated with the primary vertex in addition to the jets originating from pileup interactions. Any mismodelling of the JVF distribution therefore
introduces a possible bias in the shape and normalisation of the distributions. A systematic uncertainty is determined after repeating the full analysis using modified JVF cuts that cover possible differences in efficiency between
data and simulation.  The JVF cuts are varied by $\pm$0.03 and the uncertainty due to JVF modelling is found to be between 0.2\% and 2.8\% in the \lowpt\ and \control\ regions, respectively.

Hard jets originating from the additional (pileup) interactions are also reconstructed in the event and any mismodelling of pileup jets in the simulation is a source of systematic uncertainty. In the central calorimeter region, 
the JVF cut removes a large fraction of these jets. In the forward calorimeter regions 
(outside the inner detector acceptance), no track-based cut can be applied to remove these pileup jets. 
To estimate the impact of a possible mismodelling of the jets originating from pileup, the analysis is repeated using the simulated samples after removing pileup jets, defined as those reconstruction-level jets that are not matched ($\Delta R \leq 0.3$) to a particle-level jet from the hard scattering process with $\pt > 10$~GeV. 
The effect of pileup on each cross section measurement is then determined by comparing the reconstruction-level event yield obtained in simulation after applying jet matching to that obtained with no matching applied.
Studies of the central jet transverse momentum in a pileup-enhanced sample (${\rm JVF} <0.1$), and the transverse energy density in the forward region of the detector~\cite{Aad:2012mfa}, indicate that the simulation
could be mismodelling the number of pileup jets by up to 35\%. 
The difference between the reconstruction-level event yields obtained with and without jet matching is
therefore scaled by 0.35 and taken as  a two-sided systematic uncertainty on the fiducial cross section. The impact on the final measurement is not large, ranging from less than 0.1\% in the \search\ region to 2.3\% in the \lowpt\ region.

In addition to the experimental uncertainties discussed above, systematic uncertainties on the correction factor, \c, due to possible event generator mismodelling are evaluated. These generator modelling uncertainties are estimated by reweighting the events, at reconstruction level and particle level, such that the kinematic distributions in the simulation match those observed in the data.  The reweighting is carried out for the two lepton transverse momenta and pseudorapidities, the two leading jet transverse momenta and pseudorapidities, and the variables used to define the fiducial regions. 
The correction factor is re-evaluated for each reweighting and the difference with respect to the nominal correction factor is taken as a theory modelling uncertainty. 
The uncertainty on the correction factor from theoretical modelling ranges from 1\% in the \lowpt\ region to 6.6\% in the \highmass\ region. 

\begin{figure}[t]
  \begin{center}
    \subfigure[] {
      \includegraphics[width=0.47\textwidth]{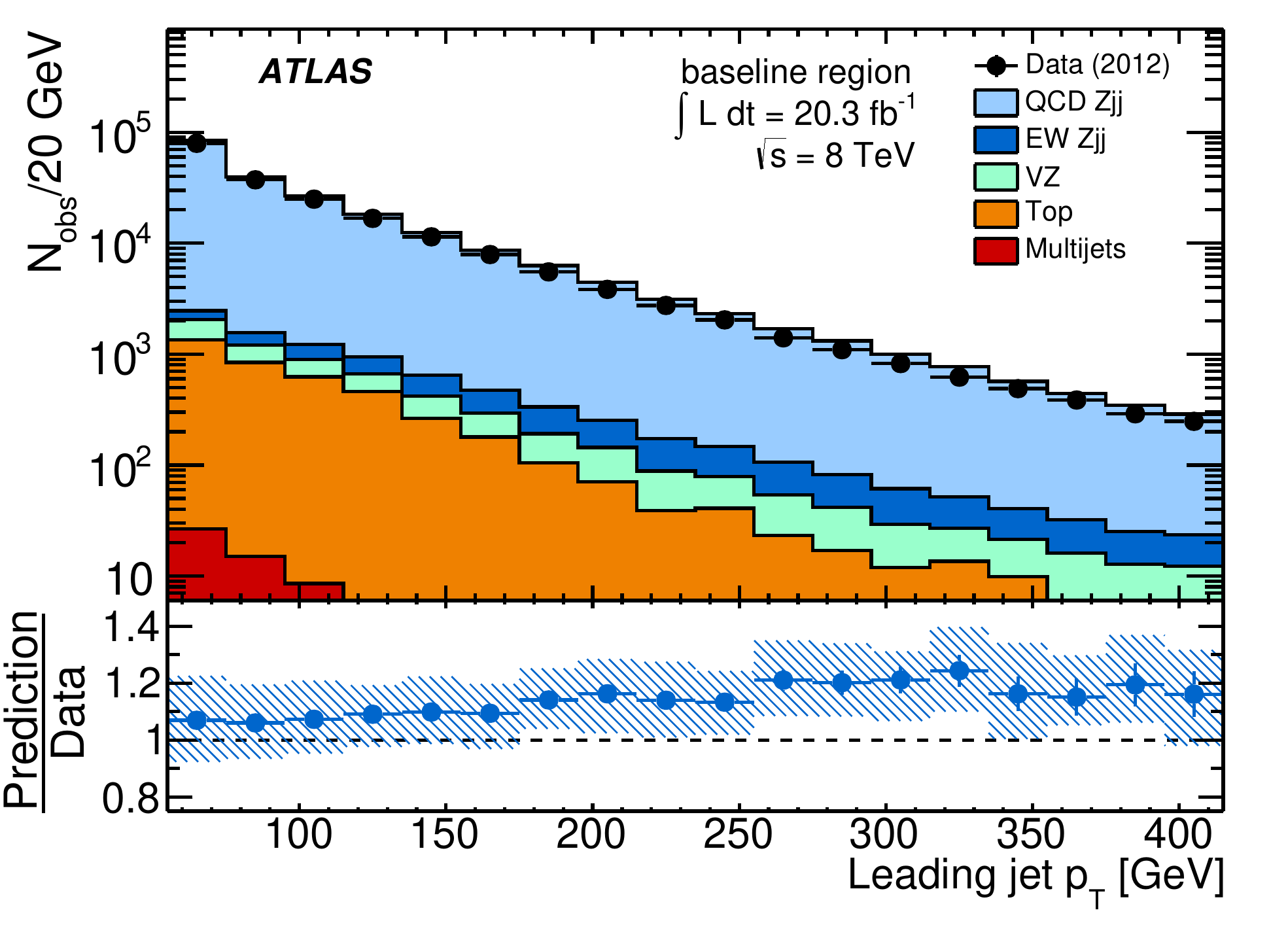}\quad
      \label{fig:jpt1}
    }
    \subfigure[] {
      \includegraphics[width=0.47\textwidth]{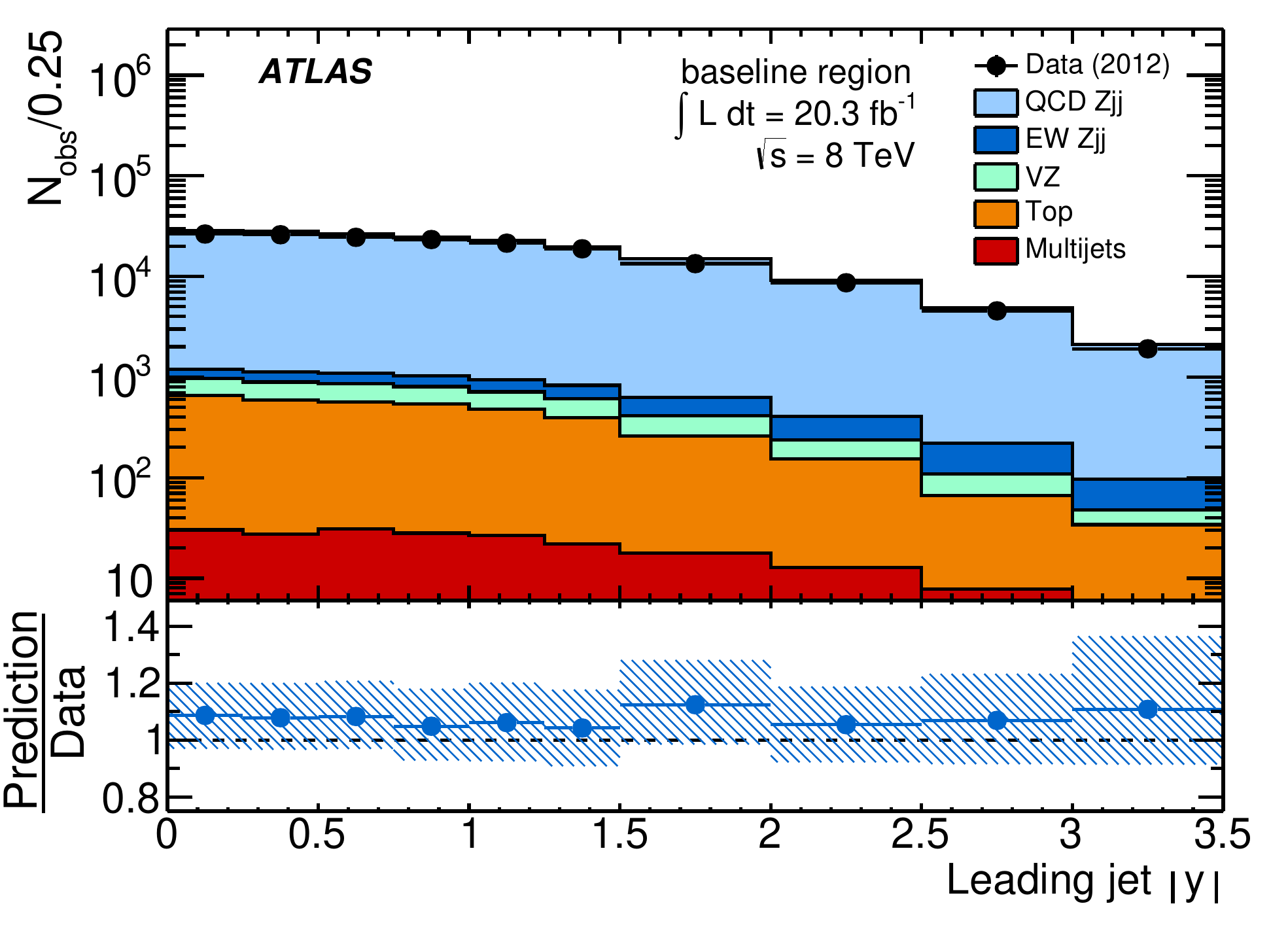}
      \label{fig:jy1}
    }
    \subfigure[] {
      \includegraphics[width=0.47\textwidth]{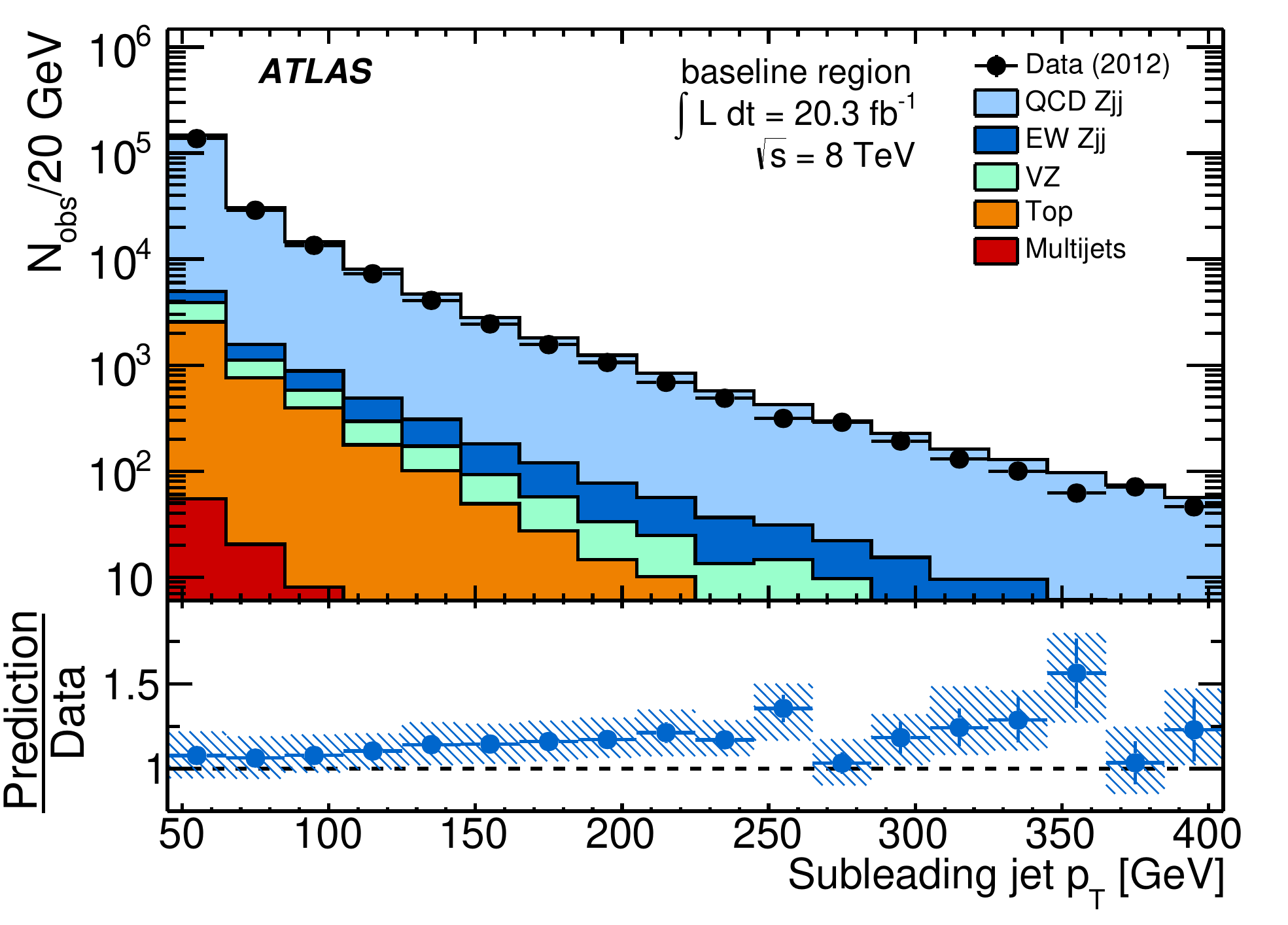}\quad
      \label{fig:jpt2}
    }
    \subfigure[] {
      \includegraphics[width=0.47\textwidth]{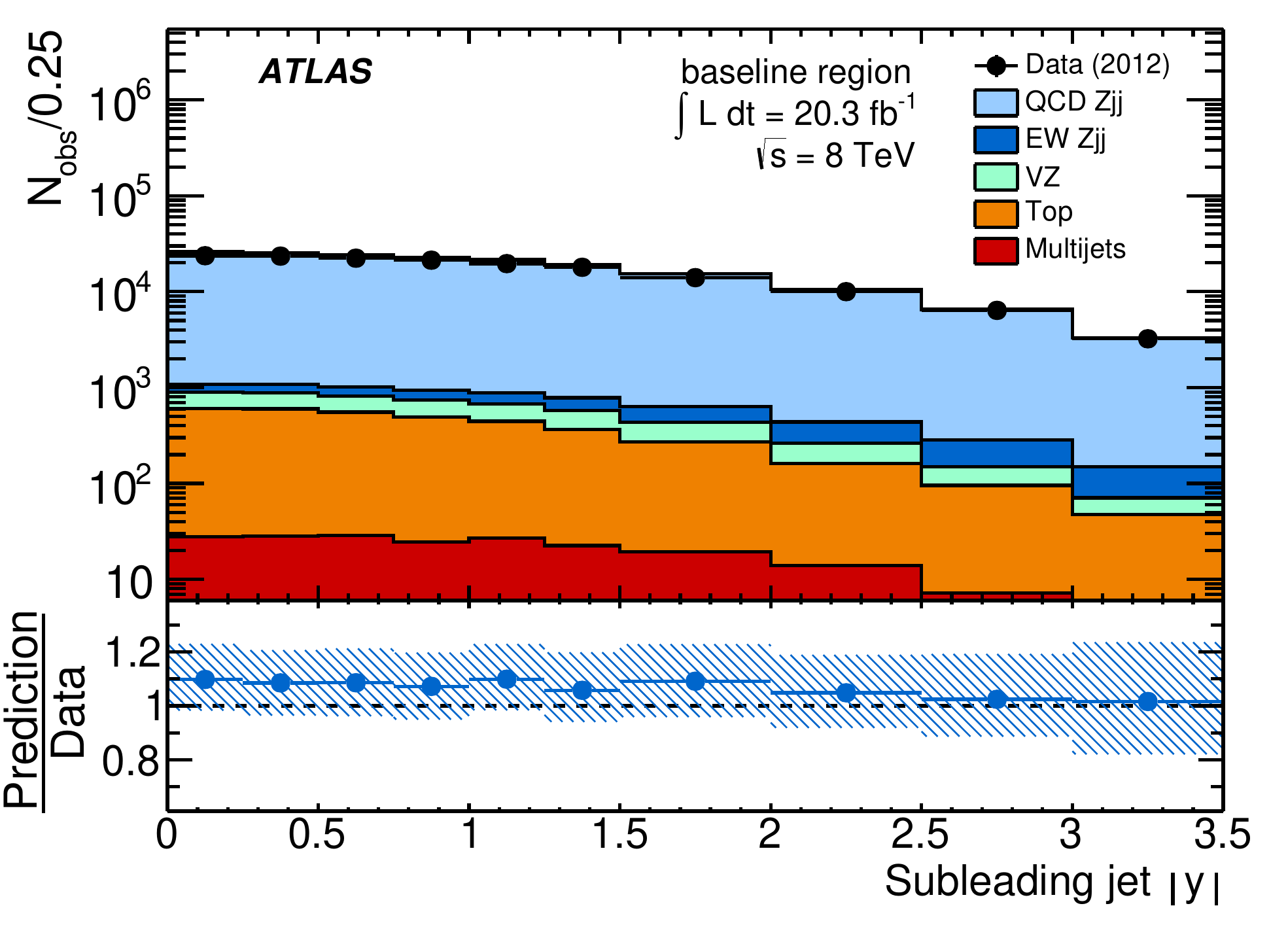}
      \label{fig:jy2}
    }
    \subfigure[] {
      \includegraphics[width=0.47\textwidth]{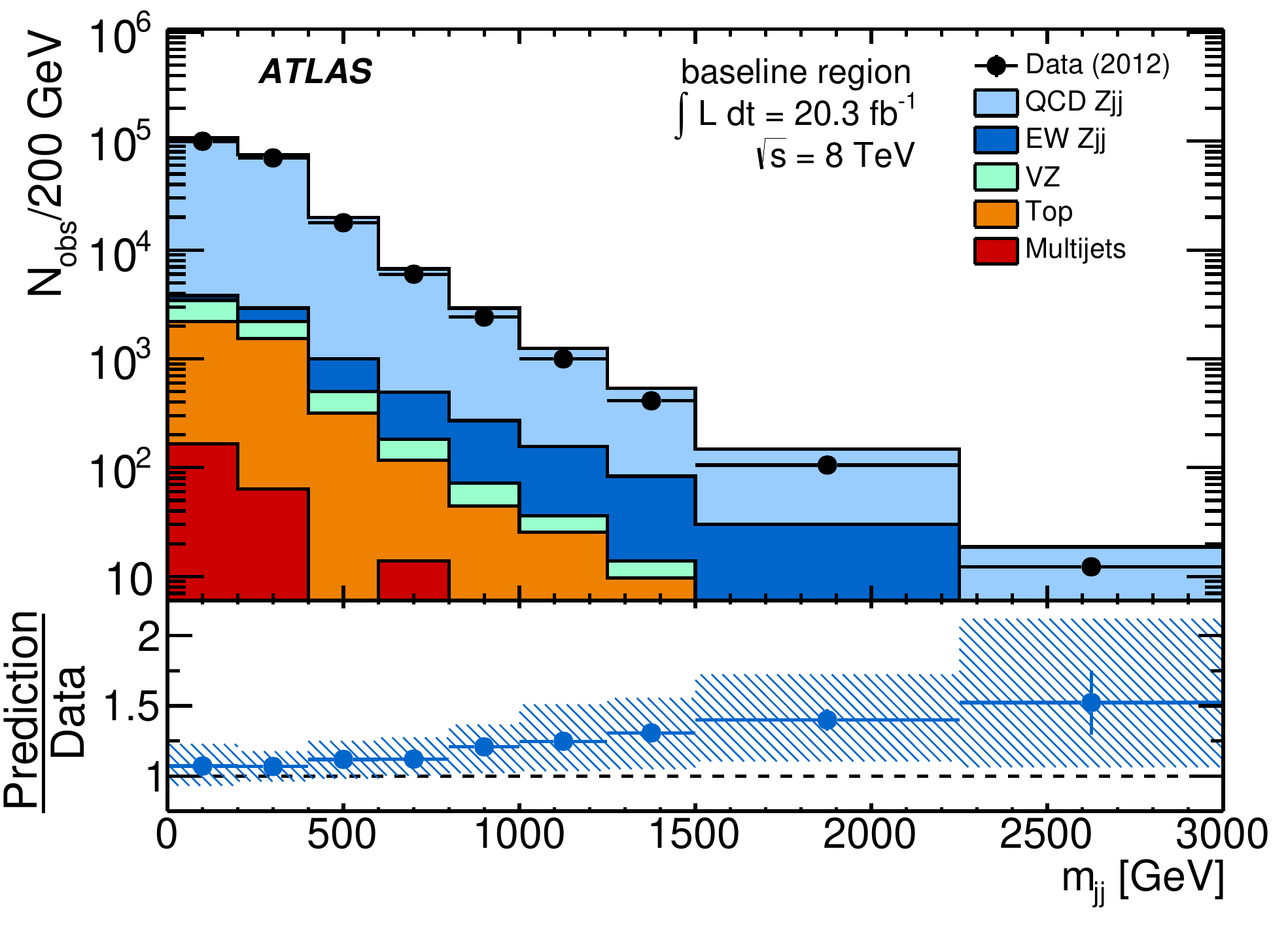}\quad
      \label{fig:reco-mjj}
    }
    \subfigure[] {
      \includegraphics[width=0.47\textwidth]{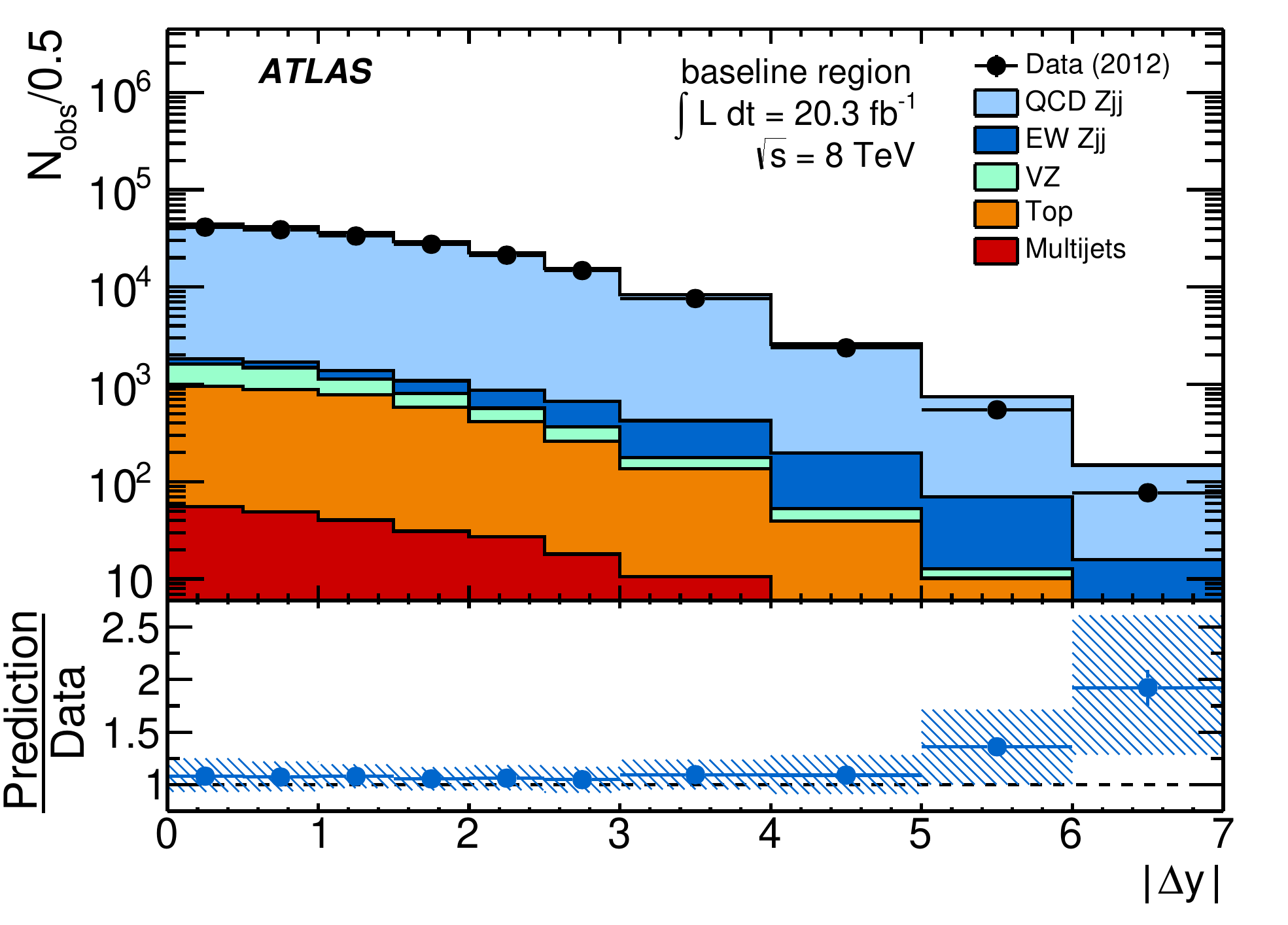}
      \label{fig:reco-dy}
    }
    \caption[]{Comparison of data and simulation in the \lowpt\ region for (a,b) the leading jet transverse momentum and rapidity, (c,d) the subleading jet transverse momentum and rapidity, (e,f) the invariant mass and rapidity span of the dijet system. The simulated samples are normalised to the cross-section predictions discussed in section \ref{sec:mc} and then stacked. The error bars reflect the statistical uncertainties of the data. The hatched band in the ratio reflects the total experimental systematic uncertainty on the simulation.} 
    \label{fig:reco-control-plots}
  \end{center}
\end{figure}

The uncertainty due to background subtraction is found to be between 0.2\% in the \search\ region and 0.5\% in the \highmass\ region. This accounts for the uncertainty in the normalisation of the inclusive \ttbar\ sample, generator modelling differences in \ttbar\ events predicted by \texttt{MC}@\texttt{NLO} and \powheg, and the uncertainty in the data-driven method used to determine the multijet background.

The total systematic uncertainty on the inclusive \Zdijet\ cross-section measurement in each fiducial region is defined as the quadrature sum of all sources of experimental and theoretical uncertainty.

\subsection{Comparison of data and simulation}

Figure \ref{fig:reco-control-plots} shows data compared to MC simulation in the \lowpt\ region, as a function of the leading jet transverse momentum and rapidity, the subleading jet transverse momentum and rapidity, and the invariant mass and rapidity separation of the two leading jets. 
The uncertainty on the simulation due to the experimental systematic uncertainties is shown in the ratio as a hatched (blue) band. In general, the simulation gives an adequate description of the data, although there are indications of generator mismodelling at high jet transverse momentum and high dijet invariant mass. The contribution from \ttbar\ and multijet events remains small in each bin of the distributions.

\subsection{Cross section determination}
\label{sec:fid-results}

\begin{table} 
\centering
\caption{Fiducial cross sections for inclusive \Zdijet\ production, measured in the $\Z \rightarrow \ell^+ \ell^-$ decay channel.}
\label{tab:crosssections}
\begin{tabular}{@{} l c c }
\hline\hline
Fiducial region \rule{0mm}{3.5mm} & $\sigma_{\rm fid}$ (pb)\rule{0mm}{4mm} \\
\hline
 \lowpt    & $5.88\hphantom{0} \pm 0.01\hphantom{0}$ (stat) $\pm 0.62\hphantom{0}$ (syst) $\pm 0.17\hphantom{0}$ (lumi) \rule{0mm}{4mm} \\
  \highpt   & $1.82\hphantom{0} \pm 0.01\hphantom{0}$ (stat) $\pm0.17\hphantom{0}$ (syst) $\pm 0.05\hphantom{0}$ (lumi) \rule{0mm}{4mm} \\
  \highmass & $0.066 \pm 0.001$ (stat) $\pm 0.012$ (syst) $\pm 0.002$ (lumi) \rule{0mm}{4mm} \\
  \search   & $1.10\hphantom{0} \pm 0.01\hphantom{0}$ (stat) $\pm 0.09\hphantom{0}$ (syst) $\pm 0.03\hphantom{0}$ (lumi) \rule{0mm}{4mm} \\
  \control  & $0.447 \pm 0.004$  (stat) $\pm0.059$ (syst) $\pm 0.013$ (lumi) \rule{0mm}{4mm} \\
\hline\hline
\end{tabular}
\end{table}

\begin{table}
\centering
\caption{Theory predictions for inclusive \Zdijet\
  production cross sections in the $\Z \rightarrow \ell^+ \ell^-$ decay channel. The strong \Zdijet\ and electroweak \Zdijet\ events are produced using \powheg. A small contribution of $ZV$ events, produced by \sherpa, is also included. The PDF uncertainty is estimated from the CT10 eigenvectors using the procedure described in ref. \cite{Lai:2010vv}. Scale and modelling uncertainties are each estimated from the envelope of  \powheg\ sample variations discussed in section \ref{sec:theory}.}
\label{tab:crosssections-theory}
\begin{tabular}{@{} l c  }
\hline\hline
Fiducial region \rule{0mm}{3.5mm} & $\sigma_{\rm theory}$ (pb)\rule{0mm}{4mm} \\ 
\hline
 \lowpt    &  $6.26\hphantom{0} \pm 0.06\hphantom{0}$ (stat)
              ${}^{+0.50\hphantom{0}}_{-0.60\hphantom{0}}$ (scale)
              ${}^{+0.29\hphantom{0}}_{-0.35\hphantom{0}}$ (PDF)
              ${}^{+0.19\hphantom{0}}_{-0.25\hphantom{0}}$ (model) \rule{0mm}{4mm} \\
 \highpt   &  $1.92\hphantom{0} \pm 0.02\hphantom{0}$ (stat)
              ${}^{+0.17\hphantom{0}}_{-0.20\hphantom{0}}$ (scale)
              ${}^{+0.09\hphantom{0}}_{-0.10\hphantom{0}} $ (PDF)
              ${}^{+0.05\hphantom{0}}_{-0.07\hphantom{0}} $ (model) \rule{0mm}{4mm} \\
 \highmass &  $0.068 \pm 0.001$ (stat)
              ${}^{+0.009}_{-0.009}$ (scale)
              ${}^{+0.004}_{-0.003}$ (PDF)
              ${}^{+0.004}_{-0.002}$ (model) \rule{0mm}{4mm} \\
 \search   &  $1.23\hphantom{0} \pm 0.01\hphantom{0}$ (stat)
              ${}^{+0.11\hphantom{0}}_{-0.13\hphantom{0}}$ (scale)
              ${}^{+0.06\hphantom{0}}_{-0.07\hphantom{0}}$ (PDF)
              ${}^{+0.03\hphantom{0}}_{-0.04\hphantom{0}}$ (model) \rule{0mm}{4mm} \\
 \control  &  $0.444 \pm 0.005$ (stat)
              ${}^{+0.051}_{-0.054}$ (scale)
              ${}^{+0.021}_{-0.025}$ (PDF)
              ${}^{+0.032}_{-0.034}$ (model) \rule{0mm}{4mm} \\
\hline\hline
\end{tabular}
\end{table}

The cross sections are measured in the muon and electron decay channels separately. The cross-section measured in each fiducial region is  found to be compatible between the two channels, with a maximum difference of 1.1$\sigma$ after accounting for those uncertainties that are uncorrelated between channels. The results are then combined\footnote{The individual- and combined-channel cross sections are defined using dressed leptons as discussed in Section \ref{sec:total}. Cross sections defined using `Born' leptons (which originate directly from the \Z-boson decay and before final state QED radiation) would differ by 2--3\%.} 
to obtain a weighted average, with each channel's weight set to the inverse squared uncorrelated uncertainty.
Table~\ref{tab:crosssections} presents the measured inclusive \Zdijet\ cross sections in the
five fiducial regions together with their statistical and systematic uncertainties. Table~\ref{tab:crosssections-theory} presents the \powheg\ prediction for strong and electroweak \Zdijet\ production, combined with the \sherpa\ prediction for the small contribution from diboson processes. Uncertainties on the theoretical predictions are broken down into statistical, scale, PDF and generator modelling uncertainties. Good agreement between data and theory is observed in all fiducial regions and a summary is shown in figure \ref{fig:xsecs}.

\begin{figure}[t]
\centering
      \includegraphics[width=0.65\textwidth]{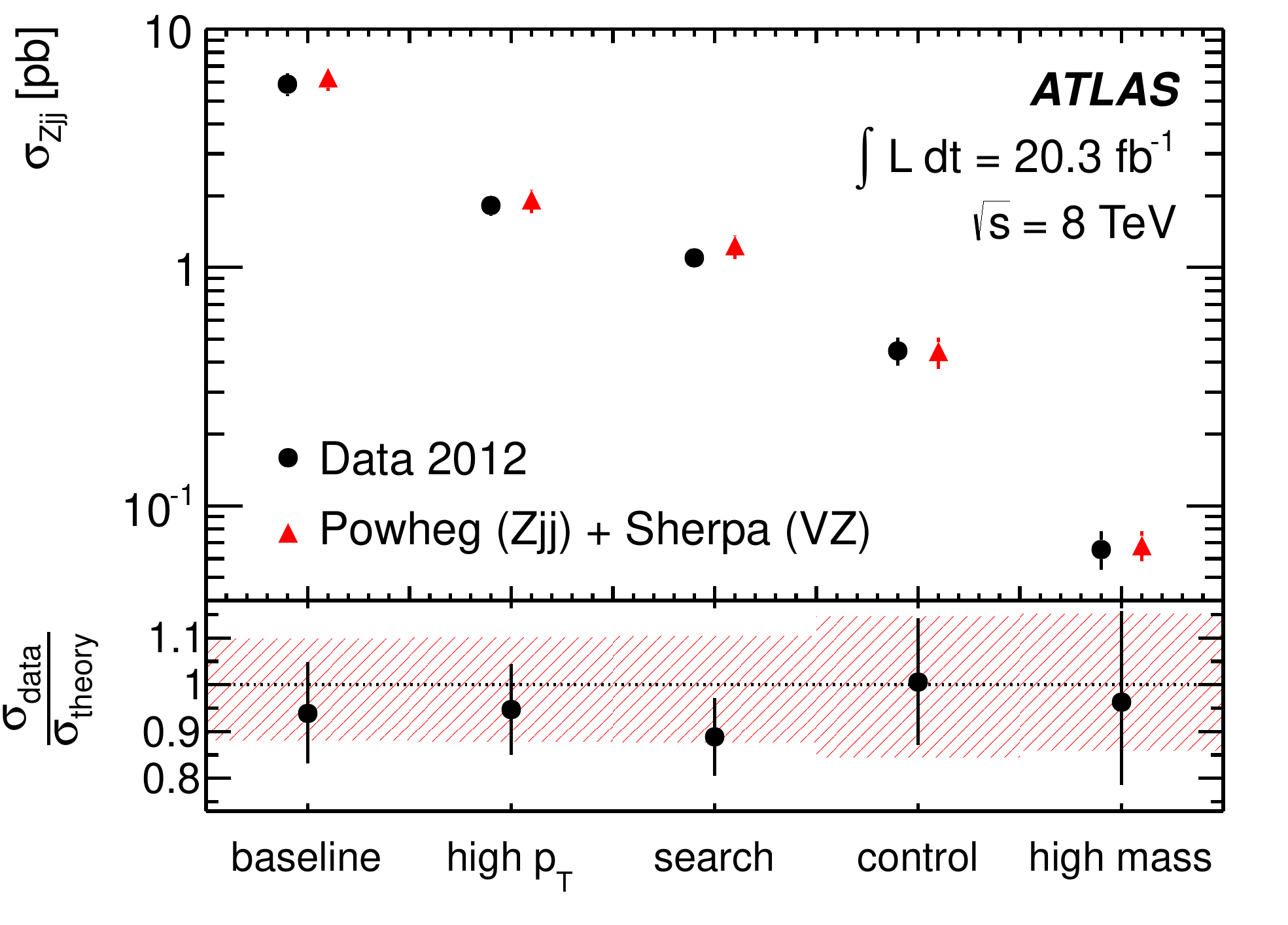}
    \caption{Fiducial cross-section measurements for inclusive \Zdijet\ production in the $\Z \rightarrow \ell^+ \ell^-$ decay channel, compared to the \powheg\ prediction for strong and electroweak \Zdijet\ production and the small contribution from $ZV$ production  predicted by \sherpa. The (black) circles represent the data and the associated error bar is the total uncertainty in the measurement. The (red) triangles represent the theoretical prediction, the associated error bar (or hatched band in the lower plot) is the total theoretical uncertainty on the prediction.}
    \label{fig:xsecs}
\end{figure}

\section{Differential distributions of inclusive \Zdijet\ production}
\label{sec:diffdist}

In this section,  inclusive \Zdijet\ differential distributions are measured in the five fiducial regions presented in the previous section. The theoretical modelling of strong \Zdijet\ production is therefore confronted in regions with differing sensitivity to the electroweak \Zdijet\ component. The data are fully corrected for detector effects and are provided in HEPDATA~\cite{hepdata} with full correlation information. 
The distributions sensitive to the kinematics of the two tagging jets are:
  \begin{itemize}
  \item{\dsdmjj: The normalised distribution of the dijet invariant mass of the two leading jets, \mjj.}
  \item{\dsddeltay: The normalised distribution of the difference in rapidity between the two leading jets, \deltay. }
  \item{\dsddphijj: The normalised distribution of the difference in azimuthal angle between the two leading jets, \dphijj}.
  \end{itemize}
The distributions sensitive to the difference in $t$-channel colour flow between electroweak and strong production of \Zdijet\ events include:
  \begin{itemize}
  \item \dnjet: The normalised distribution of the number of jets, \njet, with
    $\pt > 25$~GeV in the rapidity interval bounded by the two highest-\pt\
    jets. 
  \item \dsdptsc: The normalised distribution of the  \pt-balancing
      distribution, \ptsc\ (see eq.~\ref{eq:ptsc}).   
  \item The fraction of events that contain no additional jets with $\pt >
      25$~GeV in the rapidity interval bounded by the two highest-\pt\
    jets (the \gapfrac) as a function of \mjj\ and \deltay. 
  \item The average number of jets with $\pt >
    25$~GeV in the rapidity interval bounded by the two highest-\pt\
    jets, \avgnjet, as a function of
    \mjj\ and \deltay. 
  \item The fraction of events with $\ptsc < 0.15$ (\ptscfrac) as a function of  \mjj\ and \deltay.
  \end{itemize}

\subsection{Analysis methodology and unfolding to particle level}
The differential distributions are normalised to unity after subtracting the small  background contributions from \ttbar\ and multijet events in each bin of the distributions. An iterative Bayesian unfolding procedure~\cite{cite:unfold,Adye:2011gm} is then applied to the data to produce distributions at the particle level.
This procedure uses a detector response matrix to reverse the bin migration caused by finite detector resolution. The response matrix is constructed from the strong and electroweak \Zdijet\ simulated samples for each distribution. Events that pass the reconstruction-level but not the particle-level selection criteria (or vice versa) are also corrected for as part of the unfolding procedure. 

The Bayesian unfolding procedure relies on knowledge of the underlying particle-level distribution. This `prior' distribution is taken to be the particle-level prediction from \sherpa. 
After the first unfolding iteration, the input prior is replaced with
the unfolded distribution from the data and the unfolding process is repeated. It
is found that two iterations are sufficient to ensure convergence of
the results.

The statistical uncertainty on the data after unfolding is computed using pseudo-experiments. 
The statistical correlation between the numerator and the denominator in the jet veto distributions is retained by unfolding two-dimensional distributions constructed from the dijet observable (\mjj, \deltay) and information as to whether events passed or failed the efficiency criterion. The \ptscfrac\  
distribution is 
unfolded in a similar way.  Correlations in the \avgnjet\ distributions are retained by unfolding a two-dimensional distribution constructed from the dijet observable and the number of jets in the rapidity interval between the two leading jets. Statistical correlations between bins from different unfolded distributions are estimated using a bootstrap method \cite{Hayes:1988xc}.

\subsection{Systematic uncertainties}

The sources of experimental and theoretical uncertainty include all of those present in the measurement of the inclusive \Zdijet\ fiducial cross section (Sec.~\ref{sec:systematics}). The impact of lepton-based and luminosity systematic uncertainties on the measured distributions is negligible and the experimental systematic uncertainties therefore arise from JES, JER, JVF, as well as pileup jet modelling. The theoretical modelling uncertainties are again estimated by reweighting the simulation, such that the kinematic distributions of the variables used to define the fiducial regions match those observed in the data. An additional uncertainty associated with the closure of the Bayesian iterative procedure is estimated by 
reweighting the simulated events such that the reconstruction-level distribution being unfolded better matches the one observed in the data. The reweighting functions applied at the particle level are taken to be the ratio of the reconstruction-level distributions in data and simulation. 

\begin{figure}[t]
  \begin{center}
    \subfigure[] {
      \includegraphics[width=0.47\textwidth]{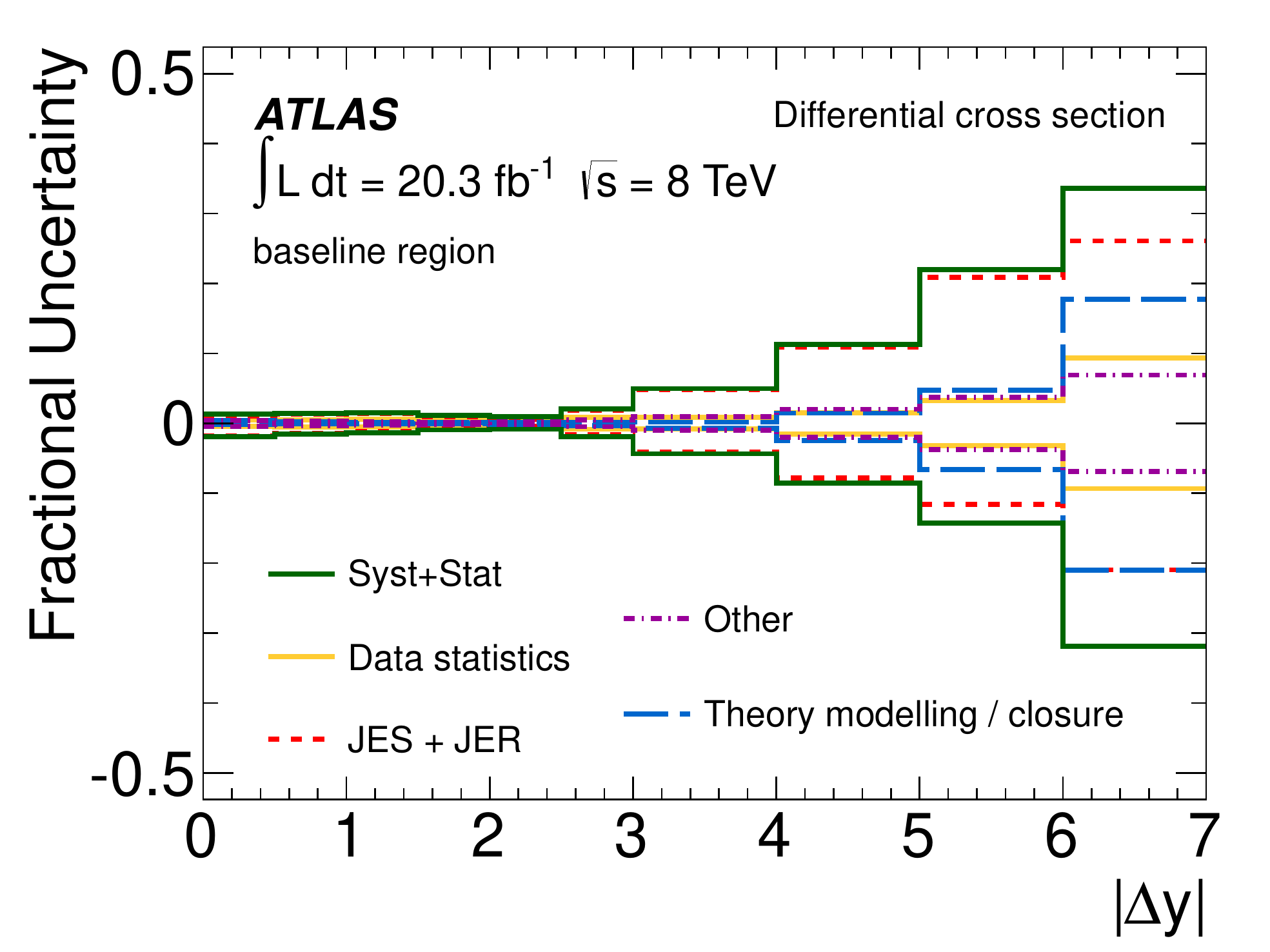}\quad
      
    }
    \subfigure[] {
      \includegraphics[width=0.47\textwidth]{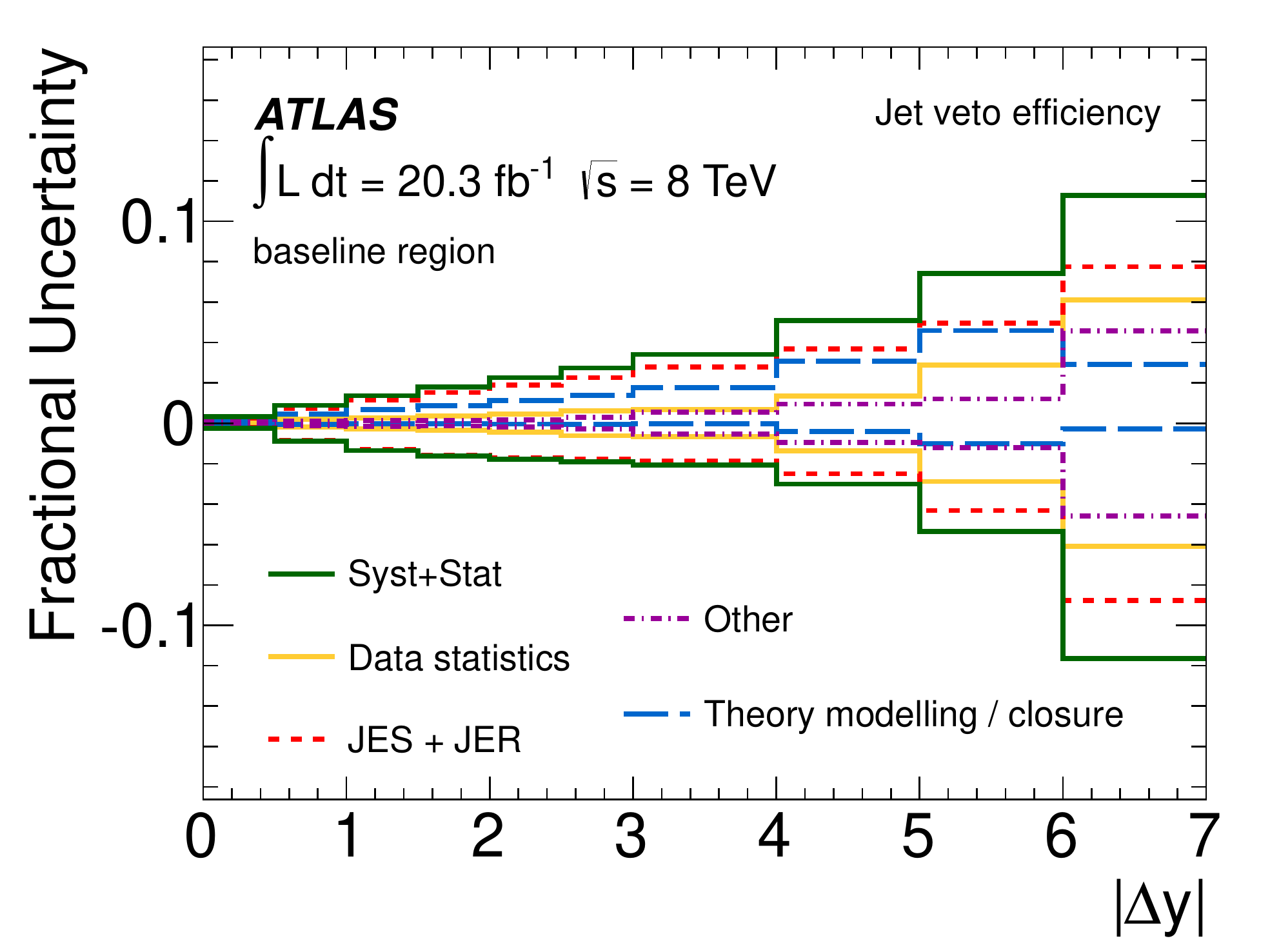}
      
    }
    \caption[]{Example systematic uncertainty breakdown for the \dsddeltay\ distribution and the jet veto efficiency as a function of \deltay\ in the \lowpt\  region. The effect of MC statistics, pileup modelling and JVF modelling are combined into one uncertainty labelled `other'. } 
    \label{fig:unfolding-Dy-syst}
  \end{center}
\end{figure}

For all sources of systematic uncertainty, the data are unfolded using a new response matrix constructed after shifting and smearing the MC events and objects. The shift in the unfolded spectrum is taken as the systematic uncertainty on the
final result. The dominant uncertainties arise from the JES and JER, with small additional uncertainties from JVF, pileup modelling and theoretical modelling. The  systematic uncertainties are presented in figure \ref{fig:unfolding-Dy-syst} for the \dsddeltay\ distribution and the jet veto efficiency as a function of \deltay, in the \lowpt\  region. The total systematic uncertainty in each bin is defined as the quadrature sum of the individual sources of experimental and theoretical uncertainty.

The unfolding procedure is cross-checked using the simulated \alpgen\ sample in place of the \sherpa\ strong \Zdijet\ sample. The data are unfolded using the new response matrix formed from these simulated events. The data unfolded using the \alpgen- and \sherpa-based response matrices are found to agree, after accounting for the larger statistical uncertainty in the \alpgen\ sample in addition to the theory modelling and closure uncertainties assigned to the nominal result.

\subsection{Unfolded differential distributions}

\begin{figure}[t]
  \begin{center}
    \subfigure[] {
      \includegraphics[width=0.47\textwidth]{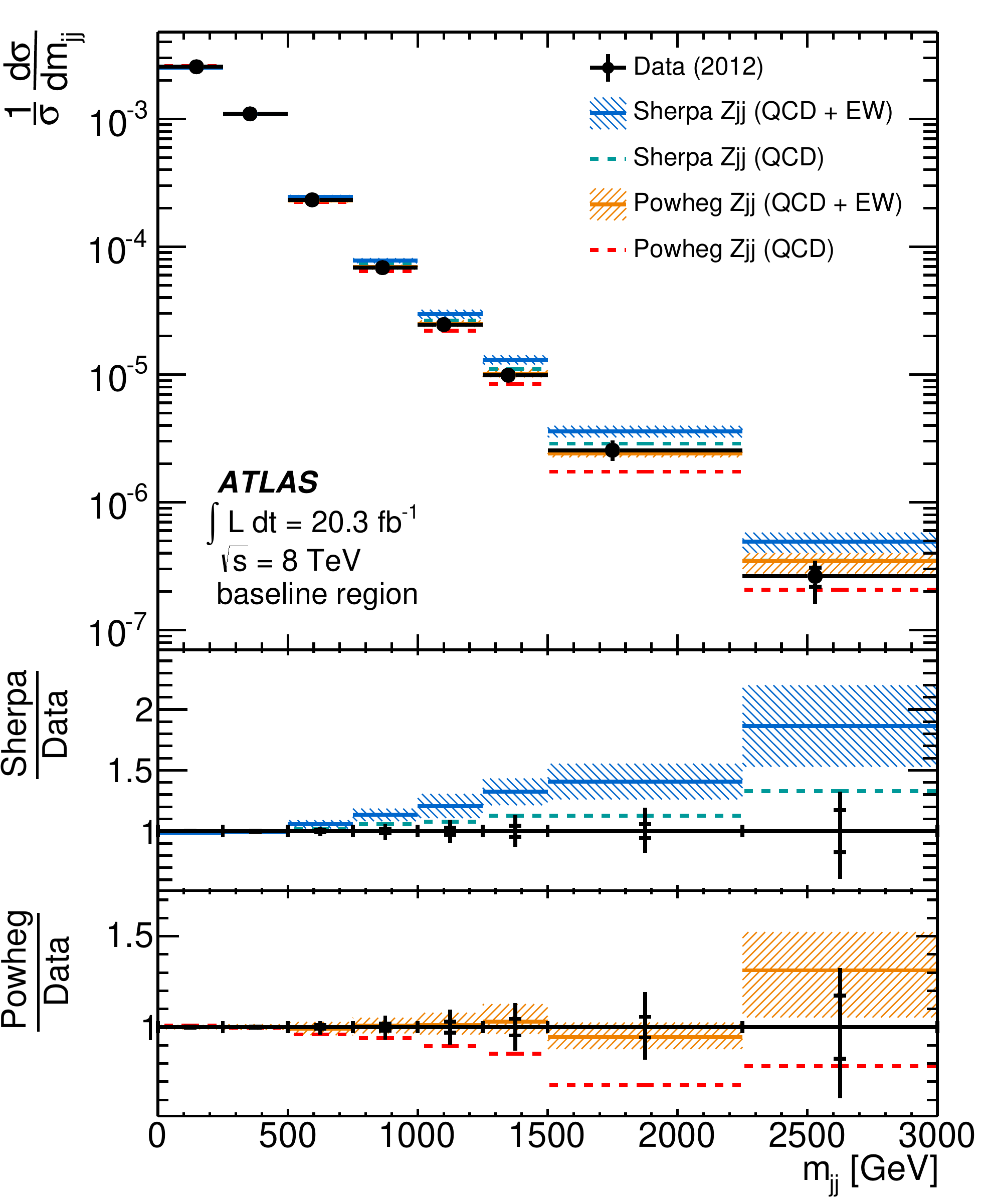}\quad
          }
    \subfigure[] {
      \includegraphics[width=0.47\textwidth]{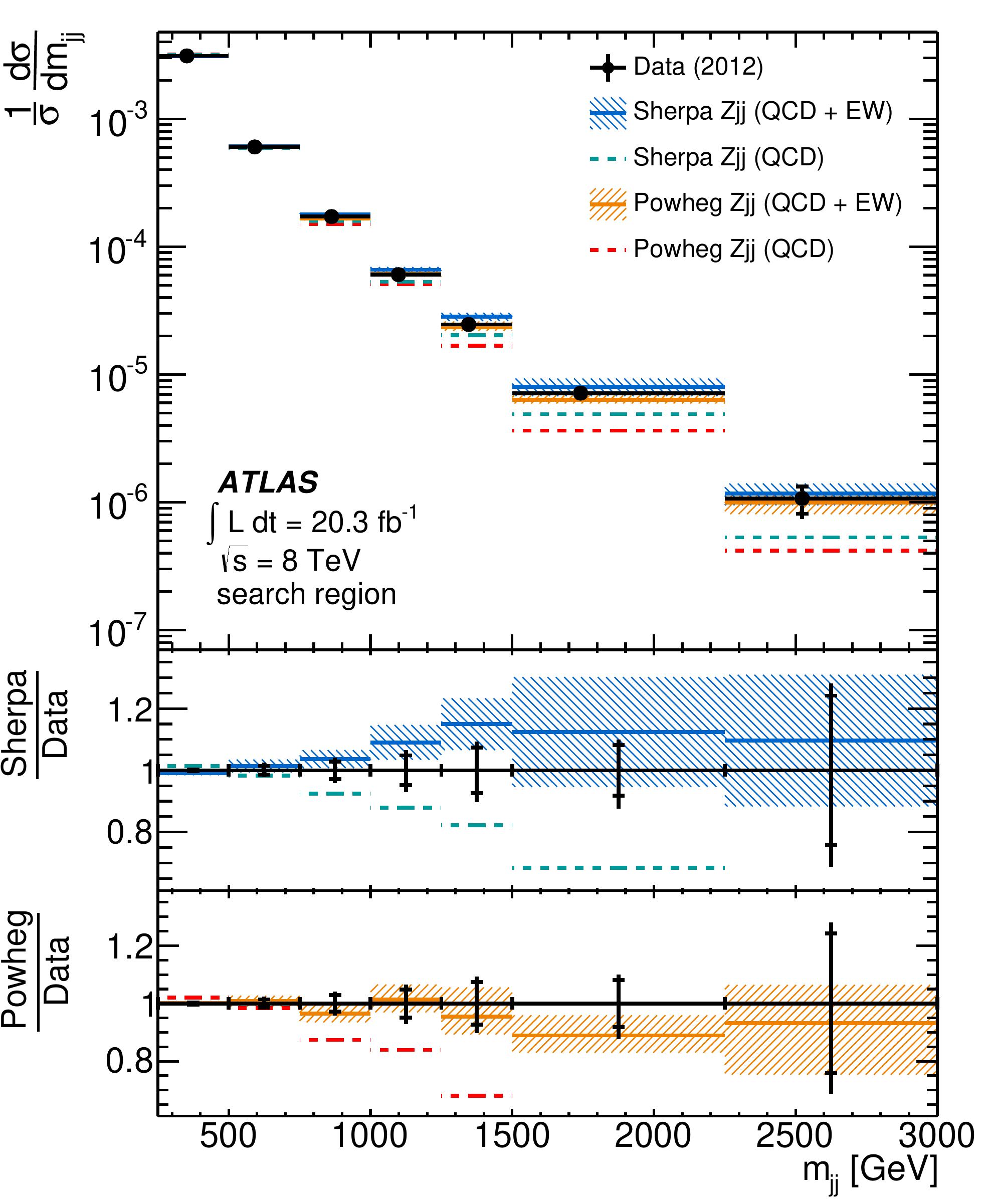}
         }
    \caption[]{Unfolded \dsdmjj\ distribution in the (a)  \lowpt\ and (b) \search\
      regions. \resultCaption} 
    \label{fig:unfolding-Mjj-Final}
  \end{center}
\end{figure}

The unfolded data are compared to particle-level predictions from \powheg\ and \sherpa\ in figures~\ref{fig:unfolding-Mjj-Final}--\ref{fig:unfolding-PtBalEff-Final}. 
The theoretical predictions are shown for combined electroweak and strong \Zdijet\ production and for strong \Zdijet\ production only. The theoretical uncertainty on the combined electroweak and strong \Zdijet\ prediction is estimated using the envelope of theory modelling uncertainties discussed in section \ref{sec:theory}. The contribution from diboson production is neglected for the theoretical predictions as the impact on the distributions is negligible.

\begin{figure}[t]
  \begin{center}
    \subfigure[] {
      \includegraphics[width=0.47\textwidth]{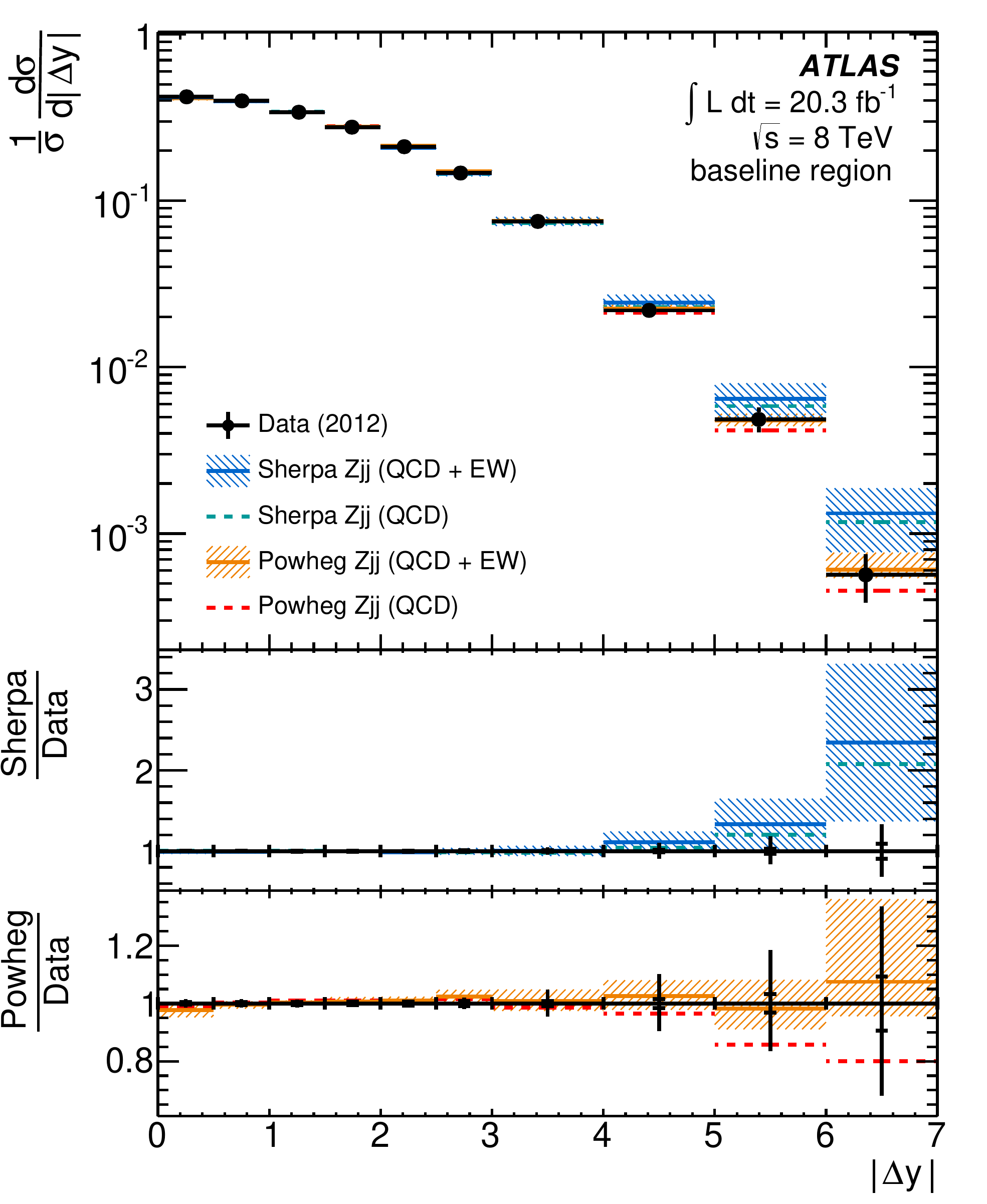}\quad
    }
    \subfigure[] {
      \includegraphics[width=0.47\textwidth]{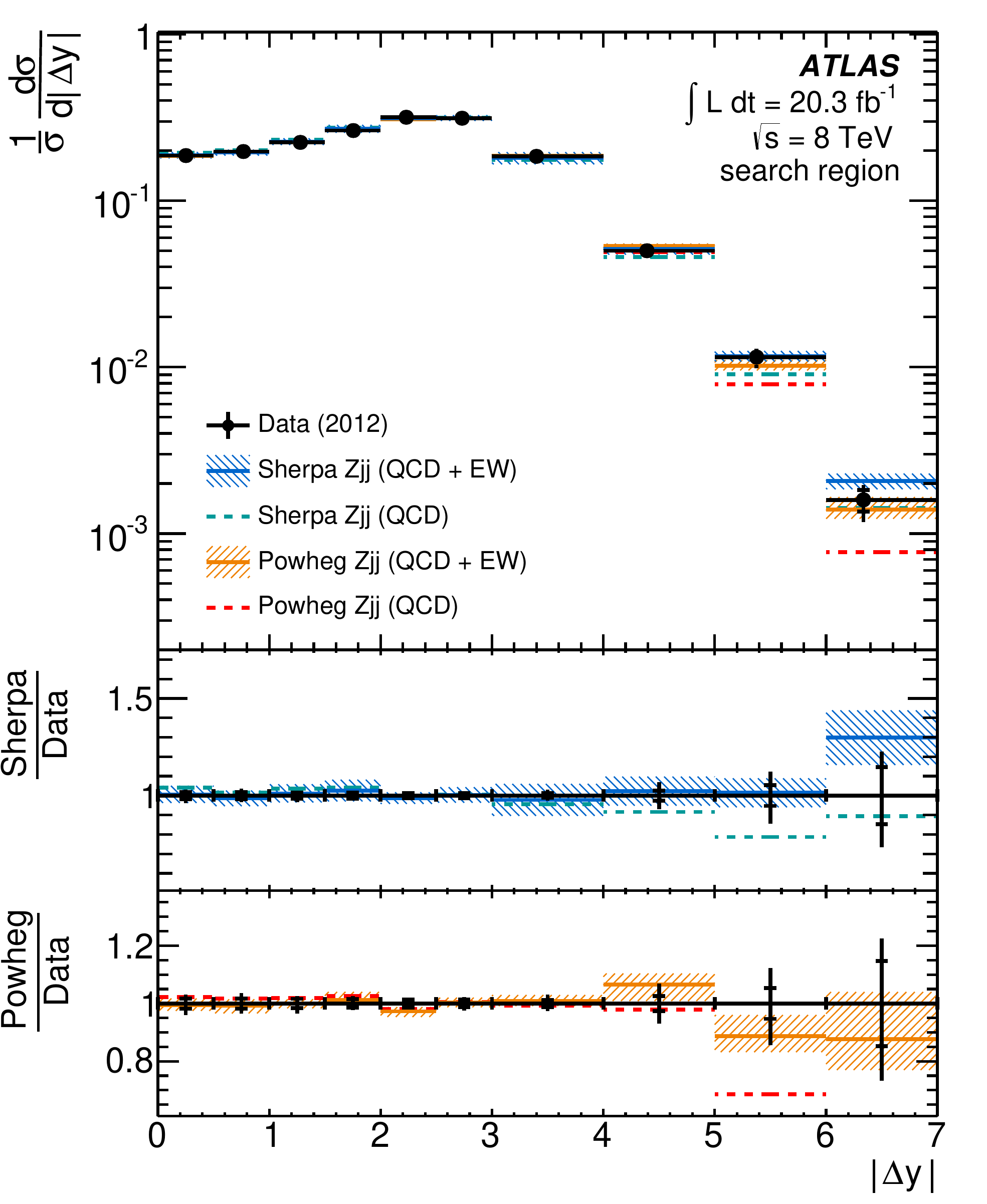}
    }
    \caption[]{Unfolded \dsddeltay\ distribution in the (a) \lowpt\ and (b) \search\ 
      regions. \resultCaptionTwo} 
    \label{fig:unfolding-Dy-Final}
  \end{center}
\end{figure}

The unfolded \dsdmjj\ and \dsddeltay\ distributions are shown in figure~\ref{fig:unfolding-Mjj-Final}~and~\ref{fig:unfolding-Dy-Final}, respectively, 
for the \lowpt\ and \search\  regions (corresponding distributions in the \highpt\ and \control\ regions are provided in appendix \ref{sec:add_data}). Both of these distributions are sensitive to the difference between electroweak and strong production of \Zdijet\ events, especially at large \mjj\ or \deltay. In the electroweak  process, the masses of the exchanged electroweak bosons lead to jets produced preferentially at large rapidities with sizeable transverse momentum. Furthermore, strong \Zdijet\ production typically involves the $t$-channel exchange of a spin-1/2 quark, which leads to steeper \mjj\ and \deltay\ spectra than the spin-1 exchange that is present in electroweak \Zdijet\ production. 

In the \lowpt\ region, the \powheg\ prediction is accurate to NLO in perturbative QCD and better describes the data at the highest values of \mjj\ and \deltay\ than \sherpa, which is accurate to LO. In particular, \sherpa\ predicts too large a fraction of events at large \mjj\ and \deltay, a feature also seen in previous measurements at the LHC and Tevatron \cite{Aad:2013ysa,Abazov:2013gpa}. In the \search\ region, the veto on additional jet activity means that both \sherpa\ and \powheg\ are accurate only to LO. Despite this, both predictions give a satisfactory description of the data if both strong and electroweak \Zdijet\ production are included. The contribution from electroweak \Zdijet\ production is evident at high \mjj\ and high \deltay\ in the \search\ region for both event generators.

\begin{figure}
  \begin{center}
    \subfigure[] {
      \includegraphics[width=0.47\textwidth]{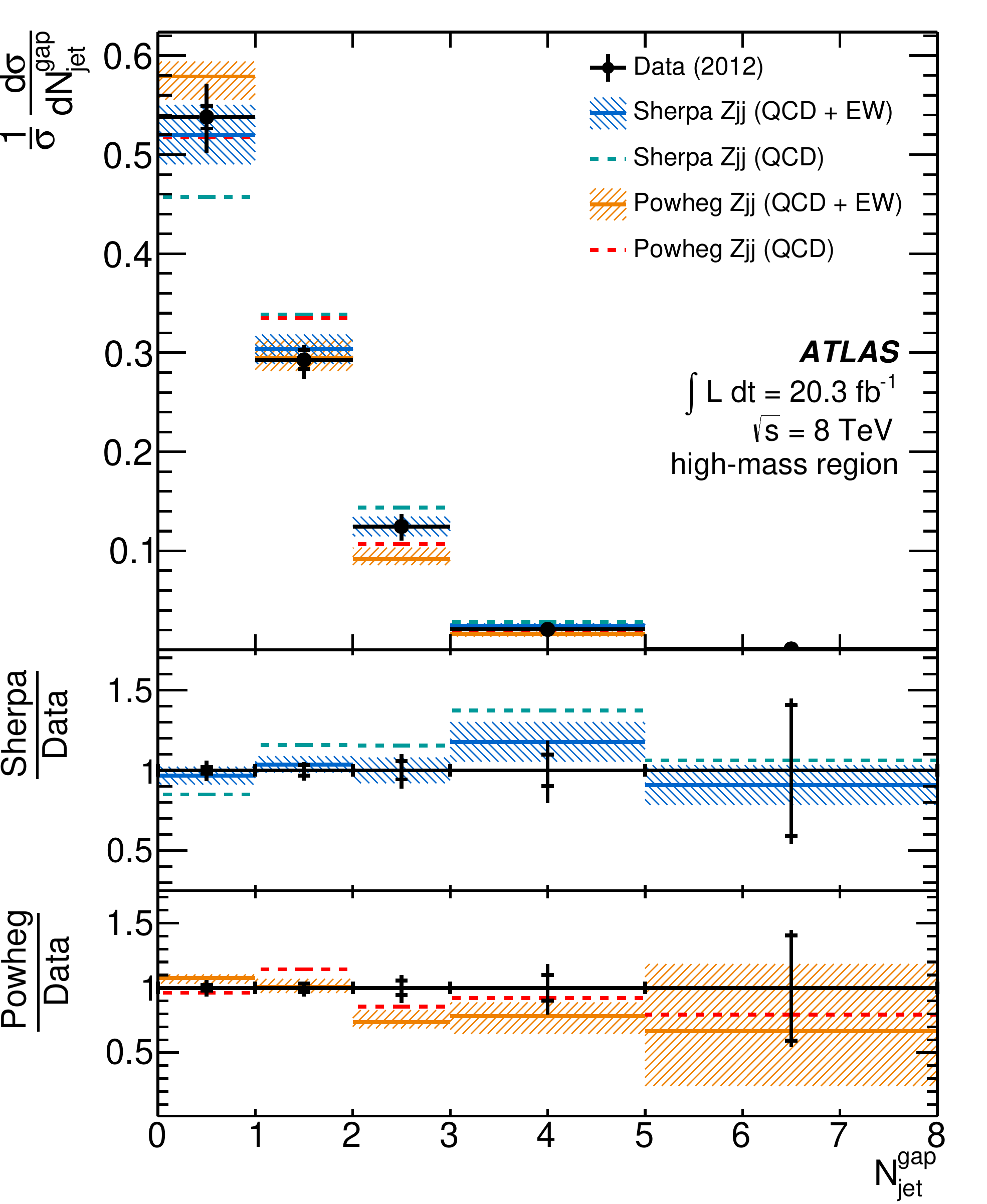}\quad
    }
    \subfigure[] {
      \includegraphics[width=0.47\textwidth]{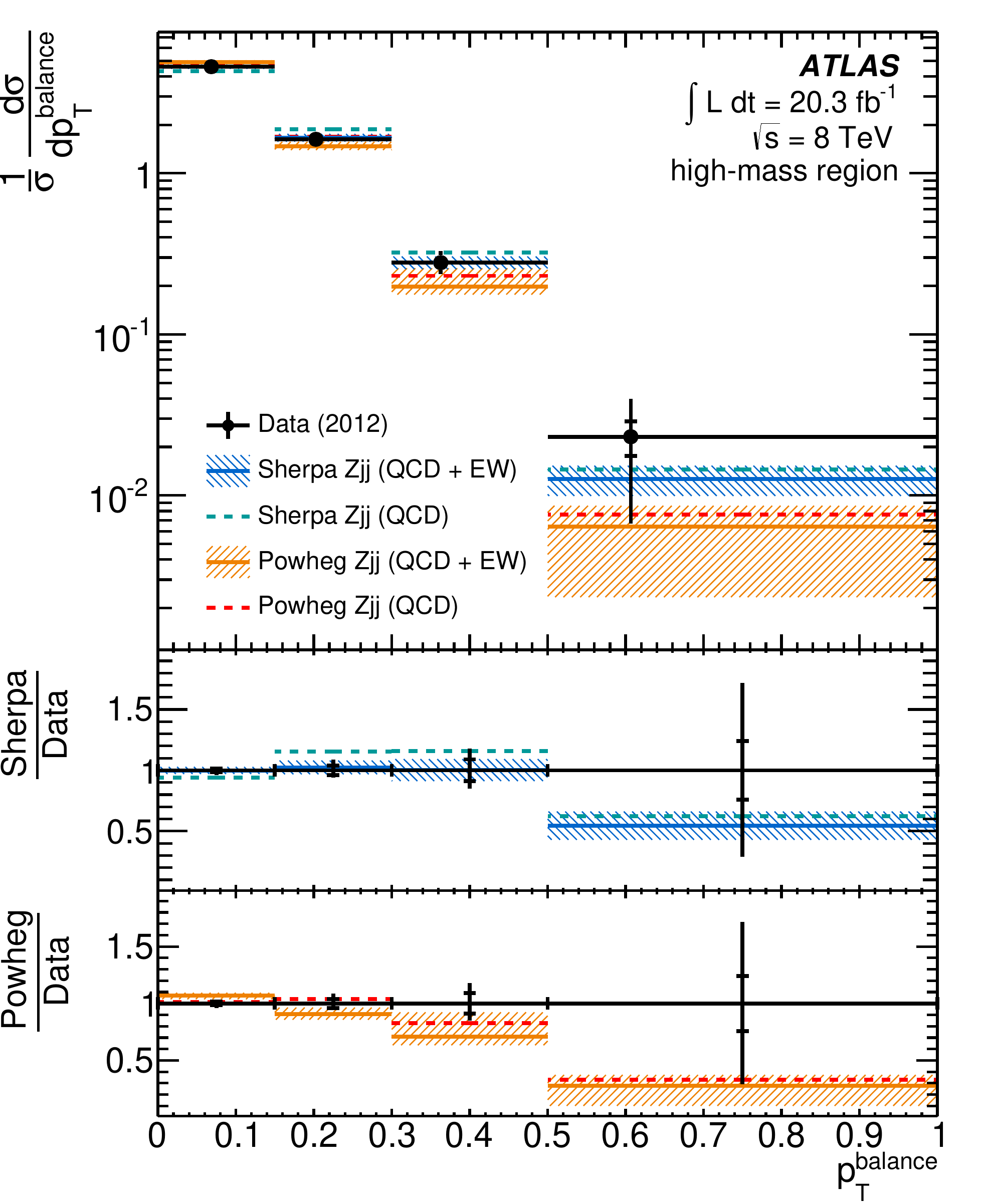}
         }
    \subfigure[] {
      \includegraphics[width=0.47\textwidth]{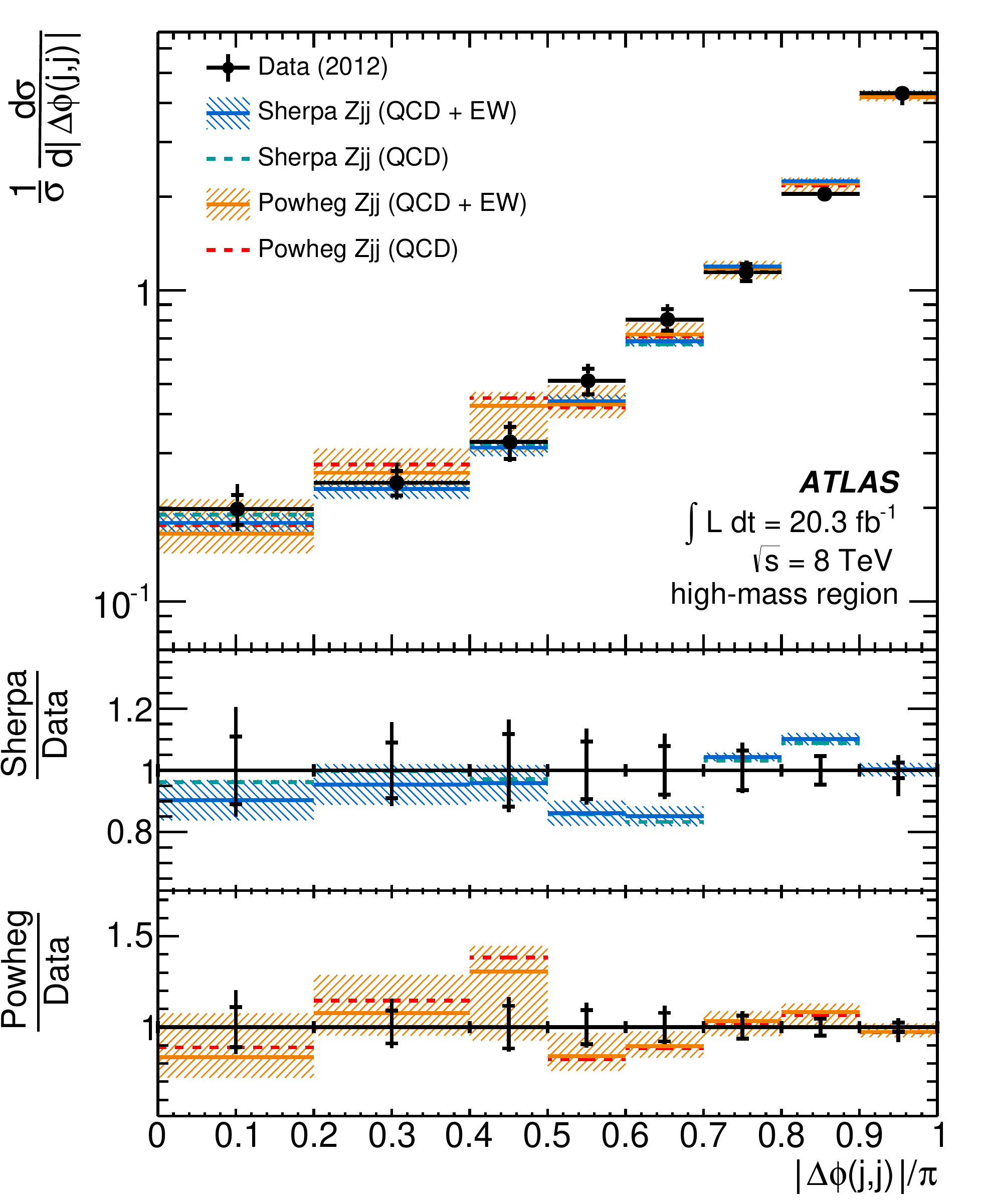}
          }
    \caption[]{Unfolded (a) \dnjet, (b) \dsdptsc\ and (c) \dsddphijj\
      distributions in the \highmass\  region. \resultCaptionTwo} 
    \label{fig:unfolding-HighMass-Final}
  \end{center}
\end{figure}

\begin{figure}
  \begin{center}
    \subfigure[] {
      \includegraphics[width=0.47\textwidth]{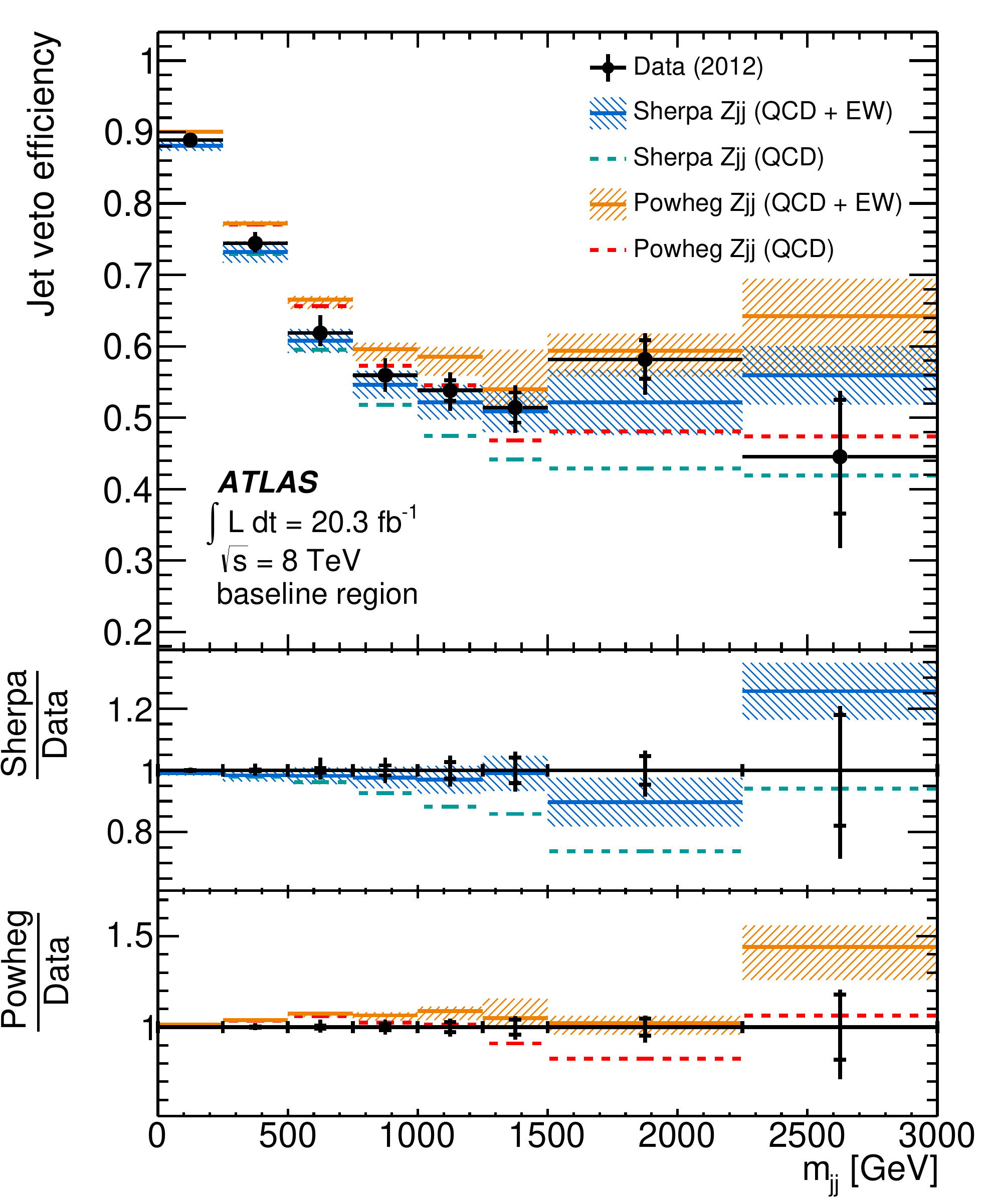}\quad
         }    \subfigure[] {
      \includegraphics[width=0.47\textwidth]{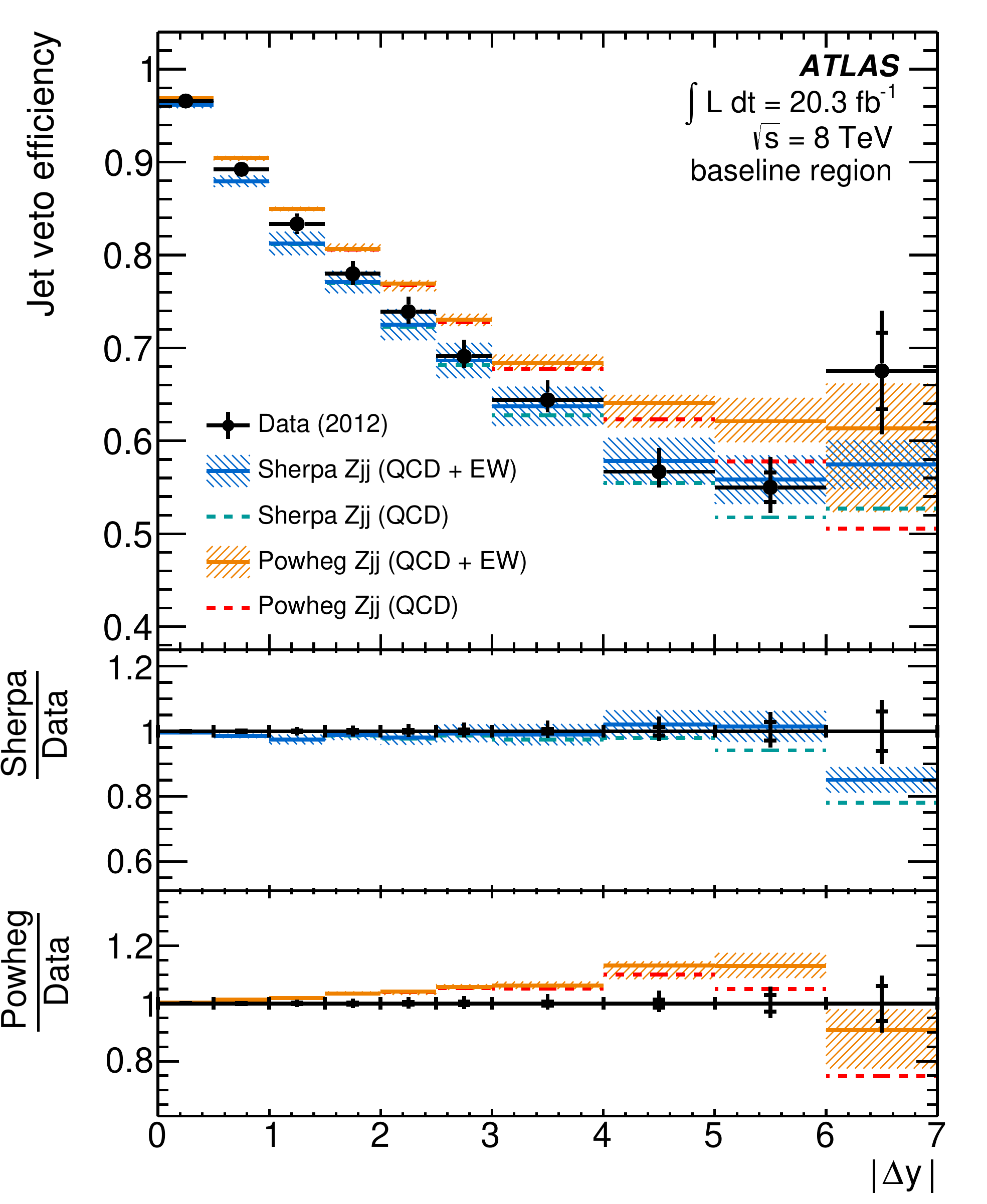}
         }
     \subfigure[] {
      \includegraphics[width=0.47\textwidth]{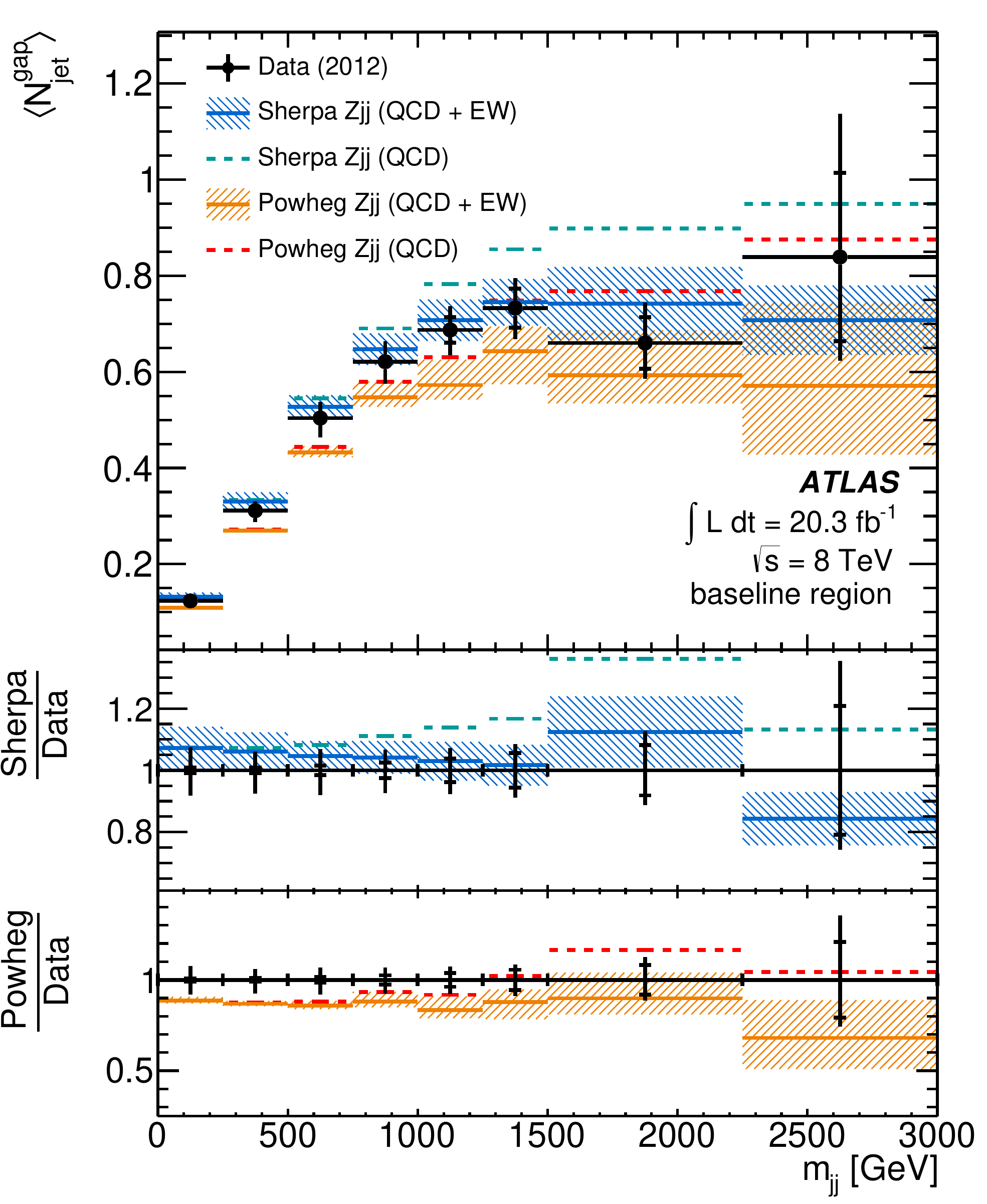}\quad
          }
    \subfigure[] {
      \includegraphics[width=0.47\textwidth]{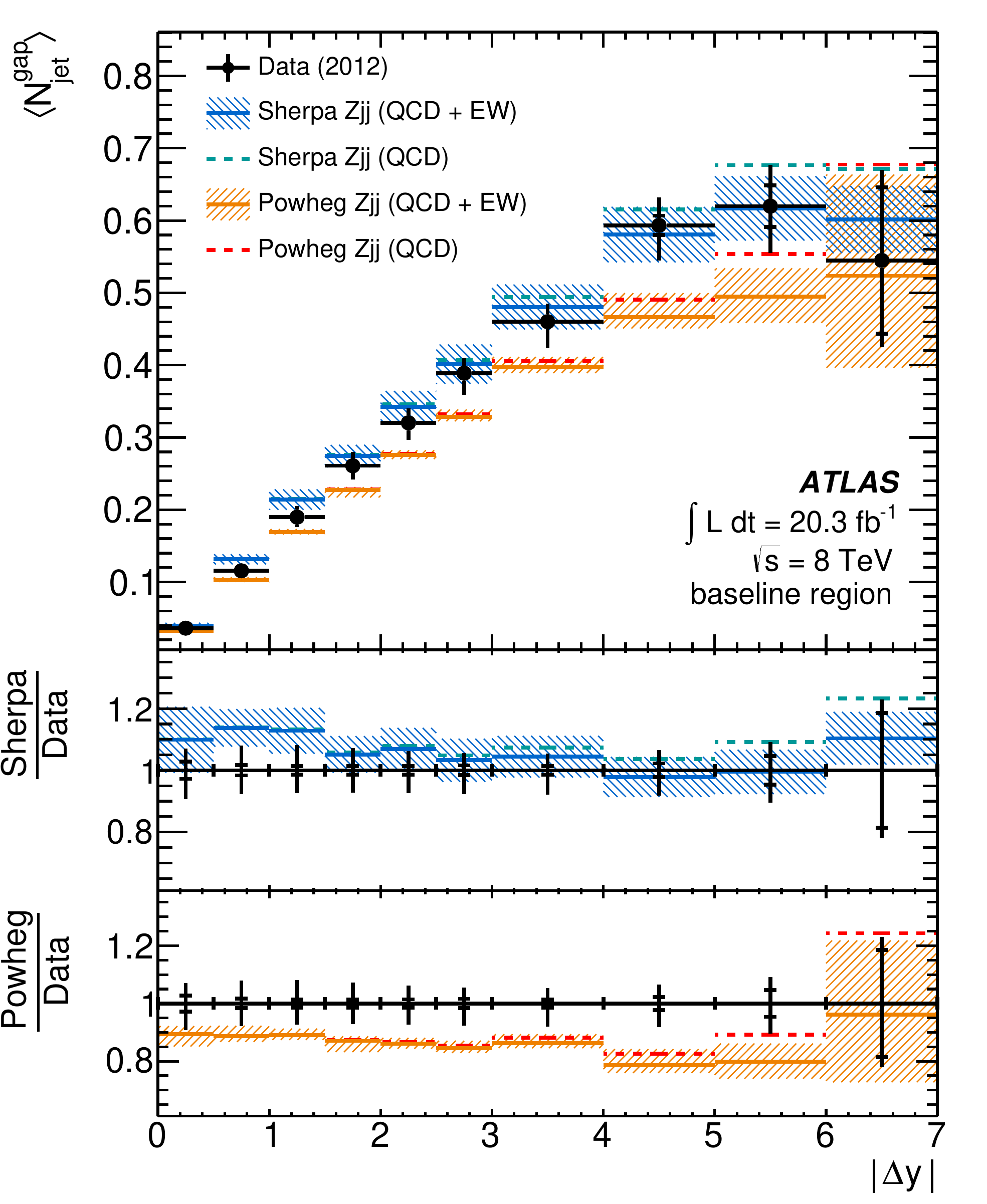}
               }
      \caption[]{Unfolded \gapfrac\ versus (a)  \mjj\  and (b) \deltay, and unfolded \avgnjet\ versus (c)  \mjj\  and (d) \deltay. All distributions are measured in  the \lowpt\  region.  \resultCaptionTwo} 
    \label{fig:unfolding-GapFraction-Final}
  \end{center}
\end{figure}

The unfolded \dnjet, \dsdptsc\ and \dsddphijj\ distributions are shown in the
\highmass\  region in figure~\ref{fig:unfolding-HighMass-Final}. Quark/gluon radiation from the electroweak \Zdijet\ process is much less likely than in the strong \Zdijet\ process because there is no colour flow between the two jets. The contribution from electroweak \Zdijet\
production is clear in the low-multiplicity region of the \dnjet\ distribution for both \powheg\ and \sherpa, demonstrating the effectiveness of the jet veto at separating the strong and electroweak components of \Zdijet\ production.  
Both \powheg\ and \sherpa\ adequately describe the data for the \dsdptsc\ and
\dsddphijj\ distributions; the latter distribution has little sensitivity to the electroweak process.\footnote{Although the azimuthal angle between the jets is not sensitive to the differences between strong and electroweak \Zdijet\ production, it is of interest in Higgs-plus-two-jet studies, as the vector boson fusion and gluon fusion production channels have very different azimuthal structure \cite{Plehn:2001nj,Klamke:2007cu,Andersen:2010zx}.}

Figure~\ref{fig:unfolding-GapFraction-Final} 
shows the unfolded jet veto efficiency and \avgnjet\ distributions as a function of \mjj\ and \deltay\ in the \lowpt\  region (corresponding distributions in the \highpt\  region are provided in appendix \ref{sec:add_data}). These variables probe the theoretical description of wide-angle quark and gluon radiation in strong \Zdijet\ events as a function of the energy scale of the dijet system. For the electroweak process, quark and gluon radiation into the rapidity interval is suppressed and little jet activity is expected. This is evident at medium-to-high values of \mjj, for which the strong \Zdijet\ prediction has more jet activity than the combined strong and electroweak \Zdijet\ prediction. 
In general, both theoretical predictions give a good description of the data (for combined strong and electroweak \Zdijet\ production), although \sherpa\ gives a slightly better description than \powheg\ when compared across both the \mjj\ and \deltay\ distributions.  \sherpa\ and \powheg\ have previously provided a good description of the jet activity in the rapidity interval bounded by a dijet system in purely dijet topologies~\cite{Aad:2011jz,Hoeche:2012fm}.

\begin{figure}
  \begin{center}
    \subfigure[] {
      \includegraphics[width=0.47\textwidth]{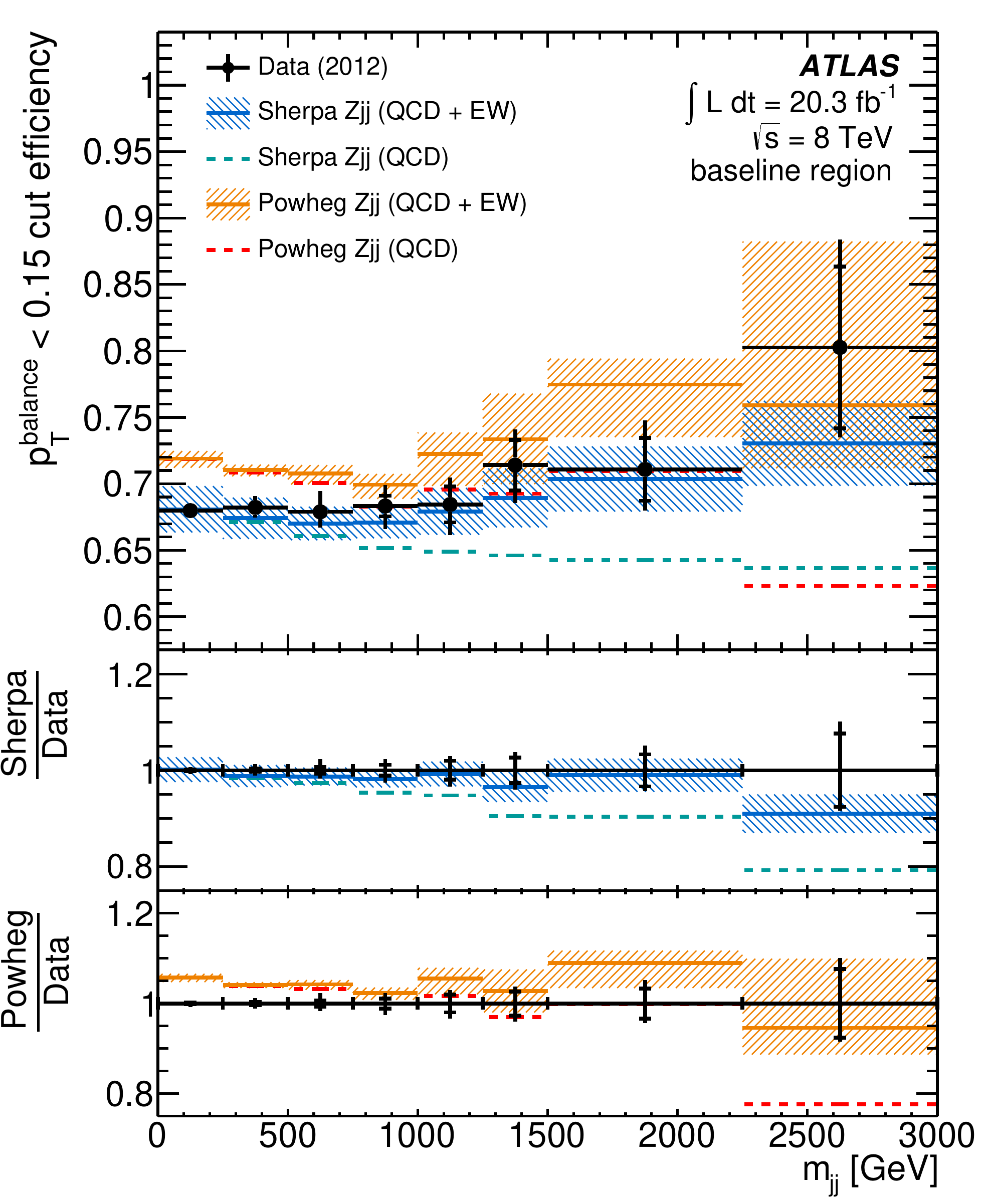}\quad
          }
    \subfigure[] {
      \includegraphics[width=0.47\textwidth]{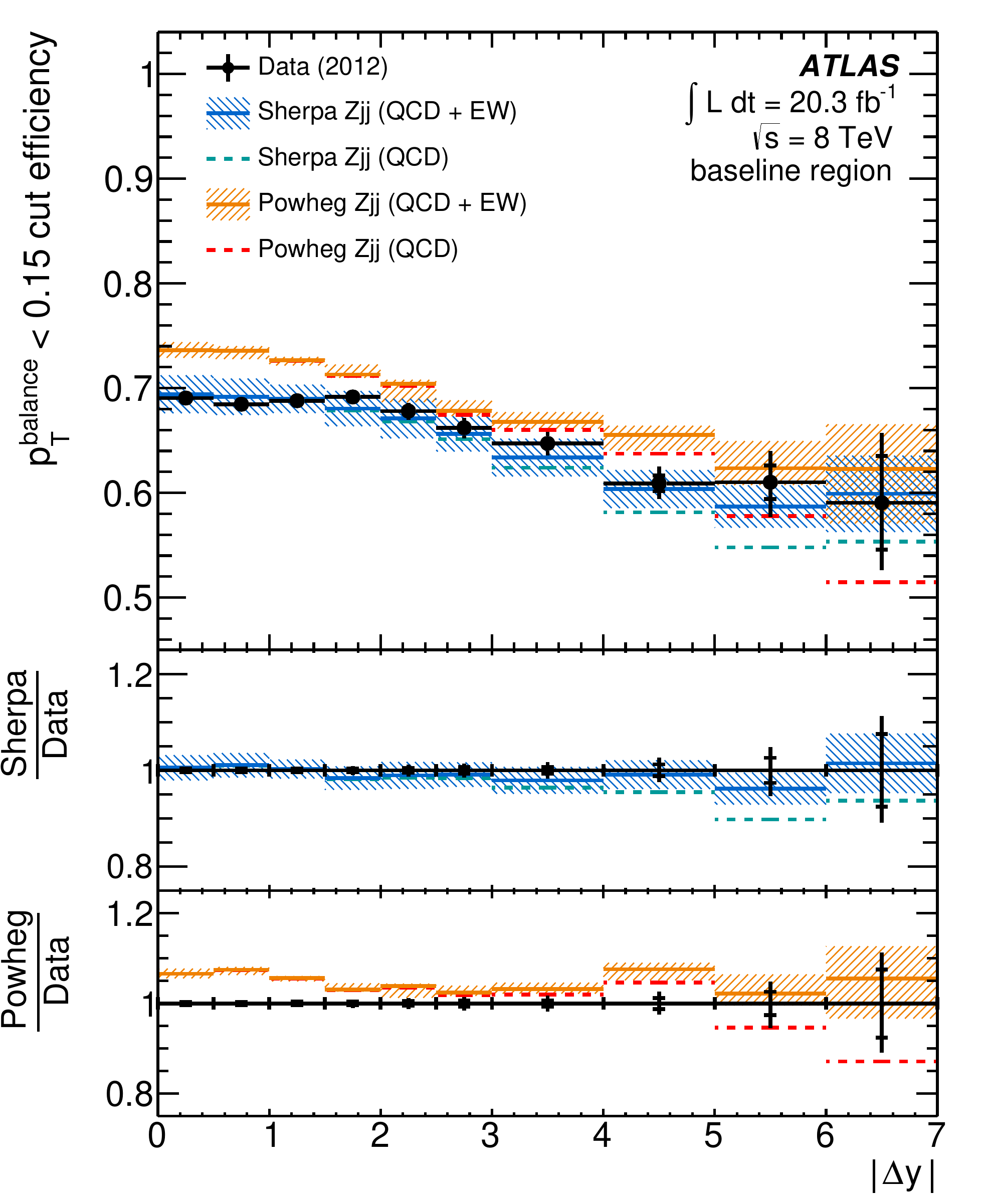}
         }
    \caption[]{Unfolded \ptsceff\ versus (a) \mjj\ and (b) \deltay\
      in the  \lowpt\  region. \resultCaptionTwo} 
    \label{fig:unfolding-PtBalEff-Final}
  \end{center}
\end{figure}

The unfolded \ptscfrac\  as a function of \mjj\ and \deltay\ in the \lowpt\  region is shown in figure~\ref{fig:unfolding-PtBalEff-Final} (the corresponding distribution in the \highpt\  region is provided in appendix \ref{sec:add_data}).  Again, with less quark/gluon radiation from the electroweak process, it is expected that the two jets are better balanced against the \Z-boson for the electroweak \Zdijet\ process than for the strong \Zdijet\ process. This is apparent at high \mjj\ and high \deltay, where the strong \Zdijet\ prediction falls below the data. For this distribution, \powheg\ describes the data poorly at low values of \mjj\ or \deltay, whereas \sherpa\ gives a good description of the data over the full range of the distributions.

In general, neither \sherpa\ nor \powheg\ is able to fully reproduce the data for all distributions in all fiducial regions. \powheg\ gives a better description of the data than \sherpa\ for the \mjj\ and \deltay\ distributions, with \sherpa\ predicting too large a cross section at the highest values of \mjj\ or \deltay. However, \sherpa\ gives a better description for variables sensitive to the additional jet activity in the event, with \powheg\ predicting too little jet activity in the rapidity interval bounded by the dijet system. The unfolded data can be used to constrain the modelling of \Zdijet\ production in the extreme phase-space regions probed in this measurement. The unfolded data are provided in HEPDATA with statistical and systematic uncertainties. Furthermore, the correlation between bins of different distributions is provided, allowing the quantitative comparison of all distributions simultaneously.

\section{Extraction of the electroweak \Zdijet\ fiducial cross section}
\label{sec:fits}

The electroweak \Zdijet\ component is extracted by
fitting the dijet invariant mass reconstructed in the \search\
 region. 
Templates are formed for the signal 
and background processes and a fit to the
dijet invariant mass distribution in the data is performed, allowing the normalisation of each template to float. 
The fit is performed using a log-likelihood maximisation~\cite{Verkerke:2003ir} and the
 number of signal and background events is extracted. 
The number of signal
events is then converted into a fiducial cross section, using a correction factor to convert from the
reconstruction-level event selection to the particle-level event
selection.

\subsection{Template construction and fit results}

The signal template is obtained from the \sherpa\ electroweak \Zdijet\ sample. The background template is constructed from the \sherpa\  strong \Zdijet\ sample plus the small contribution from the diboson and \ttbar\ samples (the other background sources are found to have negligible impact on the results). The background template is then constrained using the following data-driven technique. 
The dijet invariant mass distributions are constructed for data and MC simulation in the \control\ region and
a reweighting function is defined by fitting the ratio of the data to MC simulation with a second-order polynomial. This reweighting function is then applied directly to the background template in the \search\ region. The data are therefore used to constrain the generator modelling of the background \mjj\ shape, and the MC simulation is used only to extrapolate this constraint 
between the \control\ and \search\ regions. This procedure has the
advantage of minimising both the experimental and theoretical systematic
uncertainties on the background template. 
Figure~\ref{fig:mjj-control-both} shows the dijet invariant mass distribution in the
\control\ region for the data and the MC simulation for the electron and muon channels combined. The reweighting function is shown in the lower panel. The use of the \control\ region to constrain the background template is validated in section~\ref{sec:valid-control} and corresponding systematic uncertainties are presented in section~\ref{sec:fit-syst}.

\begin{figure}
  \begin{center}
    \subfigure[]{
      \includegraphics[width=0.47\textwidth]{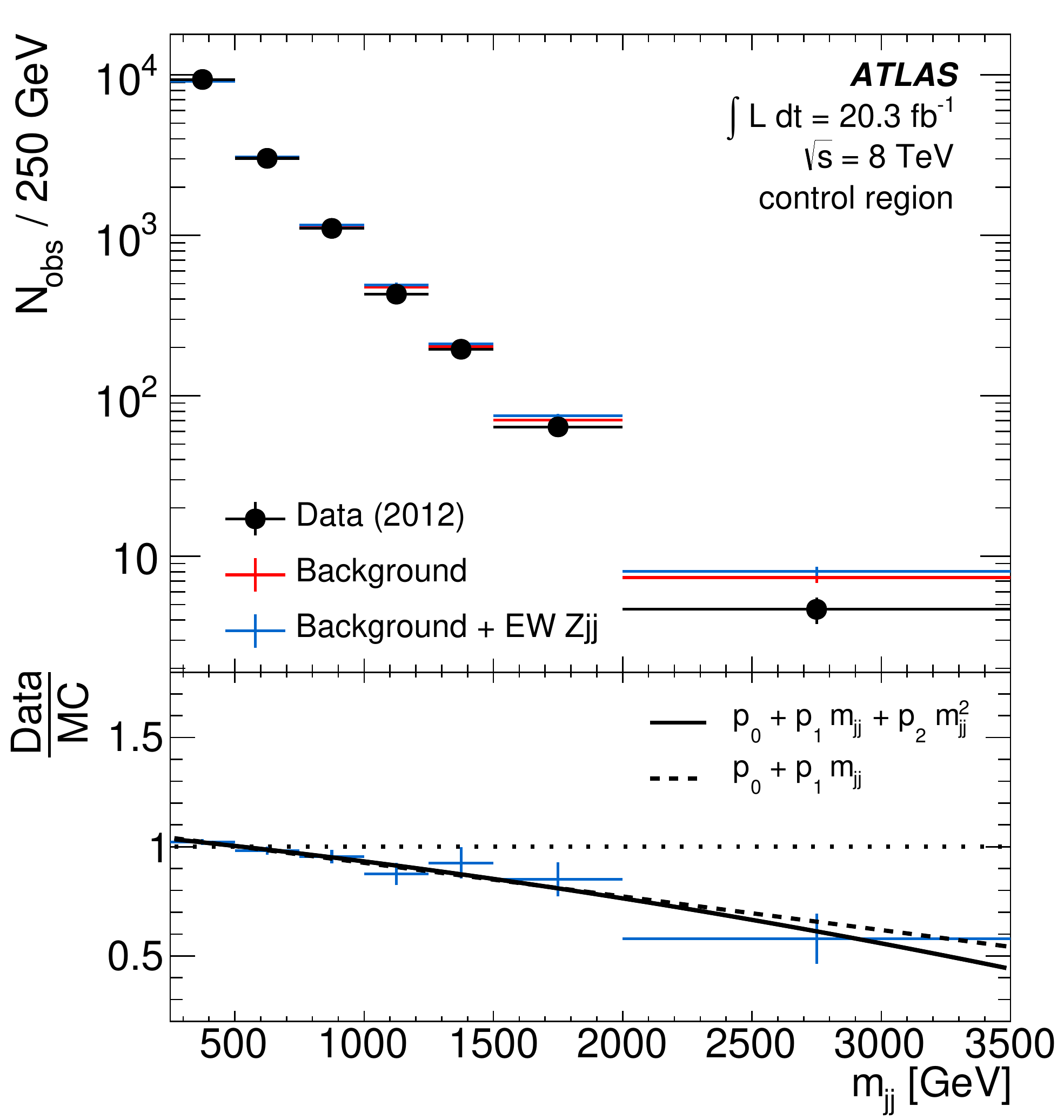}
      \label{fig:mjj-control-both}
    }
    \subfigure[]{
      \includegraphics[width=0.47\textwidth]{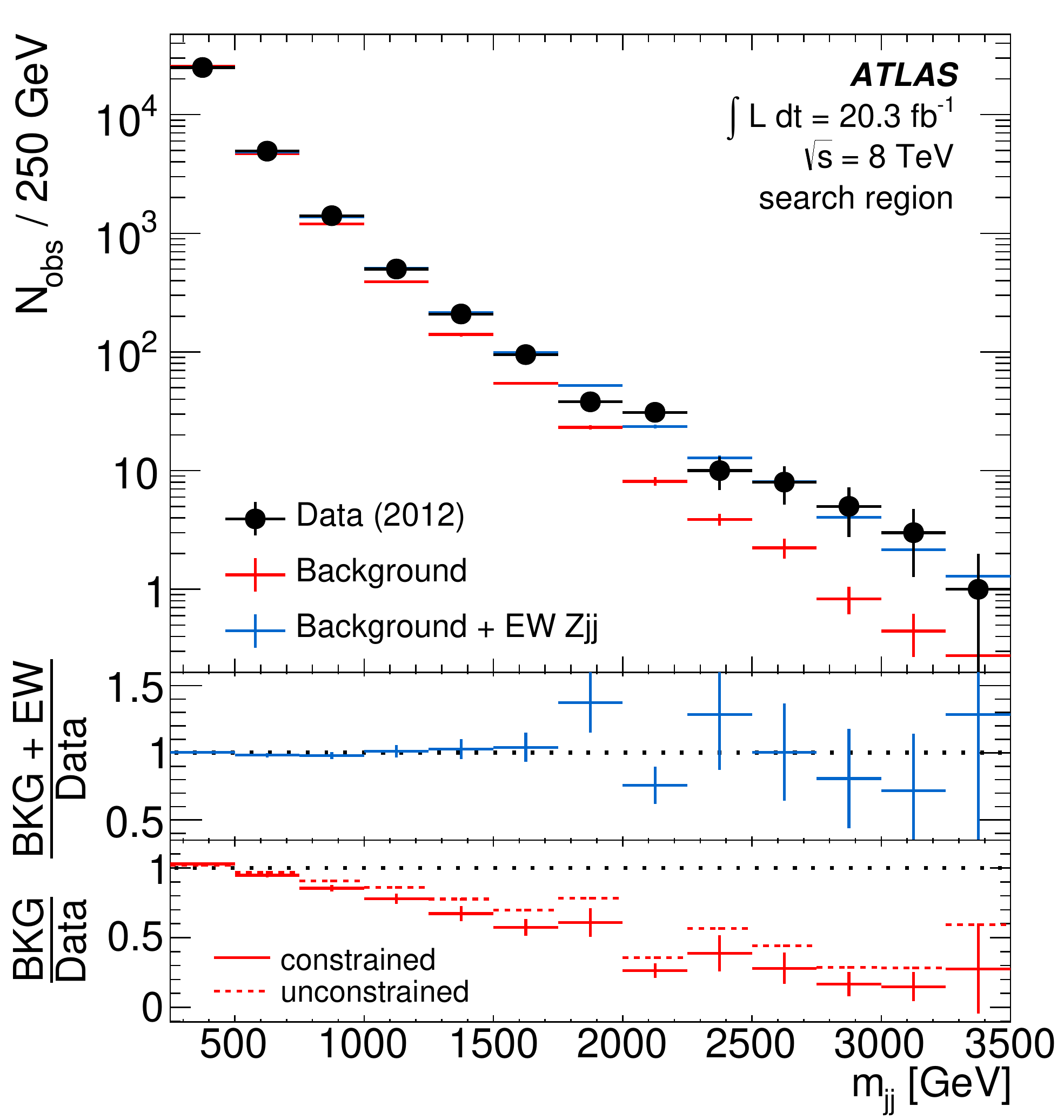}
      \label{fig:fit-result-both}
    }
    \caption{(a) The dijet invariant mass distribution in the \control\
      region. The simulation has been normalised to match the number of events observed in the data. The lower panel shows the reweighting function used to constrain the shape of the background template. 
      (b) The dijet invariant mass distribution in the \search\
      region. The signal and (constrained) background templates 
      are scaled to match the number of events obtained in the
      fit. The lowest panel shows the ratio of constrained and unconstrained background templates to the data.}
    \label{fig:mjj-fits}
  \end{center}
\end{figure}

Figure~\ref{fig:fit-result-both} shows the dijet invariant mass  distribution in the
\search\ region for the electron and muon channels combined.  The signal and background templates are normalised to the values obtained from the fit. The background template is presented after the data-driven reweighting using the second-order polynomial in figure \ref{fig:mjj-fits}~(a). The unconstrained background template is also compared to the data in the lowest panel, demonstrating that the background-only prediction always falls below the data at high-\mjj.

Table~\ref{tab:fit-result} summarises the fit results, giving the
number of signal (\newk) and background (\nbck) events expected by the MC simulation and the
number obtained from the fit, together with the statistical
uncertainties from the data (first uncertainty) and MC
templates (second uncertainty). 
The results are shown for electrons and muons separately and also with both channels combined, where the latter result is obtained by combining the two channels for the data and for the MC templates before fitting. 
For the purpose of measuring the fiducial cross section, the yields from the fits to electrons and muons are used.
For the purpose of determining systematic uncertainties on $\newk$, which are correlated between the two channels, the fractional
shift in the number of events obtained from the fit combining both channels is used. 

\begin{table}
\begin{center}
  \caption{The number of strong (\nbck) and electroweak (\newk) \Zdijet\ events as predicted by the MC simulation and obtained from a fit to the data. The number of events in data is also given. 
    The first and second uncertainties on the fitted yields are due to statistical uncertainties in data and simulation, respectively. 
  The first and second uncertainties in the MC prediction are the experimental and theoretical systematic uncertainties, respectively. 
        }
  \label{tab:fit-result}
\begin{tabular}{l r@{\,}l r@{\,}l r@{\,}l}
 \hline \hline
  & \multicolumn{2}{c}{Electron} & \multicolumn{2}{c}{Muon} & \multicolumn{2}{c}{Electron$+$muon} \rule{0mm}{4mm} \\
 \hline
 Data & \multicolumn{2}{c}{14248} & \multicolumn{2}{c}{17938} & \multicolumn{2}{c}{32186} \rule{0mm}{4mm} \\
 \hline
 MC predicted \nbck & 13700 & $\pm \, 1200\,{}^{+1400}_{-1700}$ & 18600 & $\pm \, 1500\,{}^{+1900}_{-2300}$ & 32600 & $\pm \, 2600\,{}^{+3400}_{-4000}$ \rule{0mm}{4mm} \\
 MC predicted \newk &   602 & $\pm \, 27 \pm18$       &   731 & $\pm \, 29 \pm 22 $       &  1333 & $\pm \, 50 \pm 40$ \rule{0mm}{4mm} \\
\hline
Fitted \nbck & 13351 & $\pm \, 144 \pm 29$ & 17201 & $\pm  \,161 \pm 31$ & 30530 & $\pm \, 216 \pm 40$ \rule{0mm}{4mm} \\
Fitted \newk &   897 & $\pm  \, 92 \pm 27$ &   737 & $\pm  \, 98 \pm 28$ &  1657 & $\pm \, 134 \pm 40$ \rule{0mm}{4mm} \\
 \hline \hline 
\end{tabular}
\end{center}
\end{table}

\subsection{Validation of the control region constraint procedure}\label{sec:valid-control}

The data-driven background constraint derived in the control region is an important component of the analysis as it improves the modelling of the background \mjj\ spectrum and constrains the impact of experimental and theoretical uncertainties. Several cross-checks are performed to validate the method.

The choice of polynomial used to describe the reweighting function is investigated by using a first-order polynomial instead of a second-order polynomial. The lower panel of figure~\ref{fig:mjj-control-both} shows that both choices of polynomial give very similar reweighting functions at low \mjj\ and differ only at the highest values of \mjj. The change in \newk\ is less than 2\% if the first-order polynomial is used to reweight the background template in place of the second-order polynomial.

The choice of event generator is examined by reweighting the simulated dijet invariant mass distribution for strong \Zdijet\ production using the ratio of the \powheg\ and \sherpa\ particle-level predictions. This reweighting is carried out in the \search\ and \control\ regions separately.  \powheg\ has been shown to give a much better description of the data for the dijet invariant mass in figure \ref{fig:unfolding-Mjj-Final} for all fiducial regions. The reweighting to \powheg\ improves the description of the data in the \control\ region. The data-driven reweighting function then becomes much flatter and repeating the full analysis procedure with the new templates produces a result consistent at 0.8\% with the analysis based on the \sherpa\ samples alone.

The choice of control region is studied by splitting it into six subregions that probe the additional jet activity in the rapidity interval between the two leading jets. The \control\ and \search\ regions are distinguished by this additional jet activity and these subregions allow the impact of any mismodelling in the simulation to be explored.   
Two subregions are defined by the transverse momentum of the leading jet in the rapidity interval ($25 < \pt \leq 38$~GeV and $\pt > 38$~GeV), two subregions are defined by the rapidity of the jet ($|y| \leq 0.8$ and $|y|>0.8$), and two subregions are defined by the number of jets in the rapidity interval ($N_{\rm jet}=1$ and $N_{\rm jet}\geq2$).  In addition to these six regions, an MPI-suppressed subregion is defined by the requirements $|\dphijj / \pi | < 0.9$ and $\pt^{jj} > 20$~GeV, where $\pt^{jj}$ is the transverse momentum of the dijet system. This region allows the impact of MPI on the control region constraint to be examined. 

\begin{figure}
  \begin{center}
    \subfigure[]{
      \includegraphics[width=0.47\textwidth]{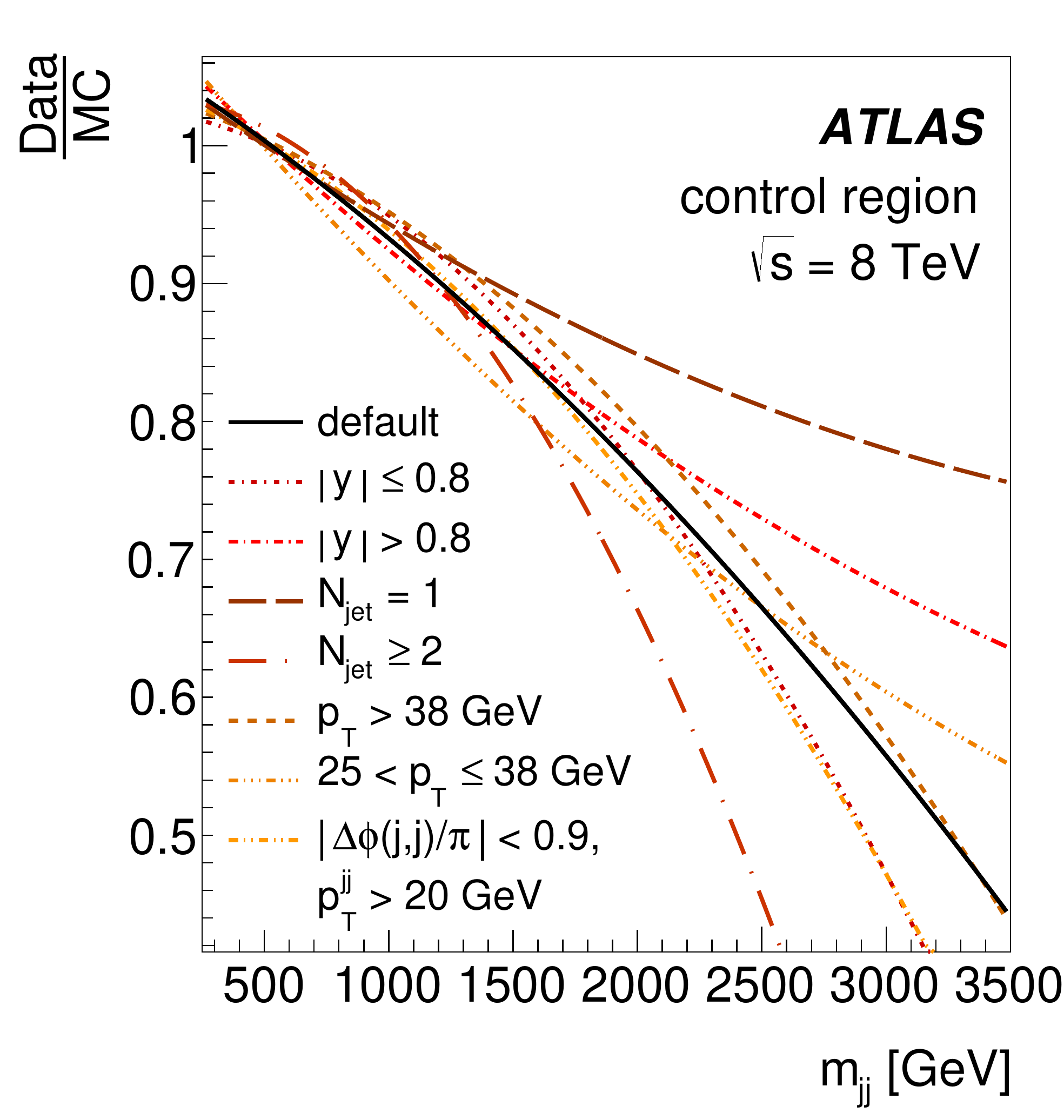}
    }
    \subfigure[]{
      \includegraphics[width=0.47\textwidth]{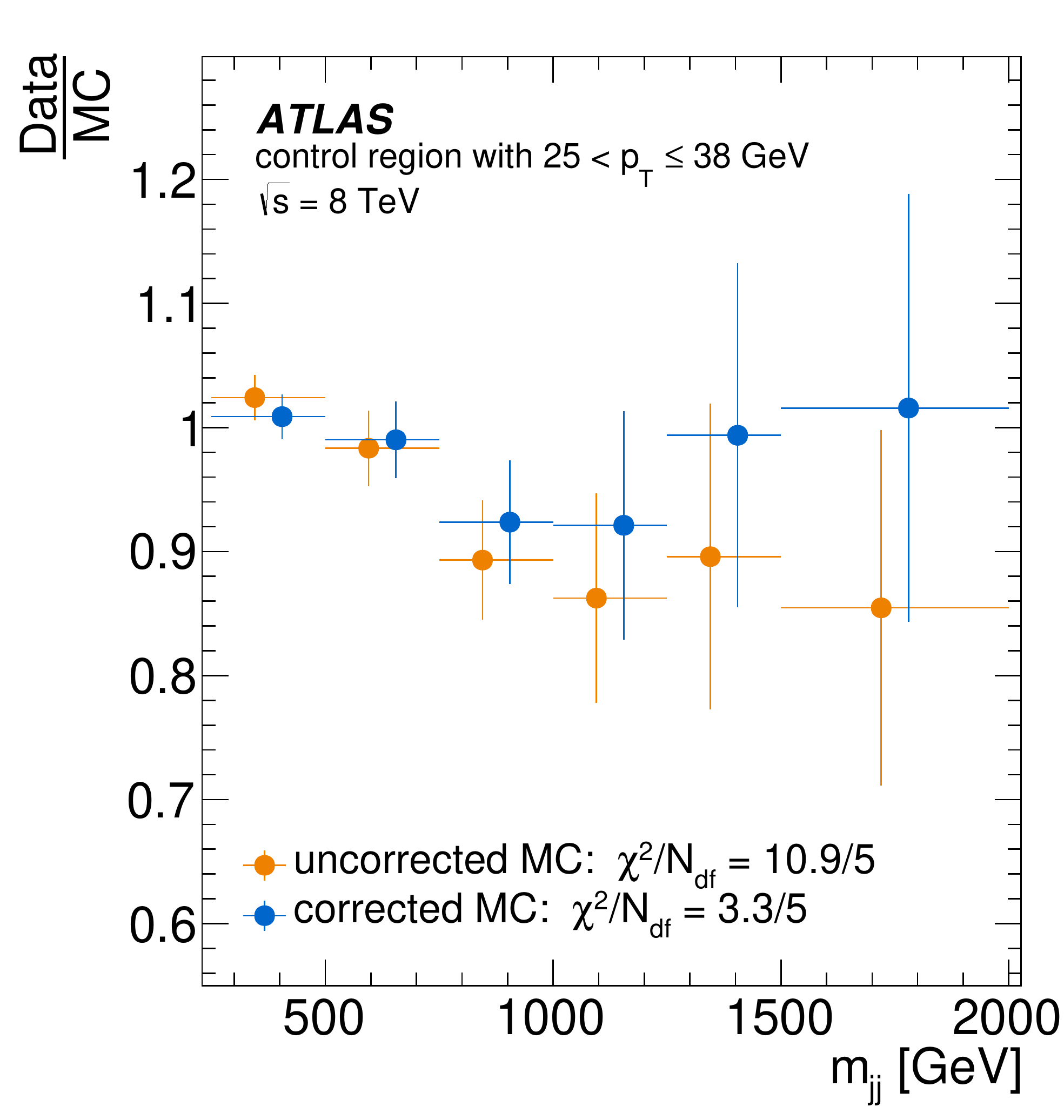}
    }
    \caption{(a) Background reweighting functions obtained for different choices of \control\ region. (b) The agreement between data and simulation in the $25 < \pt \leq 38 $~GeV subregion both before and after applying a background reweighting function derived in the $ \pt > 38$~GeV subregion.}
    \label{fig:control-valid}
  \end{center}
\end{figure}

Figure \ref{fig:control-valid}(a) shows the background reweighting functions obtained from these subregions, compared to the default function obtained from the default \control\ region. The extraction of the electroweak signal is cross-checked using each of these constraints. The values of \newk\ are consistent, with a maximum 5\% spread between subregions. 
This spread is likely to be statistical in origin, as the values of \newk\ obtained from reweighting functions derived in orthogonal subregions are found to agree to better than $1\sigma$ when considering only the statistical uncertainty associated with the reweighting functions. Although the spread of reweighting functions in figure \ref{fig:control-valid}(a) is large at high \mjj, the background modelling in this region has only a small impact on the extracted number of electroweak \Zdijet\ events. The background modelling shape has most impact at values of \mjj\ around 1--1.5~TeV, for which the spread of reweighting functions is just a few percent.

The orthogonal subregions are also used to test the agreement between data and the corrected simulation directly. The reweighting function derived in the $ \pt > 38$~GeV subregion  is used to correct the simulation in the  $25 < \pt \leq 38$~GeV subregion, as shown in figure \ref{fig:control-valid}(b). The corrected simulation gives a better description of the data than the uncorrected simulation. Similar tests are performed for the subregions split by jet rapidity or jet multiplicity. In all cases, the corrected simulation gives a better description of data than the uncorrected simulation.

\subsection{Systematic uncertainties on the fit procedure}\label{sec:fit-syst}

Systematic uncertainties on \newk\ arise from the background template reweighting function, the jet-based experimental systematic uncertainties, and the theoretical modelling uncertainties on the \Zdijet\ samples. The uncertainty due
to the lepton-based systematic uncertainties is negligible. A summary of the systematic uncertainties discussed in this section is presented
in table~\ref{tab:syst-summary}.  
The systematic uncertainty due to the limited number of events in the control region is obtained using pseudo-experiments, and is found to be 8.9\% and 11.2\% in the electron and muon channels, respectively. 
The remaining experimental systematic uncertainties affect the extracted value of \newk\ by changing the shape of the signal template and/or the
shape of the background template. The experimental systematic uncertainties that change the template shape are due to JES, JER, JVF, as well as pileup jet modelling, as discussed in section~\ref{sec:systematics}. The effect on the number of fitted events due to each source of uncertainty is evaluated simultaneously for signal and background templates in order to account for correlations.

Systematic variations in the signal template are evaluated by 
taking the ratio of the template formed with a systematic shift to the nominal template, fitting that
ratio 
with a second-order polynomial, applying that polynomial as
a reweighting function to the signal template, and repeating the fit for the number of electroweak events. The use of the polynomial to estimate the systematic shift reduces the impact of statistical fluctuations at large \mjj.

For the systematic variations in the background template, the data provide a constraint in the \control\ region,
meaning that only the effect of each systematic variation on the extrapolation between the \control\ and
\search\ regions needs to be evaluated. 
A double ratio is formed from the systematic-shifted to nominal ratios in the
\search\ and \control\ regions and fitted
with a first-order polynomial function. If the gradient of the fitted function is statistically significant, defined as the parameter value being greater than 1.64 times the parameter uncertainty, then this component is considered as a significant source of
systematic uncertainty. This significance requirement is chosen to remove 90\%\ of
statistical fluctuations and avoid double counting statistical uncertainties in the simulated samples.\footnote{The choice of significance requirement was investigated by changing the requirement to 1.0 or 2.0. The resultant systematic uncertainties were unchanged from the nominal choice of 1.64.} 
For each significant source of systematic uncertainty, the first-order polynomial is 
applied as an additional reweighting to the background template in the \search\ region and the fit is repeated. 

The dominant systematic uncertainty on the extracted value of \newk\ from experimental sources is from the JES (5.6\%). This uncertainty comes  almost entirely  from the uncertainty on the signal template shape, because the shape of the background template is constrained using the \control\ region. The uncertainty due to the JVF is modest (1.1\%), whereas the uncertainty from JER and pileup modelling is effectively negligible (0.4\% and 0.3\%, respectively).

Additional uncertainties on the extraction of the electroweak component arise from the theoretical modelling in the
MC generators. Again, these affect the signal template as well as
the extrapolation between the \control\ and \search\ region for the
background template. The uncertainties due to theoretical modelling are split into two
components: PDF modelling and generator modelling. 

Uncertainties due to PDF modelling are obtained as follows. The nominal value of \newk\ is obtained using the \ctten\ PDF set. The full analysis is then repeated using simulated samples created using (i) the \ctten\ uncertainties and (ii) the central values and uncertainties of two other PDF sets, \mstw~\cite{Martin:2009iq} and
\nnpdf~\cite{Ball:2012cx}. Each PDF variation is applied to the signal and background simultaneously. For each PDF set, the uncertainty on \newk\ is then calculated using the recommended procedure from each
collaboration~\cite{Campbell:2006wx,Martin:2009bu}, with the \ctten\ results scaled to reflect 68\% probability. The $\alpha_s$ uncertainty is found to be negligible. 
The overall uncertainty due to PDF modelling is found to be ${}^{+1.5}_{-3.9}\%$ from the envelope of uncertainties obtained from each PDF set. 

The generator modelling uncertainties are determined using the dedicated \sherpa\ sample variations discussed in section \ref{sec:theory}, by varying the factorisation and renormalisation scale, varying the activity from multiple parton interactions (MPI), and changing the parton-shower scheme or CKKW matching parameters.  
To evaluate the generator modelling uncertainty, the analysis is repeated for each sample variation independently to obtain a shift in \newk. The standard deviation in the shifted values for each sample variation is obtained using a pseudo-experiment approach.  The effect of the signal modelling uncertainty on \newk\  is found to be 8.9\% from the envelope of the shifts (mean plus standard deviation) produced from the eight dedicated signal templates. A separate uncertainty on \newk\ is obtained from the envelope of the shifts (mean plus standard deviation) produced from the eight dedicated background templates. The uncertainty due to background modelling is found to be 7.5\%. The uncertainties in the signal and background generator modelling are taken as uncorrelated.

The uncertainties on the signal and background modelling are cross-checked by reweighting the simulated dijet invariant mass distribution for strong and electroweak \Zdijet\ production, using the ratio of the \powheg\ and \sherpa\ particle-level predictions. As discussed in section \ref{sec:valid-control}, reweighting the strong \Zdijet\ sample produces a change in \newk\ of just 0.8\%. This is  covered by the background modelling uncertainty determined from the \sherpa\ systematic variations. Reweighting the electroweak \Zdijet\ sample produces a change in \newk\ of 4.6\%, which is also covered by the signal modelling uncertainty assigned from the \sherpa\ systematic variations.

The systematic uncertainty associated with possible interference between electroweak and strong \Zdijet\ production is estimated by reweighting the background template to account for the interference contribution. The interference is determined using the dedicated \sherpa\ samples discussed in section \ref{sec:theory}. These samples use only leading-order matrix elements for \Zdijet\ production and the change in the background template is therefore estimated prior to applying the jet veto. The impact of interference is determined by repeating the full fitting procedure after reweighting the background template in either the \search\ region or the \control\ region alone. This approach assumes the interference affects only one of the two regions and therefore has a maximal impact on the analysis. If the background template is reweighted only in the \search\ region, the extracted value of \newk\ is reduced by 6.2\%. Alternatively, if the background template is reweighted only in the \control\ region, the value of \newk\ increases by 6.2\%. A conservative systematic uncertainty of $\pm 6.2\%$ is assigned to the final measurement. 

\begin{table}
\centering
\caption{Systematic uncertainties, expressed in percentages, on (i) the number of fitted signal events in the \search\ region, $\newk$, and (ii) the correction factor to the particle-level, $\c_{\rm EW}$. The uncertainties are anti-correlated between \newk\ and $\c_{\rm EW}$.}
  \label{tab:syst-summary}
  \def\mylen{\hphantom{1}}
  \begin{tabular}{l | r@{~}l r@{~}l}
  \hline \hline
   Source                     & \multicolumn{2}{c|}{$\Delta\newk$} 
                              & \multicolumn{2}{c}{$\Delta\c_{\rm EW}$}\rule{0mm}{4mm} \\
                              & \multicolumn{1}{c}{Electrons} & \multicolumn{1}{c|}{~~Muons~~}
                              & \multicolumn{1}{c}{Electrons} & \multicolumn{1}{c }{~~Muons~~} \\
  \hline
   Lepton systematics         & \multicolumn{1}{c}{---}
                              & \multicolumn{1}{c|}{---} 
                              & \multicolumn{1}{c}{$\pm 3.2~\%$} 
                              & \multicolumn{1}{c }{$\pm 2.5\%$\rule{0mm}{4mm}} \\
   Control region statistics  & \multicolumn{1}{c}{$\pm 8.9~\%$} 
                              & \multicolumn{1}{c|}{$\pm 11.2~\%$}
                              & \multicolumn{1}{c}{---}
                              & \multicolumn{1}{c }{---\rule{0mm}{4mm}} \\
   \hline
   JES                        & \multicolumn{2}{c|}{$\pm 5.6~\%$}
                              & \multicolumn{2}{c }{${}^{\mylen+2.7}_{\mylen-3.4}~\%$\rule{0mm}{4mm}} \\
   JER                        & \multicolumn{2}{c|}{$\pm 0.4~\%$}
                              & \multicolumn{2}{c }{$\pm 0.8~\%$\rule{0mm}{4mm}} \\
   Pileup jet modelling       & \multicolumn{2}{c|}{$\pm 0.3~\%$}
                              & \multicolumn{2}{c}{$\pm 0.3~\%$\rule{0mm}{4mm}} \\
   JVF                        & \multicolumn{2}{c|}{$\pm 1.1~\%$}
                              & \multicolumn{2}{c}{${}^{\mylen+0.4}_{\mylen-1.0}~\%$\rule{0mm}{4mm}} \\
   Signal modelling           & \multicolumn{2}{c|}{$\pm 8.9~\%$}
                              & \multicolumn{2}{c}{${}^{\mylen+0.6}_{\mylen-1.0}~\%$\rule{0mm}{4mm}} \\
   Background modelling       & \multicolumn{2}{c|}{$\pm 7.5~\%$}
                              & \multicolumn{2}{c}{---\rule{0mm}{4mm}} \\
   Signal/background interference & \multicolumn{2}{c|}{$\pm 6.2~\%$}
   			  & \multicolumn{2}{c}{---\rule{0mm}{4mm}} \\
   PDF                        & \multicolumn{2}{c|}{${}^{\mylen+1.5}_{\mylen-3.9}~\%$}
                              & \multicolumn{2}{c}{$\pm 0.1~\%$\rule{0mm}{4mm}} \\
                              
  \hline\hline
  \end{tabular}
\end{table}

\subsection{Measurement of fiducial cross section}
\label{sec:convert}
The fitted values of \newk\ for the electron and muon channels are
converted to a fiducial cross section, defined as:
\begin{equation}
\sewk = \frac{\newk}{{\lint \, \, {\cdot} \, \, \cal{C}_{\rm EW}}}
\label{eqn:EWKxsecDefn}
\end{equation}
where  $\c_{\rm EW}$ is a
correction factor based on the reconstruction- to particle-level ratio of the \sherpa\ prediction
for electroweak \Zdijet\ production in the \search\   region.

The correction factors are 0.80 and 0.66 in the muon and electron channels, respectively. The difference in correction factor between the two channels arises primarily from the different reconstruction and identification efficiencies for muons and electrons.
The systematic
uncertainties on the correction factor are divided into those that are uncorrelated between the
electron and muon channels (MC
sample statistics, lepton reconstruction, identification, trigger, energy scale
and energy smearing) and those that are correlated (JES, JER, JVF,
pileup jet modelling, generator modelling and PDFs). 
The generator modelling can affect the correction factor due to differences in the
kinematics of final-state particles. This uncertainty is 
determined by reweighting the nominal MC simulation such that the particle-level distributions match those of the dedicated \sherpa\ model variations discussed in section \ref{sec:theory}. This is carried out for all kinematic distributions for which a cut is made in defining the \search\ region and the resulting uncertainties are added in quadrature. 
A breakdown of the uncertainties on the correction factor is given in table \ref{tab:syst-summary}. The JES and lepton identification are the largest sources of uncertainty.

For each source of systematic uncertainty, the impact on \newk\ and $\c_{\rm EW}$ is found to be anti-correlated, and the fractional uncertainty on the measured cross section is therefore obtained from a linear combination of the fractional uncertainties on \newk\ and $\c_{\rm EW}$. The total systematic uncertainty on the measured cross section is then taken to be the quadrature sum of the individual sources of systematic uncertainty.

The fiducial cross sections in the electron and muon channels are
\def\mylen{\hphantom{1}}
\begin{eqnarray*}
\sewk^{ee}     &= 67.2 \,\pm 6.9\,{\rm (stat)}\,{}^{+12.7}_{-13.4}\,{\rm (syst)}\,\pm 1.9\,{\rm
(lumi)\,fb}\hphantom{.} \qquad \rm{and}  \\
\sewk^{\mu\mu} &= 45.6 \,\pm 6.1\,{\rm (stat)}\,{}^{\mylen+9.1}_{\mylen-9.6}\,{\rm (syst)}\,\pm
1.3\,{\rm (lumi)\,fb}. \hphantom{\qquad \rm{and}}
\end{eqnarray*}
These measurements are consistent at the 1.7$\sigma$ level, 
accounting for only those uncertainties that are uncorrelated 
between the two channels. The channels are then combined using a weighted 
average, with the weight of each channel defined as the squared inverse of the 
uncorrelated uncertainties. The combined fiducial cross section is 
\begin{equation*}
\sewk = 54.7 \,\pm 4.6\,{\rm (stat)}\,{}^{+9.8}_{-10.4}\,{\rm (syst)}\,\pm 1.5\,{\rm (lumi)\,fb}.
\end{equation*}
The theoretical prediction from \powheg\ for the electroweak \Zdijet\ cross section is \linebreak
$46.1 \pm 0.2\,{\rm (stat)}\,{}^{+0.3}_{-0.2}\,{\rm (scale)}\,\pm 0.8 \,{\rm (PDF)}\,
\pm 0.5 \,{\rm (model)~fb}$, which is in good agreement with the data.

A detector-corrected fiducial cross section for electroweak \Zdijet\ production is 
also determined for the \search\ region with $\mjj > 1$~TeV, using the integral of the 
fitted signal template. In this region, electroweak production accounts for approximately 35\% 
of the events. The region at large dijet invariant mass is therefore the part of the spectrum that is most sensitive 
to the electroweak \Zdijet\ component and the least sensitive to the background normalisation. The measured 
cross section for electroweak \Zdijet\ production in the \search\ region with $\mjj > 1$~TeV is 
\begin{equation*}
\sewk\,(\mjj > 1\,{\rm TeV}) = 10.7\pm 0.9\,{\rm (stat)}\,\pm 1.9\,{\rm (syst)} \,\pm 0.3\,{\rm (lumi)\,fb},
\end{equation*}
which is again in good agreement with the theoretical prediction from \powheg,\linebreak
$9.38 \pm 0.05\,{\rm (stat)}\, {}^{+0.15}_{-0.24} \,{\rm (scale)}\, \pm 0.24 \,{\rm (PDF)}\,
\pm 0.09 \,{\rm (model)\,fb}$.

\subsection{Estimate of signal significance}

The significance of the measurement is estimated using pseudo-experiments. Pseudo-data are created for the \search\ and \control\ regions from the constrained background templates, after scaling the simulation such that the integral of the template in the \control\ region matches the number of events observed in the data. Each bin in the pseudo-data is randomly generated from a Poisson distribution with its mean set to the expected number of events in the normalised constrained templates. Signal and background templates are constructed from the nominal templates by smearing the template shape according to experimental and theoretical systematic uncertainties, which are taken to be Gaussian-distributed. The complete analysis procedure is then performed, including the use of the pseudo-data in the \control\ region to construct the reweighting function to apply to the background template in the \search\ region. The pseudo-data in the \search\ region are subsequently fitted with the new signal and background templates and a value of \newk\ is extracted. The process is repeated one billion times and none of the pseudo-experiments produce a value of \newk\ greater than (or equal to) the 1657 events observed in data. The background-only hypothesis is therefore rejected at greater than 5$\sigma$ significance.\footnote{To cross-check the possible impact of non-Gaussian tails in the systematic uncertainties, the pseudo-experiments are repeated using templates smeared (for each source of systematic uncertainty) according to a uniform distribution in the range $-5$ to $+5$ times the systematic uncertainty. 
In this extremely conservative approach, the background-only hypothesis is still rejected at greater than 5$\sigma$ significance.}

\subsection{Limits on anomalous triple gauge couplings}

The observation of electroweak \Zdijet\ production allows limits to be placed on anomalous triple gauge couplings (aTGCs). The potential benefits of using the electroweak \Zdijet\ channel as a probe of aTGCs have been discussed previously in the literature \cite{Baur:1993fv}. In the standard hadron collider analyses, aTGC limits are set by measuring vector boson pair production, for which all three gauge bosons entering the $WWZ$ vertex have time-like four-momentum. In the VBF diagram, however, two of the gauge bosons entering the $WWZ$ vertex have space-like four-momentum transfer. Electroweak  \Zdijet\ production therefore offers a complementary test of aTGCs, because the effects of boson propagators present in electroweak \Zdijet\ production are different from those in vector boson pair production. Reference \cite{Baur:1993fv} emphasises that full information on triple gauge boson couplings can be obtained only if electroweak vector boson production is measured in addition to vector boson pair production.

The effective Lagrangian, $\mathcal{L}$, for aTGCs can be written as
\begin{equation} 
\frac{\mathcal{L}}{g_{WWZ}} = i \left[ g_{1,Z} \left( W^{\dagger}_{\mu\nu} W^{\mu} Z^{\nu}  - W_{\mu\nu} W^{\dagger \mu} Z^{\nu} \right) +   \kappa_Z W^{\dagger}_{\mu} W_{\nu} Z^{\mu\nu} + \frac{\lambda_Z}{m_W^2} W^{\dagger}_{\rho\mu} W_{\nu}^{\mu} Z^{\nu\rho} \right]
\end{equation}
if only those terms that conserve charge conjugation and parity are retained from the general expression \cite{Hagiwara:1986vm}. Here, $g_{WWZ} = -e\, \rm{cot}\, \theta_W$, $e$ is the electric charge, $\theta_W$ is the weak mixing angle, $W^{\mu}$ and $Z^{\mu}$ are the $W$-boson and $Z$-boson fields, $X_{\mu\nu} = \partial_{\mu}X_{\nu} - \partial_{\nu}X_{\mu}$ for $X=W$ or $Z$, and $g_{1,Z}$, $\kappa_Z$ and $\lambda_{Z}$ are dimensionless couplings. The SM values of these dimensionless couplings are $g_{1,Z}^{SM}=1$, $\kappa_Z^{SM}=1$ and $\lambda_{Z}^{SM}=0$. 

The tree-level $S$-matrix for this effective Lagrangian violates unitarity at large energy scales. Unitarity is restored in the full theory by propagator (form factor) effects. A typical approach is to modify the couplings by a dipole form factor
\begin{equation}
a(\hat{s}) = \frac{a_0}{(1 + \hat{s}/\Lambda^2)^2}
\end{equation}
where $a_0$ is the bare coupling, $\hat{s}$ is the partonic centre-of-mass energy and $\Lambda$ is a unitarisation scale. In this measurement, limits are placed on the aTGC parameters for unitarisation scales set to $\Lambda=6$ TeV and $\Lambda=\infty$. The unitarisation scale of 6~TeV is chosen as the largest common value allowed by unitarity considerations \cite{Baur:1987mt,Aihara:1995iq} for all anomalous couplings probed in the measurement.

The extracted number of events in the \search\ region with $\mjj > 1$~TeV is used to place limits on the aTGCs as this region is the least affected by the background normalisation and signal template shape. There are 900 events observed in the data in this region. The expected number of electroweak \Zdijet\ events in the SM is 261, estimated using the \sherpa\ sample. The expected number of background events is 592, estimated after fitting the \mjj\ spectrum with signal and background templates. Note that the normalisation of the background template is effectively governed by the low-\mjj\ region, thus limiting the impact of background modelling in the aTGC limit setting in the high \mjj\ tail. The aTGC parameters (and form factors) are varied within the \sherpa\ event generator, allowing the change in the number of electroweak \Zdijet\ events to be estimated. 

Frequentist confidence intervals are set for the anomalous couplings by performing a profile likelihood test \cite{Cowan:2010js}. A Poisson likelihood function is constructed from the observed number of events, the expected signal as a function of the anomalous couplings and the estimated number of background events. The systematic uncertainties are included in the likelihood function as nuisance parameters with correlated Gaussian constraints. A given aTGC parameter point is rejected at 95\% confidence level if more than 95\% of randomly generated pseudo-experiments exhibit a value of the profile likelihood ratio larger than that observed in data.   

The limits presented in this paper are sensitive to the momentum transfer in the $t$-channel due to the high-\mjj\ requirement and, therefore, most sensitive to the terms in the Lagrangian that contain a derivative of the $W$-boson field. Table \ref{tab:aTGC} shows the 95\% confidence intervals  obtained on the anomalous coupling parameters  $\lambda_{Z}$ and $g_{1,Z}$. The limits are not as stringent as those set in $WZ$ production \cite{Aad:2012twa}, which are approximately a factor of three smaller in the $\lambda_{Z}$ coupling, for example. 

\begin{center}
\begin{table}
\centering
\caption{The 95\% confidence intervals obtained on the aTGC parameters from counting the number of events with $\mjj > 1$~TeV in the \search\ region. Observed and expected intervals, labelled `obs' and `exp' respectively, are presented for unitarisation scales of $\Lambda=6$ TeV and $\Lambda=\infty$. The parameter $\Delta g_{1,Z}$ refers to the deviation of $g_{1,Z}$ from the SM value.}
\label{tab:aTGC}
\begin{tabular*}{0.9\textwidth}{l c c c c}
\hline\hline
aTGC  & $\Lambda=6$~TeV (obs) & $\Lambda=6$~TeV (exp) & $\Lambda=\infty$~(obs) & $\Lambda=\infty$~(exp)  \\
\hline
$\Delta g_{1,Z} $  & [$-0.65$, $0.33$] & [$-0.58$, $0.27$] &  [$-0.50$, $0.26$] & [$-0.45$, $0.22$] \\
 $\lambda_{Z}$  &  [$-0.22$, $0.19$]  & [$-0.19$, $0.16$] & [$-0.15$, $0.13$] & [$-0.14$, $0.11$] \\
\hline\hline
\end{tabular*}
\end{table}
\end{center}

\section{Summary}
\label{sec:conclusions}
Fiducial cross sections for electroweak \Zdijet\ production have been presented for proton-proton collisions at $\sqrt{s}=$ 8~TeV, using a dataset corresponding to an integrated luminosity of 20.3~fb$^{-1}$ collected by the ATLAS experiment at the Large Hadron Collider. The background-only model has been rejected above the 5$\sigma$ level and these measurements  constitute 
 observation of the electroweak \Zdijet\ process. 
The measured cross sections are in good agreement with the Standard Model expectation and limits have been set on anomalous triple gauge couplings. 
In addition, cross sections and differential distributions have been
measured for inclusive \Zdijet\ production in five fiducial regions. The cross-section measurements are all in good
agreement with the prediction from \powheg\ for \Zdijet\ production. The differential
distributions are sensitive to the electroweak component of \Zdijet\ production, as well as the modelling of strong \Zdijet\ production in the extreme phase-space regions probed. The data are compared to theoretical predictions
from the \sherpa\ and \powheg\ event generators. Neither prediction is able to fully reproduce the data for all distributions and the data can be used to constrain the theoretical modelling in these extreme phase-space regions.

\section*{Acknowledgements}


We thank CERN for the very successful operation of the LHC, as well as the
support staff from our institutions without whom ATLAS could not be
operated efficiently. We also thank  Stefan Hoeche and Frank Krauss for insight and cross-checks related to the interference between electroweak and strong Zjj production as predicted by the Sherpa event generator.

We acknowledge the support of ANPCyT, Argentina; YerPhI, Armenia; ARC,
Australia; BMWF and FWF, Austria; ANAS, Azerbaijan; SSTC, Belarus; CNPq and FAPESP,
Brazil; NSERC, NRC and CFI, Canada; CERN; CONICYT, Chile; CAS, MOST and NSFC,
China; COLCIENCIAS, Colombia; MSMT CR, MPO CR and VSC CR, Czech Republic;
DNRF, DNSRC and Lundbeck Foundation, Denmark; EPLANET, ERC and NSRF, European Union;
IN2P3-CNRS, CEA-DSM/IRFU, France; GNSF, Georgia; BMBF, DFG, HGF, MPG and AvH
Foundation, Germany; GSRT and NSRF, Greece; ISF, MINERVA, GIF, DIP and Benoziyo Center,
Israel; INFN, Italy; MEXT and JSPS, Japan; CNRST, Morocco; FOM and NWO,
Netherlands; BRF and RCN, Norway; MNiSW and NCN, Poland; GRICES and FCT, Portugal; MNE/IFA, Romania; MES of Russia and ROSATOM, Russian Federation; JINR; MSTD,
Serbia; MSSR, Slovakia; ARRS and MIZ\v{S}, Slovenia; DST/NRF, South Africa;
MINECO, Spain; SRC and Wallenberg Foundation, Sweden; SER, SNSF and Cantons of
Bern and Geneva, Switzerland; NSC, Taiwan; TAEK, Turkey; STFC, the Royal
Society and Leverhulme Trust, United Kingdom; DOE and NSF, United States of
America.

The crucial computing support from all WLCG partners is acknowledged
gratefully, in particular from CERN and the ATLAS Tier-1 facilities at
TRIUMF (Canada), NDGF (Denmark, Norway, Sweden), CC-IN2P3 (France),
KIT/GridKA (Germany), INFN-CNAF (Italy), NL-T1 (Netherlands), PIC (Spain),
ASGC (Taiwan), RAL (UK) and BNL (USA) and in the Tier-2 facilities
worldwide.

\appendix
\section{Additional inclusive \Zdijet\ differential distributions}
\label{sec:add_data}

In this section, unfolded inclusive \Zdijet\ distributions are presented in fiducial regions that complement the data presented in Sec.~\ref{sec:diffdist}. These  additional data are fully corrected for detector effects and available in HEPDATA. The unfolded \dsdmjj\ and \dsddeltay\ distributions are shown in figure~\ref{fig:unfolding-Mjj-Final-aux}~and~\ref{fig:unfolding-Dy-Final-aux}, respectively, for the \highpt\ and \control\ regions. Figure~\ref{fig:unfolding-GapFraction-Final-aux} shows the unfolded jet veto efficiency and \avgnjet\ distributions as a function of \mjj\ and \deltay\ in the \highpt\  region. Finally, the unfolded \ptscfrac\  as a function of \mjj\ and \deltay\ in the \highpt\  region is shown in figure~\ref{fig:unfolding-PtBalEff-Final-aux}. 

\begin{figure}[h]
  \begin{center}
    \subfigure[] {
      \includegraphics[width=0.47\textwidth]{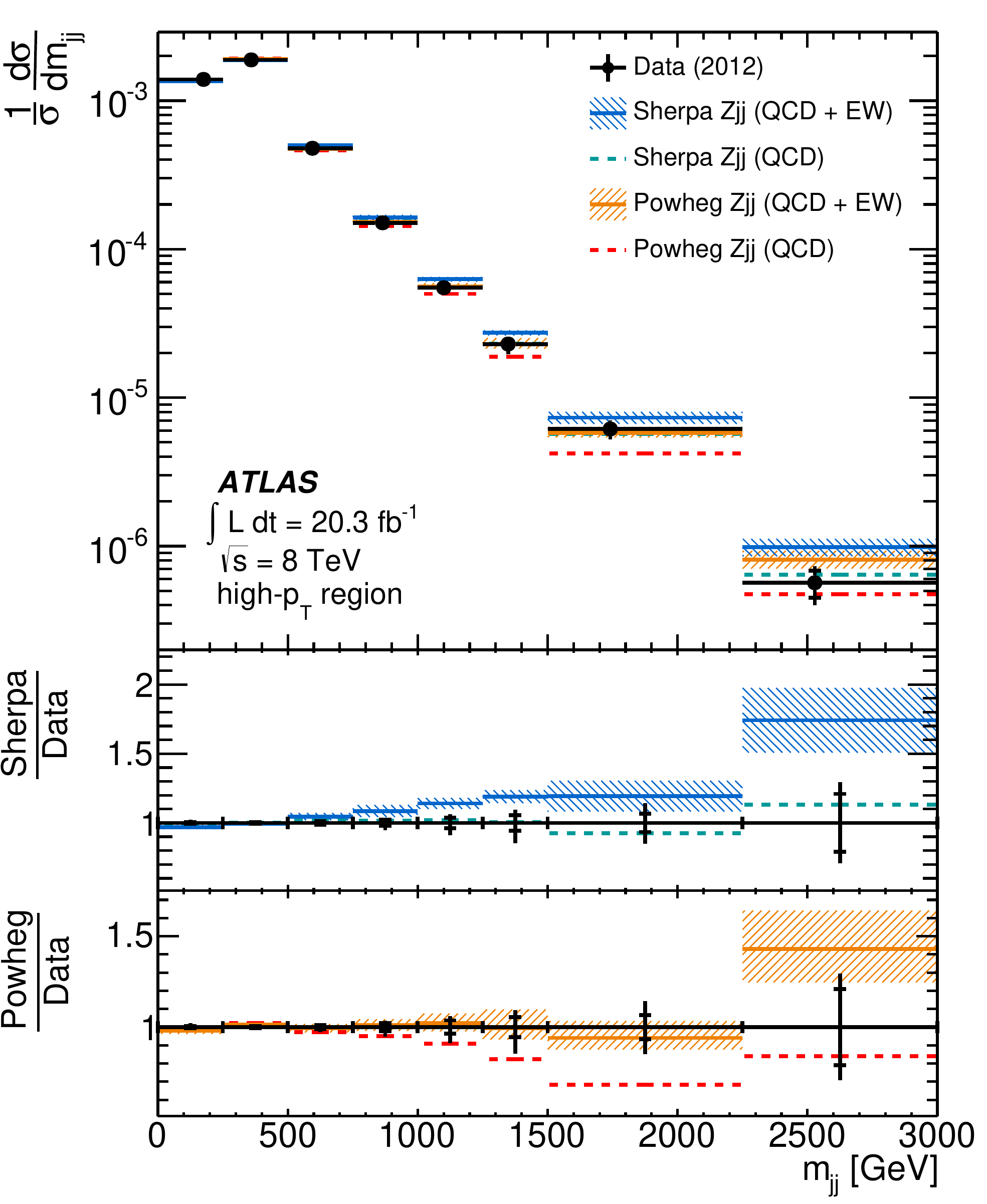}\quad
    }
    \subfigure[] {
      \includegraphics[width=0.47\textwidth]{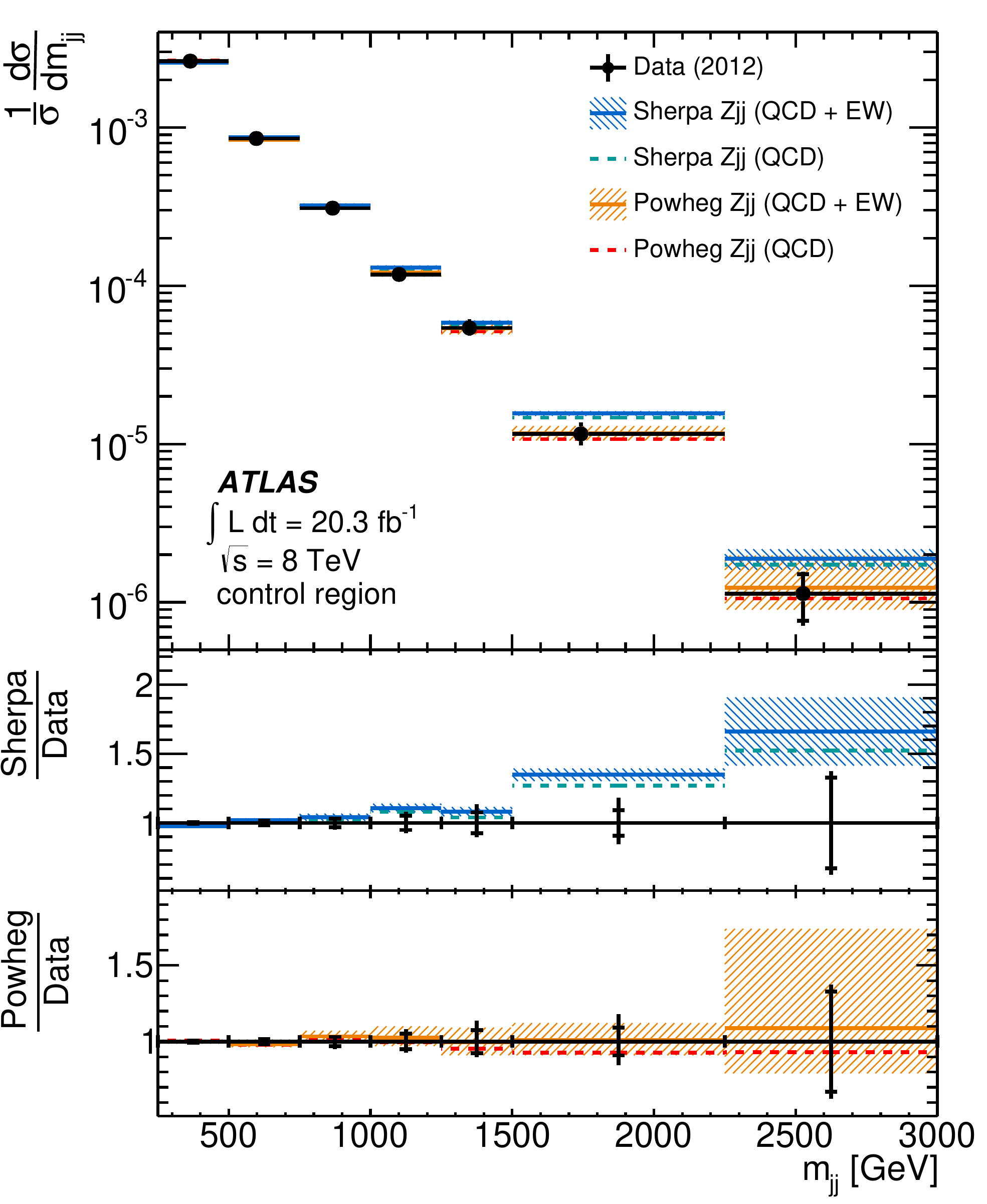}
    }
        \caption[]{Unfolded \dsdmjj\ distribution in (a) the \highpt\ and (b) \control\ 
      regions. \resultCaptionTwo} 
    \label{fig:unfolding-Mjj-Final-aux}
  \end{center}
\end{figure}

\begin{figure}[t]
  \begin{center}
    \subfigure[] {
      \includegraphics[width=0.47\textwidth]{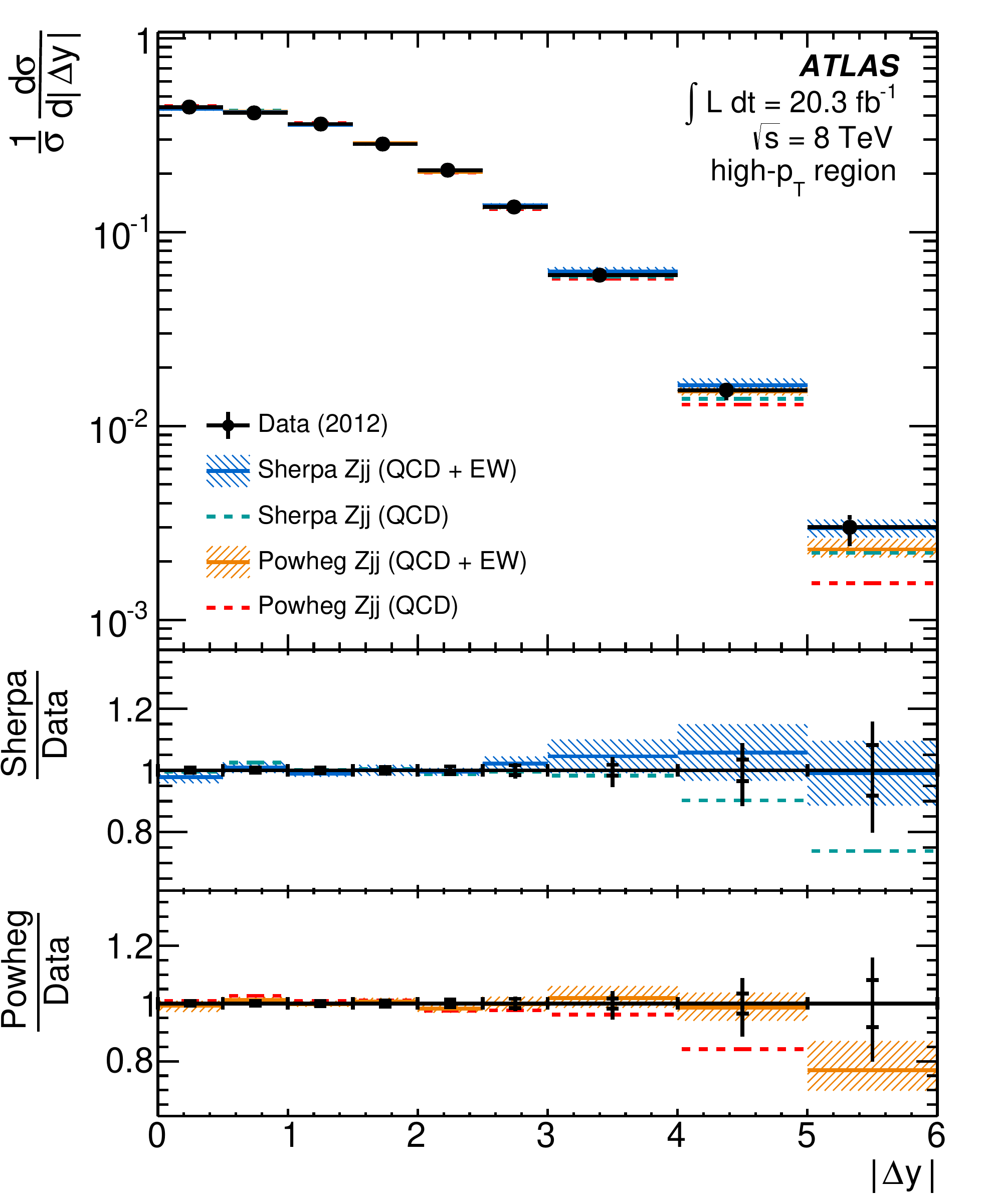}\quad
    }
    \subfigure[] {
      \includegraphics[width=0.47\textwidth]{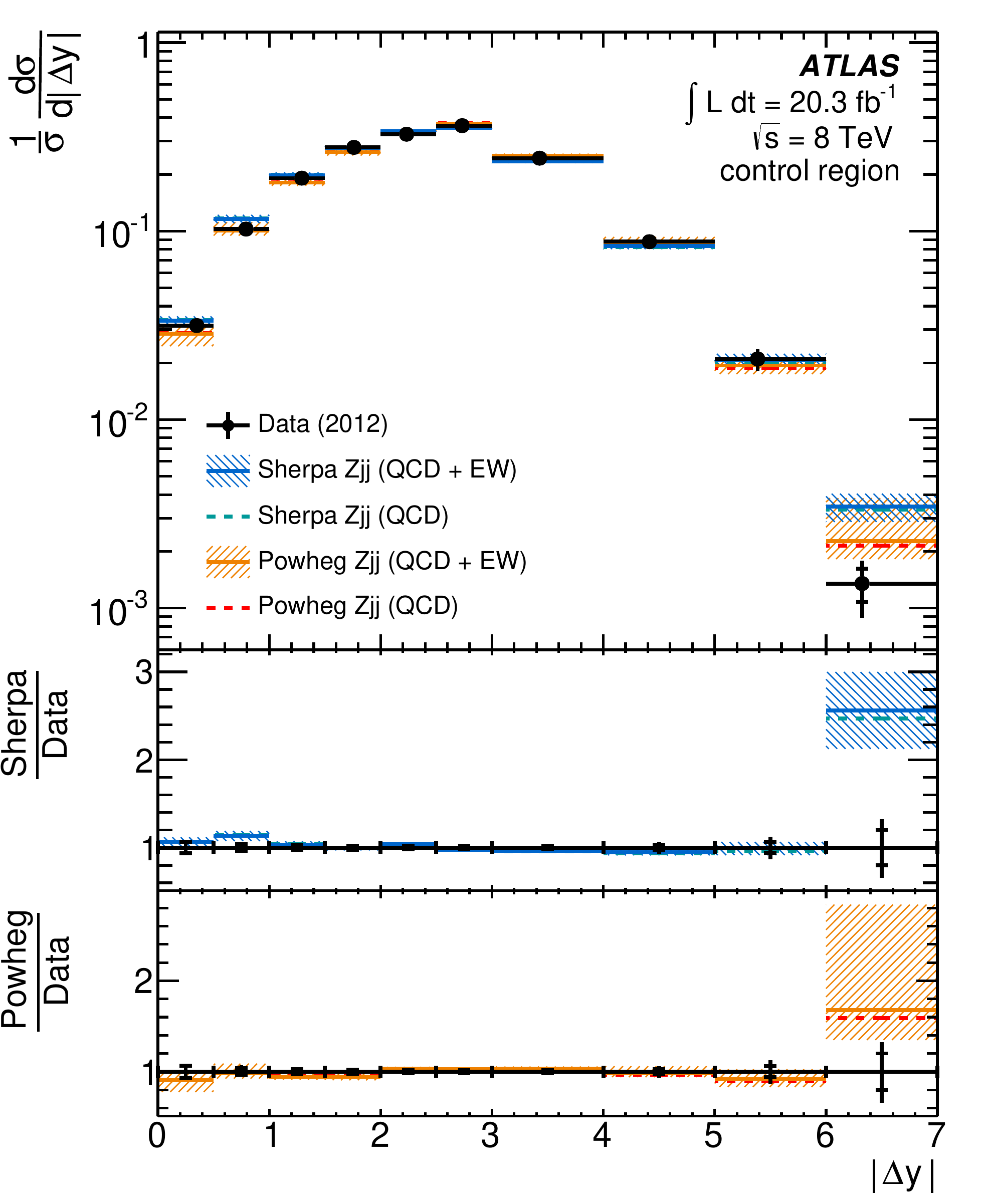}
    }
    \caption[]{Unfolded \dsddeltay\ distribution in (a) the \highpt\ and (b) \control\ 
      regions. \resultCaptionTwo} 
    \label{fig:unfolding-Dy-Final-aux}
  \end{center}
\end{figure}

\begin{figure}
  \begin{center}
    \subfigure[] {
      \includegraphics[width=0.47\textwidth]{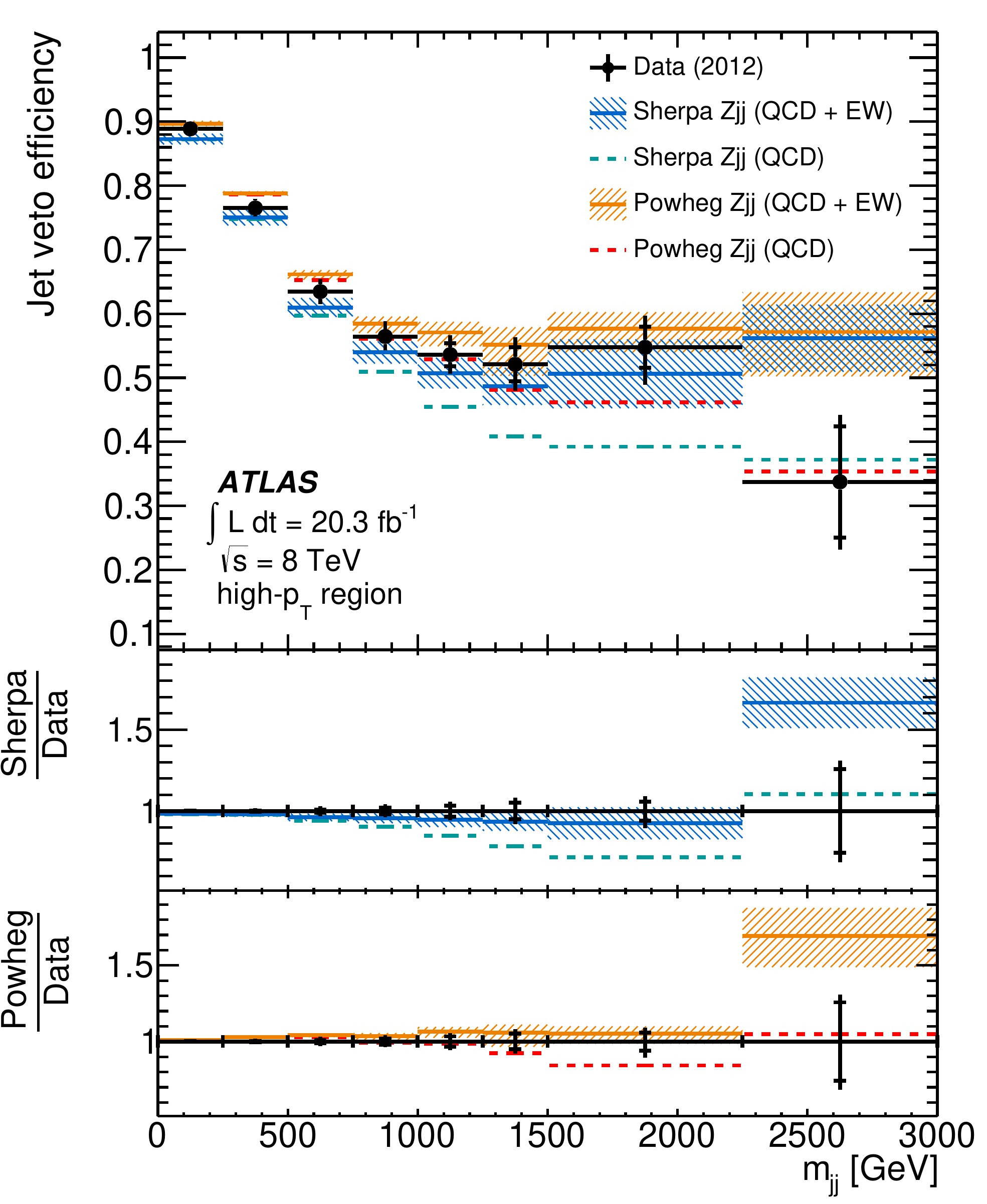}\quad
    }   
     \subfigure[] {
      \includegraphics[width=0.47\textwidth]{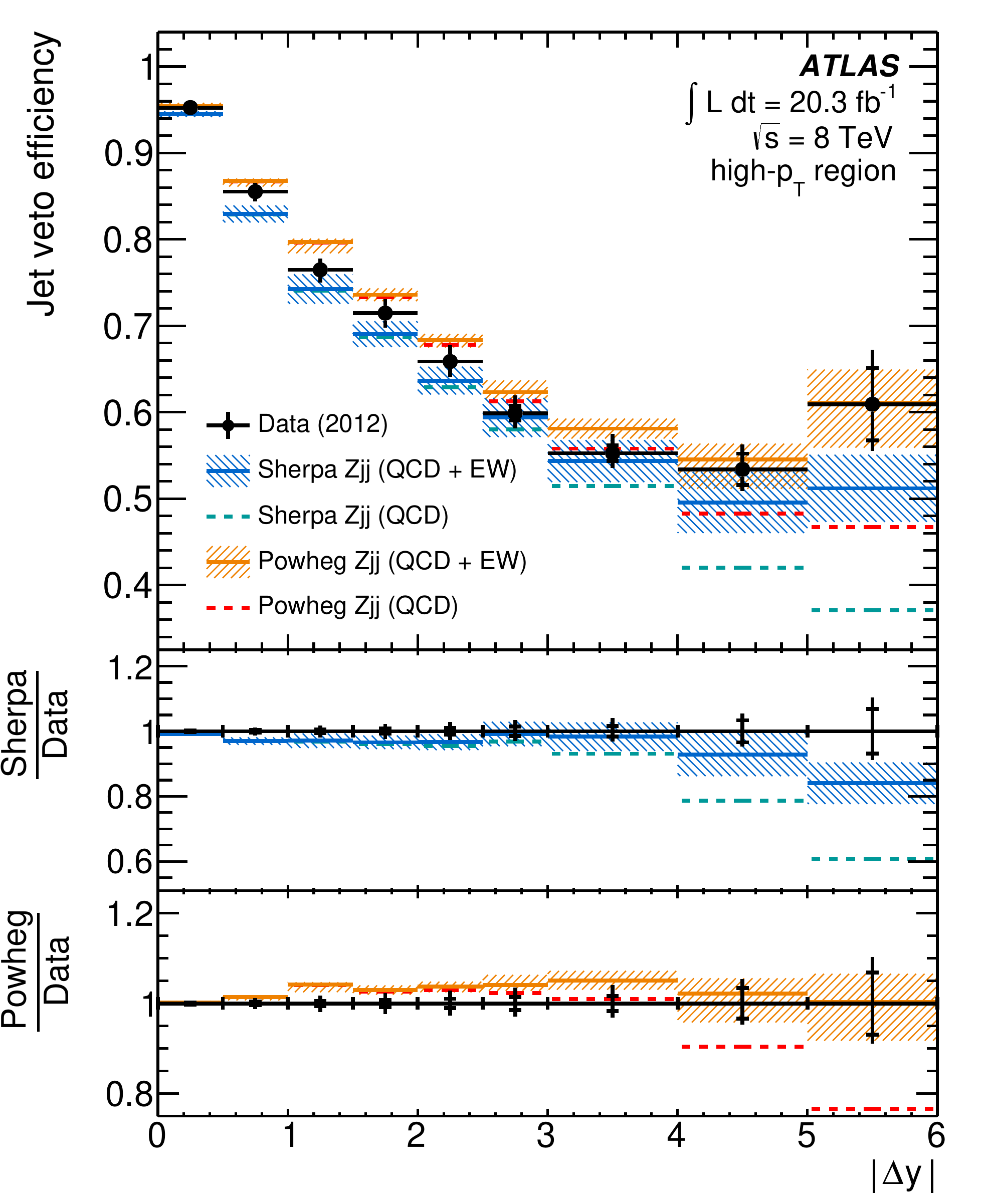}
    }
     \subfigure[] {
      \includegraphics[width=0.47\textwidth]{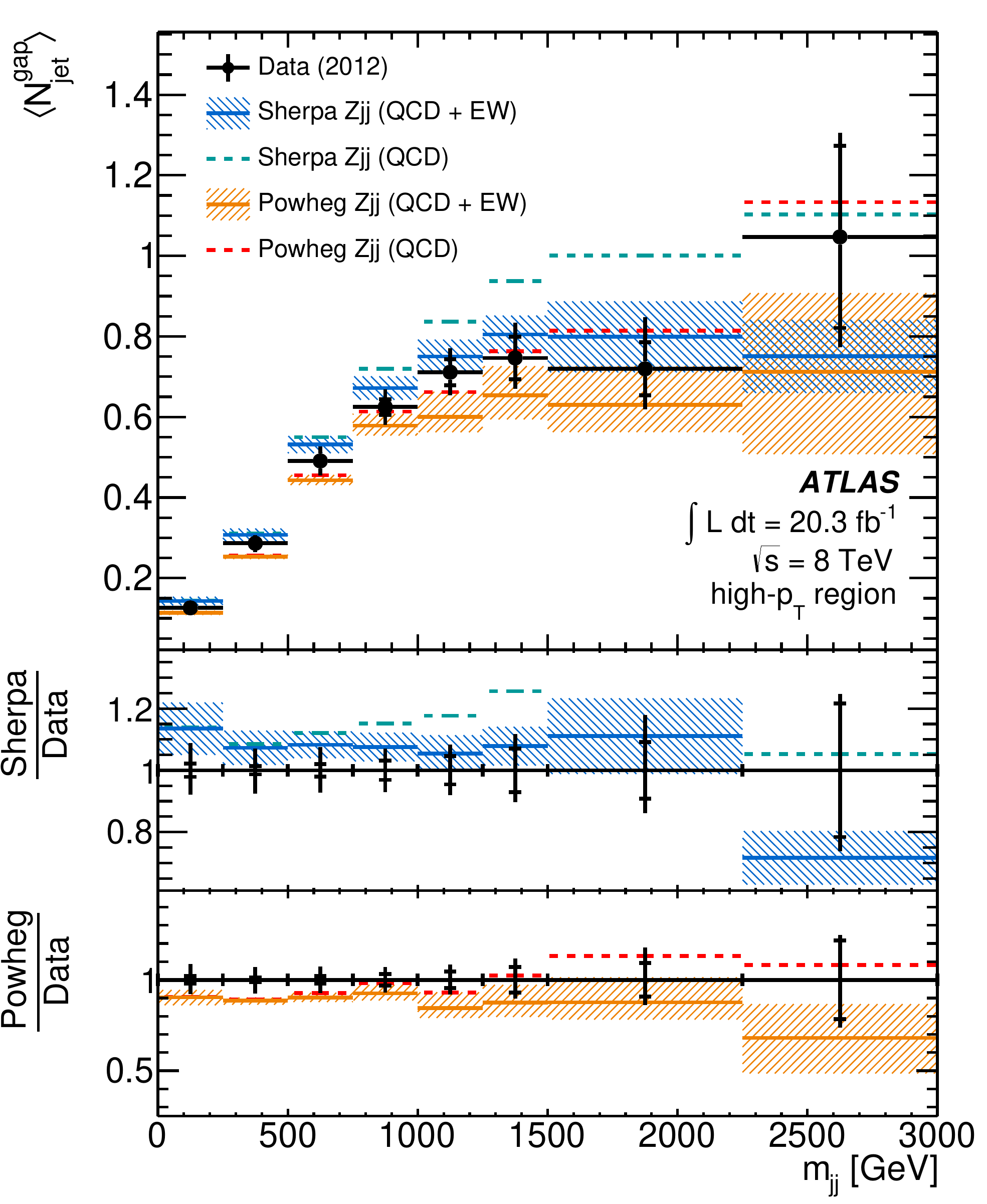}\quad
    }
    \subfigure[] {
      \includegraphics[width=0.47\textwidth]{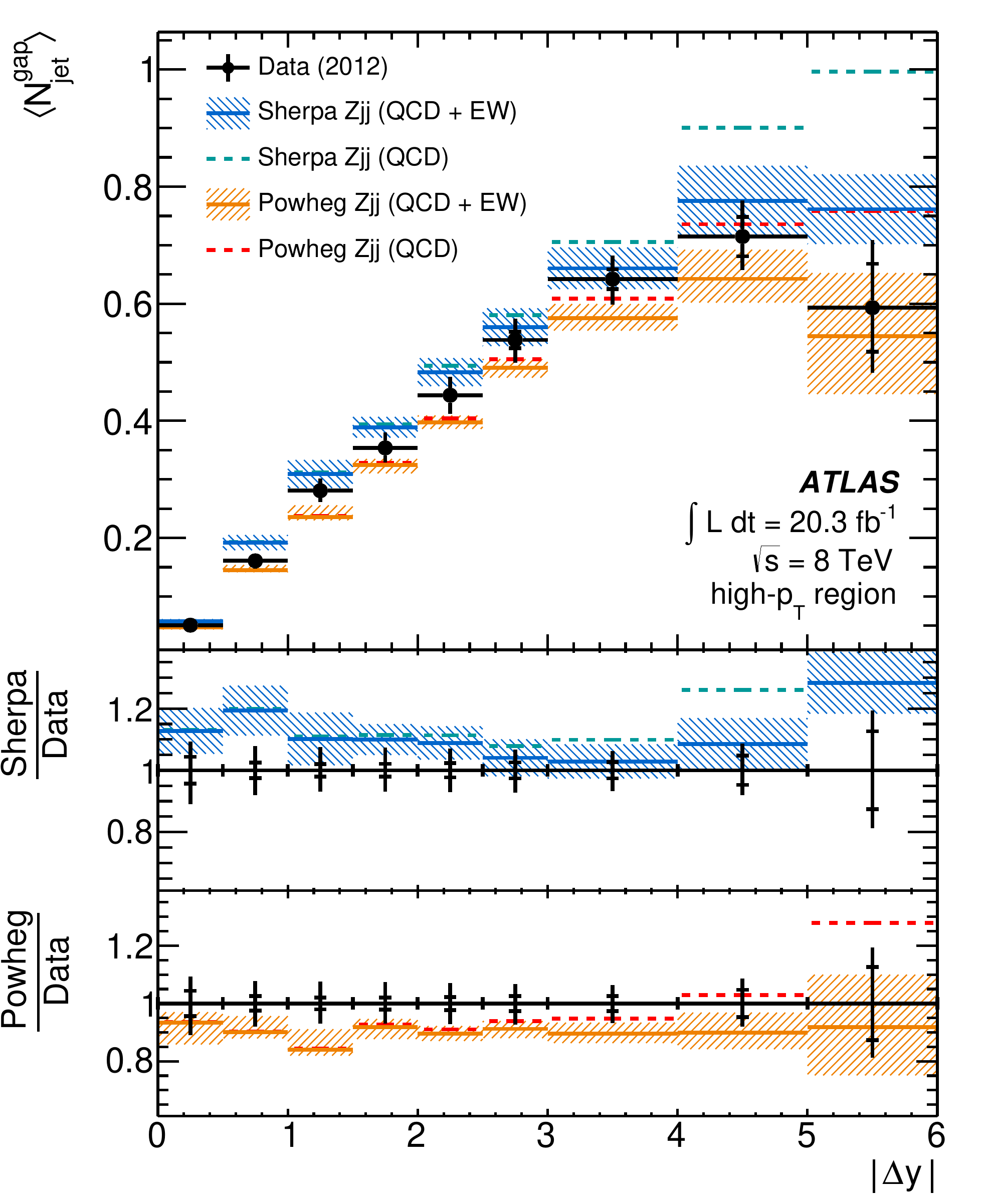}
          }
      \caption[]{Unfolded \gapfrac\ versus (a)  \mjj\  and (b) \deltay\ 
      in the \highpt\ region. Unfolded \avgnjet\ versus (c)  \mjj\  and (d) \deltay\ in the \highpt\  region.  \resultCaptionTwo} 
    \label{fig:unfolding-GapFraction-Final-aux}
  \end{center}
\end{figure}

\begin{figure}
  \begin{center}
   \graphicspath{{figures/unfolding/}}
    \subfigure[] {
      \includegraphics[width=0.47\textwidth]{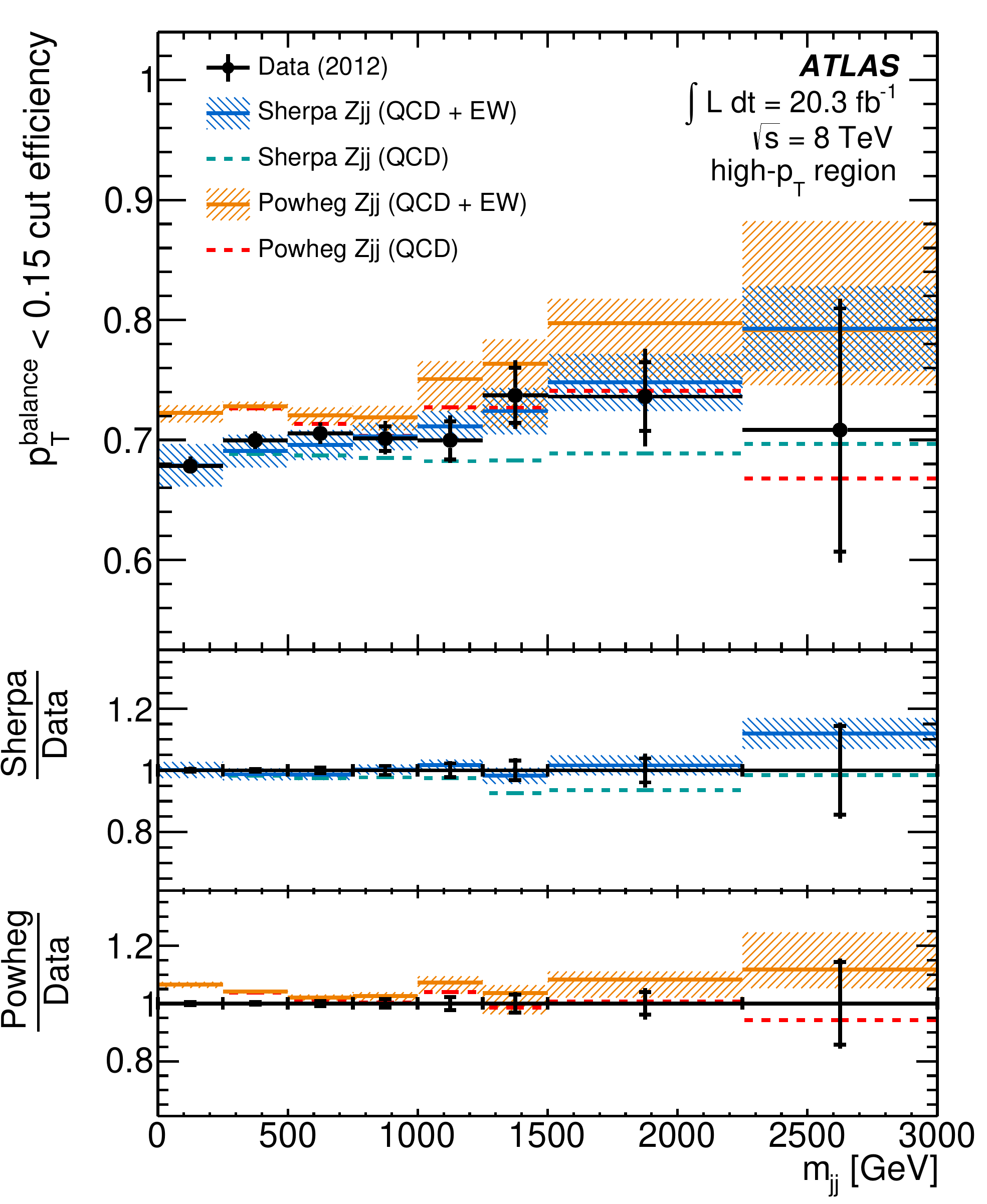}\quad
    }
    \subfigure[] {
      \includegraphics[width=0.47\textwidth]{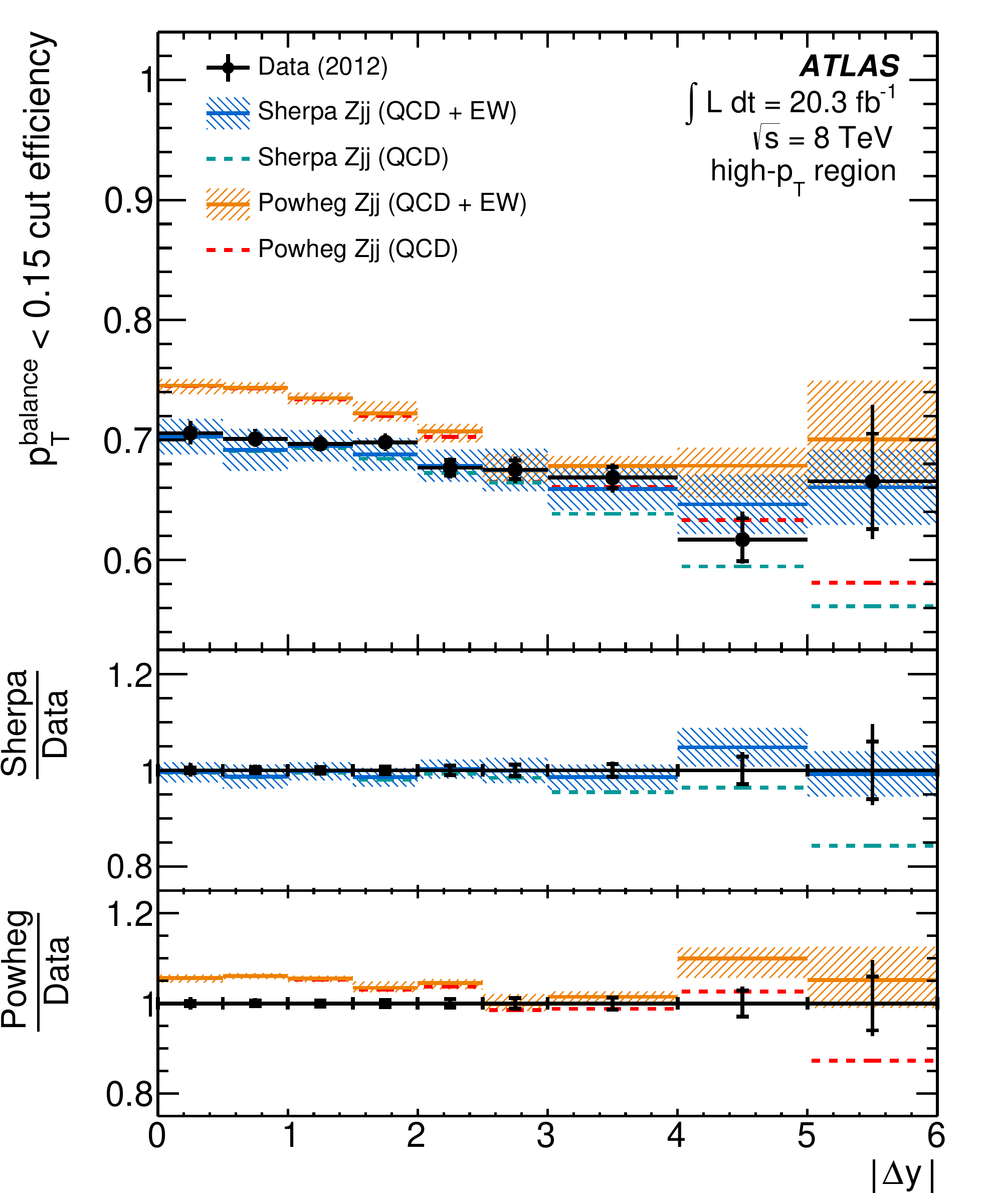}
    }
    \caption[]{Unfolded \ptsceff\ versus (a) \mjj\ and (b) \deltay\
      in the  \highpt\  region. \resultCaptionTwo} 
    \label{fig:unfolding-PtBalEff-Final-aux}
  \end{center}
\end{figure}

\clearpage

\bibliographystyle{JHEP}
\providecommand{\href}[2]{#2}\begingroup\raggedright\endgroup

\clearpage
\begin{flushleft}
{\Large The ATLAS Collaboration}

\bigskip

G.~Aad$^{\rm 84}$,
T.~Abajyan$^{\rm 21}$,
B.~Abbott$^{\rm 112}$,
J.~Abdallah$^{\rm 152}$,
S.~Abdel~Khalek$^{\rm 116}$,
O.~Abdinov$^{\rm 11}$,
R.~Aben$^{\rm 106}$,
B.~Abi$^{\rm 113}$,
M.~Abolins$^{\rm 89}$,
O.S.~AbouZeid$^{\rm 159}$
% ...
\bigskip
\\
% ...
$^{11}$ Institute of Physics, Azerbaijan Academy of Sciences, Baku, Azerbaijan\\
% ...
$^{21}$ Physikalisches Institut, University of Bonn, Bonn, Germany\\
% ...
$^{84}$ CPPM, Aix-Marseille Universit{\'e} and CNRS/IN2P3, Marseille, France\\
% ...
$^{89}$ Department of Physics and Astronomy, Michigan State University, East Lansing MI, United States of America\\
% ...
$^{106}$ Nikhef National Institute for Subatomic Physics and University of Amsterdam, Amsterdam, Netherlands\\
% ...
$^{112}$ Homer L. Dodge Department of Physics and Astronomy, University of Oklahoma, Norman OK, United States of America\\
$^{113}$ Department of Physics, Oklahoma State University, Stillwater OK, United States of America\\
% ...
$^{116}$ LAL, Universit{\'e} Paris-Sud and CNRS/IN2P3, Orsay, France\\
% ...
$^{152}$ Institute of Physics, Academia Sinica, Taipei, Taiwan\\
% ...
$^{159}$ Department of Physics, University of Toronto, Toronto ON, Canada\\
% ...
\end{flushleft}



\end{document}